\documentclass[PAPER, atlasdraft=false, cernpreprint, texlive=2023, UKenglish, texmf, orcidlogo]{atlasdoc}
 
\usepackage{atlaspackage}
\usepackage{atlasbiblatex}

\newcommand{\PbPb}{\mbox{Pb+Pb}}
\newcommand{\pbpb}{\mbox{Pb+Pb}}
\newcommand{\dR}{\mbox{$\Delta R$}}
\newcommand{\dNdxJ}{\mbox{$\frac{1}{N_{\textrm{pair}}}\frac{dN_{\textrm{pair}}}{dx_{\mathrm{J}}}$}}
\newcommand{\xJ}{\mbox{$x_{\mathrm{J}}$}}
\newcommand{\xj}{\mbox{$x_{\mathrm{J}}$}}
\newcommand{\ptone}{\mbox{$p_{\textrm{T,1}}$}}
\newcommand{\pttwo}{\mbox{$p_{\textrm{T,2}}$}}
\newcommand{\pta}{\mbox{$p_{\textrm{T,1}}$}}
\newcommand{\ptb}{\mbox{$p_{\textrm{T,2}}$}}

\newcommand{\ptlead}{\mbox{$p_{\mathrm{T,leading}}$}}

\newcommand{\Ncoll}{\mbox{$N_{\mathrm{coll}}$}}
\newcommand{\Npair}{\mbox{$N_{\mathrm{pair}}$}}

\newcommand{\Njet}{\mbox{$N_{\mathrm{jet}}$}}

\newcommand{\Npart}{\mbox{$N_{\mathrm{part}}$}}

\newcommand{\sqrtsnn}{\mbox{$\sqrt{s_{_\text{NN}}}$}}
\newcommand{\sqrts}{\mbox{$\sqrt{s}$}}

\newcommand{\pythiaeight}{\textsc{Pythia}\,8}      

\newcommand{\glauber}{\textsc{TGlauberMC} v3.2}

\newcommand{\geantfour}{\textsc{Geant}4}
\newcommand{\herwig}{{\textsc Herwig++}}

\newcommand{\RAA}{\mbox{$R_{\mathrm{AA}}$}}
\newcommand{\RAApair}{\mbox{$R_{\mathrm{AA}}^{\mathrm{pair}}$}}
\newcommand{\RAAlead}{\mbox{$R_{\mathrm{AA}}^{\mathrm{pair}}(\pta)$}}
\newcommand{\RAAsublead}{\mbox{$R_{\mathrm{AA}}^{\mathrm{pair}}(\ptb)$}}

\newcommand{\RFour}{\mbox{$R = 0.4$}}

\newcommand{\ETfcal}{\mbox{$\Sigma E_{\mathrm{T}}^{\mathrm{FCal}}$}}

\newcommand{\invnb}{\mbox{${\mathrm{nb}^{-1}}$}}
\newcommand{\invpb}{\mbox{${\mathrm{pb}^{-1}}$}}

\newcommand{\pTjet}{\mbox{$p_{{\mathrm{T}}}$}}

\newcommand{\pTtrk}{\mbox{$p_{{\mathrm{T}}}^{\mathrm{ch}}$}}

\newcommand{\pTtrue}{\mbox{$p_{\mathrm{T}}^{\mathrm{truth}}$}}

\newcommand{\ETjet}{\mbox{$\ET^{\mathrm{jet}}$}}

\newcommand{\ptjet}{\mbox{$p_{\mathrm{T}}^{\mathrm{jet}}$}}

\newcommand{\pttruth}{\mbox{$p_{\mathrm{T}}^{\mathrm{truth}}$}}

\newcommand{\ptjettruth}{\mbox{$p_{{\mathrm{T}}}^{\mathrm{truth}}$}}
\newcommand{\ptjettrutha}{\mbox{$p_{{\mathrm{T,1}}}^{\mathrm{truth}}$}}
\newcommand{\ptjettruthb}{\mbox{$p_{{\mathrm{T,2}}}^{\mathrm{truth}}$}}
\newcommand{\ptjetreco}{\mbox{$p_{{\mathrm{T}}}^{\mathrm{reco}}$}}
\newcommand{\ptjetrecoa}{\mbox{$p_{{\mathrm{T,1}}}^{\mathrm{reco}}$}}
\newcommand{\ptjetrecob}{\mbox{$p_{{\mathrm{T,2}}}^{\mathrm{reco}}$}}

\newcommand*{\TAA}{\ensuremath{T_{\text{AA}}}\xspace}
\newcommand*{\avgTAA}{\ensuremath{\langle\TAA\rangle}\xspace}

\usepackage{atlasphysics}

\graphicspath{{logos/}{figures/}}

\AtlasTitle{Measurements of the suppression and correlations of dijets in \PbPb\ collisions at $\sqrtsnn = 5.02$~\TeV}
 
\AtlasVersion{2.2}
\AtlasAbstract{
Studies of the correlations of the two highest transverse momentum (leading) jets in individual \pbpb\ collision events can provide information about the mechanism of jet quenching by the hot and dense matter created in such collisions.
In \pbpb\ and \pp\ collisions at  $\sqrtsnn = 5.02$~\TeV, measurements of the leading dijet transverse momentum (\pT) correlations are presented.
Additionally, measurements in \pbpb\ collisions of the dijet pair nuclear modification factors
projected along leading and subleading jet \pT\ are made.
The measurements are performed using the ATLAS detector at the LHC with 260~\invpb\
of \pp\ data collected in 2017
and 2.2~\invnb\ of \pbpb\ data collected in 2015 and 2018.
An unfolding procedure is applied to the two-dimensional leading and subleading jet \pt\ distributions to account for experimental effects in the measurement of both jets.
Results are provided for dijets with leading jet \pt\ greater than 100~\GeV.
Measurements of the dijet-yield-normalized \xJ\ distributions in \pbpb\
collisions show an increased fraction of imbalanced jets compared to \pp\ collisions;
these measurements are in agreement with previous measurements of the same quantity
at 2.76~\TeV\ in the overlapping kinematic range.
Measurements of the absolutely-normalized dijet rate in \pbpb\ and \pp\ collisions are
also presented, and show that balanced dijets are significantly
more suppressed than imbalanced dijets in \pbpb\ collisions.
It is observed in the measurements of the pair nuclear modification
factors that the subleading jets are significantly suppressed
relative to leading jets with \pt\
between 100 and 316~\GeV\ for all centralities in \pbpb\ collisions.
}
 
\author{The ATLAS Collaboration}
\PreprintIdNumber{CERN-EP-2022-046}
 
\AtlasRefCode{HION-2019-02}
\AtlasJournal{PRC}
\AtlasJournalRef{Phys. Rev. C. 107 (2023) 054908}
\AtlasDOI{10.1103/PhysRevC.107.054908}

\hypersetup{pdftitle={ATLAS document},pdfauthor={The ATLAS Collaboration}}
 
\begin{document}
 
\maketitle

\section{Introduction}
\label{sec:intro}
 
The principal physics aim of the heavy-ion program at the Large Hadron
Collider (LHC) is to produce and measure the properties of the
quark--gluon plasma (QGP). In quantum chromodynamics (QCD), the QGP
is a high-temperature state of matter in which quarks and gluons are no
longer confined in color-neutral hadrons such as protons and
neutrons (for a recent review, see Ref.~\cite{Busza:2018rrf}).  In order
to understand the properties of the QGP at short distances,
high transverse momenta (\pt) probes such as jets of correlated
particles are used.
 
The overall rate of
jets in a given centrality\footnote{Centrality characterizes the degree to which
the nuclei overlap.  The most central collisions have full overlap and the
highest multiplicities, while the most peripheral collisions have only
minimal overlap and have multiplicity closer to that in \pp\ collisions at
the same nucleon--nucleon collision energy.} interval in Pb+Pb collisions at a given \pt\ is
characterized by the nuclear modification factor,
\begin{equation}
\RAA = \left. \frac{1}{N_{\textrm{evt}}} \frac{dN_{\text{jet}}}{d\pt} \middle/
\Big( \right. \avgTAA \frac{d\sigma_{\pp}}{d\pT}\Big),
\label{eq:RAA}
\end{equation}
where $N_{\text{jet}}$ and $\sigma_{\pp}$ are the jet yield in \pbpb\ collisions and the jet cross-section
in \pp\ collisions,
respectively, measured as a function of the jet \pt, and where $N_{\textrm{evt}}$ is
the total number of \pbpb\ events  within the
chosen centrality interval. Additionally, \avgTAA
is the mean nuclear thickness function~\cite{Miller:2007ri} for the centrality interval.
The \RAA\ value is one in the absence of nuclear effects, however in the most central \pbpb\ collisions,
\RAA\ is observed to be about a factor of two lower than that in
\pp\ collisions scaled by \avgTAA, up to a \pt of approximately
1~\TeV~\cite{ATLAS:2018gwx,CMS-HIN-18-014,ALICE:2019qyj}.  This suppression,
referred to as \textit{jet quenching}, is generally explained by the interplay of
radiative and collisional energy loss in reducing the jet \pt\ by
moving energy associated with the initial parton to wider angles, with some of
it being outside the jet cone.
Given the  steeply falling jet \pt\ spectrum, this  leads to fewer
jets being measured at a given jet \pt, giving rise to \RAA\ values below one.
For  recent reviews, see Refs.~\cite{Cao:2020wlm,Cunqueiro:2021wls}.

Jets are mostly produced in pairs in 2~$\to$~2 scattering processes.
The QCD evolution of the initial partons gives rise to back-to-back jets, referred to here as ``dijets''.
Compared to single-jet suppression, dijets in \pbpb\ collisions provide a cleaner probe for
studying jet quenching.
The two jets are expected to experience asymmetric energy loss due to traversing unequal path lengths
in the QGP~\cite{Qin:2010mn}, driven by the geometry of the overlapping nuclei and the relative
orientation of the jet trajectories through the evolving medium.
Measurements of the azimuthal anisotropy of jets~\cite{HION-2020-09}
have already shown that the geometry of the overlapping nuclei affects the relative rates
of jets measured in \pbpb\ collisions.
Additionally, jets are expected to experience jet-by-jet fluctuations in the energy-loss process~\cite{Milhano:2015mng}.
The measurement of the \pt\ balance of dijets provides a way to constrain the relative importance of
fluctuations and geometry in jet quenching.
 
In order to compare the final transverse momenta of the two jets which define a dijet, the leading
dijet momentum balance:
\begin{equation}
\xJ \equiv \pttwo/\ptone
\label{eq:xJ}
\end{equation}
is measured.
Here, the leading dijet is constructed using the highest two \pT\ jets out of the set of jets in an event, \pta\ is the
transverse momentum of the highest-\pt\
(leading) jet, and \pttwo\ is
the transverse momentum of the second-highest-\pt\
(subleading) jet.

In \pp\ collisions, the showering process in vacuum (in the absence of a significant QGP),
as well as higher-order scattering processes,
can lead to imbalanced dijet transverse momenta.
However, the most probable situation is that the jets are nearly balanced in \pt~\cite{ATLAS:2017xfa}.
Previous dijet measurements in $\pbpb$ collisions have observed that
jets are more likely to have a larger momentum imbalance in
$\pbpb$ collisions than in
$\pp$ collisions~\cite{ATLAS:2010isq,ATLAS:2017xfa,CMS:2011iwn}.

Previous publications reported only the dijet imbalance
normalized by the measured dijet yields, in order to study the changes in the shape
of the \xJ\ distribution as a function of the heavy-ion collision centrality.
Past studies did not address the absolute rate at which dijets are produced
in \pbpb\ collisions, which could assess whether leading dijets are suppressed at levels similar to those for
``inclusive jets'' (all jets in the event which meet \pT\ and rapidity selections)~\cite{ATLAS:2018gwx}.
This paper thus presents both types of measurements with leading dijets in \pbpb\ and
\pp\ collisions at $\sqrtsnn = 5.02$~\TeV.  These measurements use 2.2~\invnb\ of
\pbpb\ collisions collected in 2015 and 2018 as well as 260~\invpb\ of
\pp\ data collected in 2017 with the ATLAS detector~\cite{PERF-2007-01} at the LHC.
 
Jets are reconstructed using the \antikt\ algorithm~\cite{Cacciari:2008gp}
with radius parameter \RFour.  Leading dijets are constructed from the two highest-\pT\ jets in the event and are
required to have the two jets nearly back-to-back in azimuth with
$\Delta\phi \equiv |\phi_{1} - \phi_{2}| \geq 7\pi/8$ and $|y| < 2.1$.\footnote{ATLAS uses a right-handed
coordinate system with its origin at the nominal interaction point (IP) in the center of the detector, and the $z$-axis along the beam pipe.
The $x$-axis points from the IP to the center of the LHC ring, and the $y$-axis points upward.
Cylindrical coordinates $(r,\phi)$ are used in the transverse plane, $\phi$ being the azimuthal angle around the $z$-axis.
The pseudorapidity is defined in terms of the polar angle $\theta$ as $\eta=-\ln\tan(\theta/2)$.
The rapidity is defined as $y = 0.5\ln[(E + p_z)/(E-p_z)]$ where $E$ and $p_z$ are the energy and $z$-component of the momentum along the beam direction, respectively.
Transverse momentum and transverse energy are defined as $\pt = p \sin\theta$ and $\ET = E \sin\theta$, respectively.
The angular distance between two objects with relative differences $\Delta \eta$ in pseudorapidity and $\Delta \phi$ in azimuth is given by $\Delta R = \sqrt{(\Delta \eta )^2 + (\Delta \phi)^2}$.}
The measurement is performed in logarithmic \pt\ bins from 100 to 562~\GeV\ for leading jets
and above 32~\GeV\ for subleading jets.
Events whose two highest-\pT\ jets do not meet the selection criteria are discarded.
 
The primary observable
for this measurement is the two-dimensional yield of leading dijets ($\Npair$) meeting the selection criteria:
\begin{equation}
\label{eq:jointdist}
\frac{d^2 \Npair}{d\ptone d\pttwo}.
\end{equation}
Projections of the two-dimensional distribution in Exp.~(\ref{eq:jointdist}) can be used to construct $\xJ$
distributions as a function of $\ptone$ and $\pttwo$.
The \xJ\ values, as defined in Eq.~(\ref{eq:xJ}),  are reported for 0.32~$<\xJ<$~1.0
for selections in \pta.
The minimum \xJ\ value is chosen to be consistent with previous measurements~\cite{ATLAS:2017xfa}.
The \xJ\ distributions are presented with two
normalizations.  The first uses the dijet yield normalization to extract the \textit{dijet-yield-normalized}
\xJ\ distributions:
\begin{equation}
\frac{1}{\Npair}\frac{d\Npair}{d\xJ }.
\label{eq:selfxj}
\end{equation}
This normalization, Exp.~(\ref{eq:selfxj}), has been used in previous dijet
measurements by ATLAS and CMS~\cite{ATLAS:2010isq,CMS:2011iwn,ATLAS:2017xfa}.
The second normalization provides \textit{absolutely normalized} \xJ\ distributions:
\begin{equation}
\frac{1}{\avgTAA N_{\mathrm{evt}}^{\mathrm{AA}}} \frac{dN_{\text{pair}}^{ \mathrm{AA}}}{d\xJ}
\label{eq:absxjpbpb}
\end{equation}
in \pbpb\ collisions and
\begin{equation}
\frac{1}{L_{\pp}}\frac{dN_{\text{pair}}^{\pp}}{d\xJ}
\label{eq:absxjpp}
\end{equation}
in \pp\ collisions. Here \avgTAA~~\cite{Miller:2007ri} is the mean nuclear thickness
function
for the events under consideration, $\Nevt^{\mathrm{AA}}$ is the number
of minimum-bias events, and $L_{\pp}$ is the integrated luminosity of the \pp\ collisions~\cite{ATLAS-CONF-2019-021}.
The absolutely normalized \xJ\ distributions allow
a direct comparison between the dijet rates measured in \pbpb\ and \pp\ collisions.
Additionally, the absolutely normalized \xJ\ distributions can be integrated over the measurement range of 0.32~$<\xJ<$~1.0
to construct the pair nuclear modification factors for dijets as a function
of the leading and subleading jet \pt.
These quantities are defined analogously to the nuclear modification
factor for inclusive jets, Eq.~(\ref{eq:RAA}), as
\begin{equation}
\RAAlead = \frac{\frac{1}{\langle \TAA \rangle N_{\mathrm{evt}}^{\mathrm{AA}}}
\int_{0.32\times\pta}^{\pta}\frac{d^{2}N_{\text{pair}}^{ \mathrm{AA}}}{d\pta d\ptb} d\ptb}{
\frac{1}{L_{pp}}
\int_{0.32\times \pta}^{\pta}\frac{d^2N_{\text{pair}}^{\pp}}{d\pta d\ptb} d\ptb}
\label{eq:raa_lead}
\end{equation}
and
\begin{equation}
\RAAsublead = \frac{\frac{1}{\langle \TAA \rangle N_{\mathrm{evt}}^{\mathrm{AA}}}
\int_{\ptb}^{\ptb/0.32}\frac{d^{2}N_{\text{pair}}^{ \mathrm{AA}}}{d\pta d\ptb} d\pta}{
\frac{1}{L_{pp}}
\int_{\ptb}^{\ptb/0.32}\frac{d^2N_{\text{pair}}^{\pp}}{d\pta d\ptb} d\pta}.
\label{eq:raa_sublead}
\end{equation}
By integrating over \ptb\ (\pta), one can access information from \RAAlead\ (\RAAsublead) about the
differential rate of dijet production in leading (subleading) jet \pT\ bins.
Comparison of these two quantities
at a fixed jet \pt\ provides information about the relative suppression of leading and subleading
jets.


\section{ATLAS detector}
\label{sec:detector}

The ATLAS detector \cite{PERF-2007-01} at the LHC covers nearly the full solid angle around the nominal interaction point. It contains an inner tracking detector surrounded by a thin superconducting solenoid, electromagnetic and hadronic calorimeters, a zero-degree calorimeter, and
a muon spectrometer which incorporates three large superconducting toroidal magnets. The inner-detector system is immersed in a 2~T axial magnetic field and provides charged-particle tracking within $|\eta| < 2.5$.
 
The ATLAS calorimeter system covers the pseudorapidity range $|\eta| < 4.9$. Within the region $|\eta| < 3.2$, electromagnetic calorimetry is provided by barrel and endcap high-granularity lead/liquid-argon (LAr) calorimeters, with an additional thin LAr presampler covering $|\eta| < 1.8$ to allow corrections for energy lost in material upstream of the calorimeters. Hadronic calorimetry is provided by a steel/scintillator-tile calorimeter which is segmented into three barrel structures within |$\eta| < 1.7$, as well as two copper/LAr hadronic endcap calorimeters. To complete the solid angle coverage, forward copper/LAr and tungsten/LAr calorimeter modules
(FCal) are used, optimized for electromagnetic and hadronic measurements respectively.
 
The zero-degree calorimeters (ZDCs) are located symmetrically at $z = \pm 140$~m and cover $|\eta| > 8.3$. They are constructed using tungsten absorber plates with quartz fibers to transmit the Cherenkov light. In \pbpb\ collisions the ZDCs primarily observe ``spectator'' neutrons, which do not interact during the primary collision.
 
A two-level trigger system is used within ATLAS to select events of interest \cite{ATLAS:2020esi}. The first-level trigger is implemented in hardware and uses a subset of the detector information to reduce the event rate to at most 100~kHz. This is followed by the software-based high-level trigger, which further reduces the event rate to several kHz of triggered events that are recorded to permanent storage.
 
An extensive software suite~\cite{ATL-SOFT-PUB-2021-001} is used in the reconstruction and analysis of real and simulated data, in detector operations, and in the trigger and data acquisition systems of the experiment.


\section{Data and Monte Carlo selection}
\label{sec:datamcsample}
 
The \pbpb\ data used to perform these measurements were collected in 2015 and 2018, and the \pp\ data used were collected in 2017 with the average number of inelastic interactions per bunch crossing ranging from 1.4 to 4.4. The events were triggered using a series of various \pt-threshold
single-jet triggers with the leading jets in the analysis being in a region where the trigger is fully efficient.
Although only a small fraction of the \pbpb\ events ($<0.5\%$) contain multiple collisions, these were suppressed utilizing the observed anti-correlation, expected from the nuclear geometry, between the total transverse energy deposited in both of the forward calorimeters, \ETfcal, and the energy in the ZDC, which is proportional to the number of observed spectator neutrons.
Pileup collisions are not rejected in \pp\ collisions.
 
The overlap area of the two colliding nuclei in \pbpb\ collisions is
characterized by the event centrality, which is estimated from the
total transverse energy deposited in the
FCal~\cite{ATLAS:2017rij}. This measurement considers five centrality
intervals as defined according to successive percentiles of the
\ETfcal\ distribution obtained from minimum-bias collisions. The
centrality intervals considered in this measurement are 0--10\%
(largest \ETfcal), 10--20\%, 20--40\%, 40--60\%, and 60--80\%
(smallest \ETfcal). The values of the mean nuclear thickness function,
\avgTAA~\cite{Miller:2007ri}, are determined using the \glauber~\cite{Loizides:2017ack}; the uncertainties in \avgTAA\ are
discussed in Ref.~\cite{ATLAS:2019ibd}.  The \avgTAA\ values and
their uncertainties are listed in Table~\ref{tab:lumi} for each centrality
selection considered in this measurement.
 
\begin{table}[ht]
\centering
\caption{\label{tab:lumi} The  \avgTAA\ values and uncertainties for the centrality
selections used in this measurement. These are the results from \glauber\ modeling of the summed transverse energy in the forward calorimeters, \ETfcal.}
\vspace{0.2cm}
\begin{tabular}{|c|c|}
\hline
Centrality selection & \avgTAA $\pm~\delta\avgTAA $ [1/mb]\\
\hline
~~0--10\% &  23.35  $\pm$ 0.20~~ \\
10--20\% & 14.33 $\pm$ 0.17~~ \\
20--40\% & 6.79  $\pm$ 0.16 \\
40--60\% & 1.96  $\pm$ 0.09 \\
60--80\% & 0.39 $\pm$ 0.03 \\
\hline
\end{tabular}
\end{table}
 
This analysis uses four Monte Carlo (MC) samples to evaluate the detector performance and correct for detector effects.
The \pp\ MC sample used in this analysis includes $3.2 \times 10^7$ \pythiaeight~\cite{SJOSTRAND2015159} \pp\ jet events generated at $\sqrts = 5.02$~\TeV\ with parameter values set according to the A14 tune~\cite{ATL-PHYS-PUB-2014-021} and the \textsc{NNPDF23lo} parton distribution functions
(PDFs)~\cite{Ball:2012cx}. Pileup due to additional inelastic \pp\ interactions is similarly generated using \pythiaeight\ with the same PDFs and utilizing the A3 tune \cite{ATL-PHYS-PUB-2016-017}, tuned for inclusive QCD processes, matching the number of extra collisions in the \pp\ data. The MC samples for \pbpb\ collisions, one for the 2015 detector conditions and one for the 2018 detector conditions, each use $3.2 \times 10^7$ \pp\ \pythiaeight\ events with the same A14 tune and PDFs as used for the generation of the \pp\ jet MC samples.
The underlying event contribution to the detector signal is accounted for by overlaying the simulated \pp\ events with dedicated \pbpb\ data events from the 2015 \pbpb\ run and 2018 \pbpb\ run respectively.
The data events from \pbpb\ collisions were combined with the signal from the \pythiaeight\
simulation of hard scattering events at the digitization stage, and then reconstructed as a combined event.
This procedure enables the ``data overlay'' sample to accurately reproduce the effects of the underlying event on the jet response.
This sample was reweighted on an event-by-event basis to ensure the same centrality distribution as is measured in the jet-triggered data samples. The detector response
in all three MC samples was simulated utilizing
\geantfour~\cite{GEANT4:2002zbu,ATLAS:2010arf}.
Finally, \pp\ \herwig~\cite{Bahr:2008pv} events
using the UEEE5 tune~\cite{Gieseke:2012ft} and the
CTEQ6L1 PDFs~\cite{Pumplin:2002vw} are used for uncertainty studies.


\section{Jet reconstruction and performance}
The jet reconstruction procedures follow those used by ATLAS for previous jet measurements in \PbPb\ collisions~\cite{ATLAS:2018gwx,HION-2020-09}.
Jets are reconstructed using
the \antikt\ algorithm~\cite{Cacciari:2008gp} implemented in the FastJet software package~\cite{Cacciari:2011ma}. In both \pp\ and \pbpb\ collisions, jets with
$R=0.2$ and \RFour\ are formed by
clustering calorimetric towers of spatial size $\Delta\eta \times\Delta\phi=0.1\times\pi/32$. The energies in the towers are obtained
by summing the energies of calorimeter cells at the electromagnetic energy scale~\cite{PERF-2011-03} within the tower boundaries.
In \pbpb\ collisions, a background
subtraction procedure is applied to estimate, within each event, the underlying event (UE) average transverse energy density,
$\rho(\eta,\phi)$, where the $\phi$ dependence is due to global azimuthal correlations in the particle production from the
hydrodynamic flow~\cite{HION-2016-06}. The modulation accounts for the contribution to the UE of the second-, third-, and fourth-order azimuthal anisotropy harmonics
characterized by values of flow coefficients $v_n^\mathrm{UE}$~\cite{HION-2016-06}.
Additionally, the
UE is also corrected for $\eta$- and $\phi$-dependent nonuniformities of the detector response
by correction factors derived in minimum-bias \PbPb\ data.
In \pp\ collisions, the same background subtraction procedure is applied to remove the pileup contribution to the jet,
but without the $\phi$-dependent modulation and without the correction for $\eta$- and $\phi$-dependent nonuniformities.
 
An iterative procedure is used to remove the impact of jets on the estimated $\rho$ and $v_n^\mathrm{UE}$
values.  The first estimate of the average transverse energy density of the UE, $\rho(\eta)$, is evaluated in 0.1 intervals of $\eta$, excluding towers
within $\Delta R =0.4 $ of ``seed'' jets. In the first subtraction step, the seeds are defined to be a combination of $R = 0.2$ jets and $R=0.4$ track-jets. Track-jets are
reconstructed by applying the \antikt\ algorithm with $R = 0.4$ to charged particles with $\pT > 4$~\GeV. The $R=0.2$ jets must pass a cut on the
minimum value of the tower $E_{\mathrm{T}}$  and on a ratio of maximum tower  $E_{\mathrm{T}}$ to
average tower  $E_{\mathrm{T}}$,
while the track-jets are required to have $\pT > 7$~\GeV. The underlying event contribution is then subtracted from each tower
constituent and the jet kinematics are recalculated. After the first iteration, the $\rho$ and $v_n$ values are updated by excluding from the UE determination
the regions within $\Delta{R} = 0.4$ of both the track-jets and the newly reconstructed $R=0.2$ jets with $\pt > 25$~\GeV\ (8~\GeV) in \pbpb\ (\pp) collisions. The updated
$\rho$ and $v_n^\mathrm{UE}$ values are used to update the jet kinematic properties in the second iteration.
Jet $\eta$- and $\pT$-dependent correction factors derived in simulations are applied to the measured jet
energy to correct for the calorimeter energy response~\cite{ATLAS:2014hvo,ATLAS:2015twa}. An additional correction based on in situ studies of jets recoiling
against photons and jets in other regions of the calorimeter is applied~\cite{HION-2018-06}. This calibration is followed by a
``cross-calibration'' which relates the jet energy scale (JES) of jets reconstructed by the procedure outlined in this section to the JES
in 13~\TeV\ \pp\ collisions~\cite{ATLAS:2015twa}.

\begin{figure}[t]
\centering
\subfloat{
\centering
\includegraphics[width=0.49\textwidth]{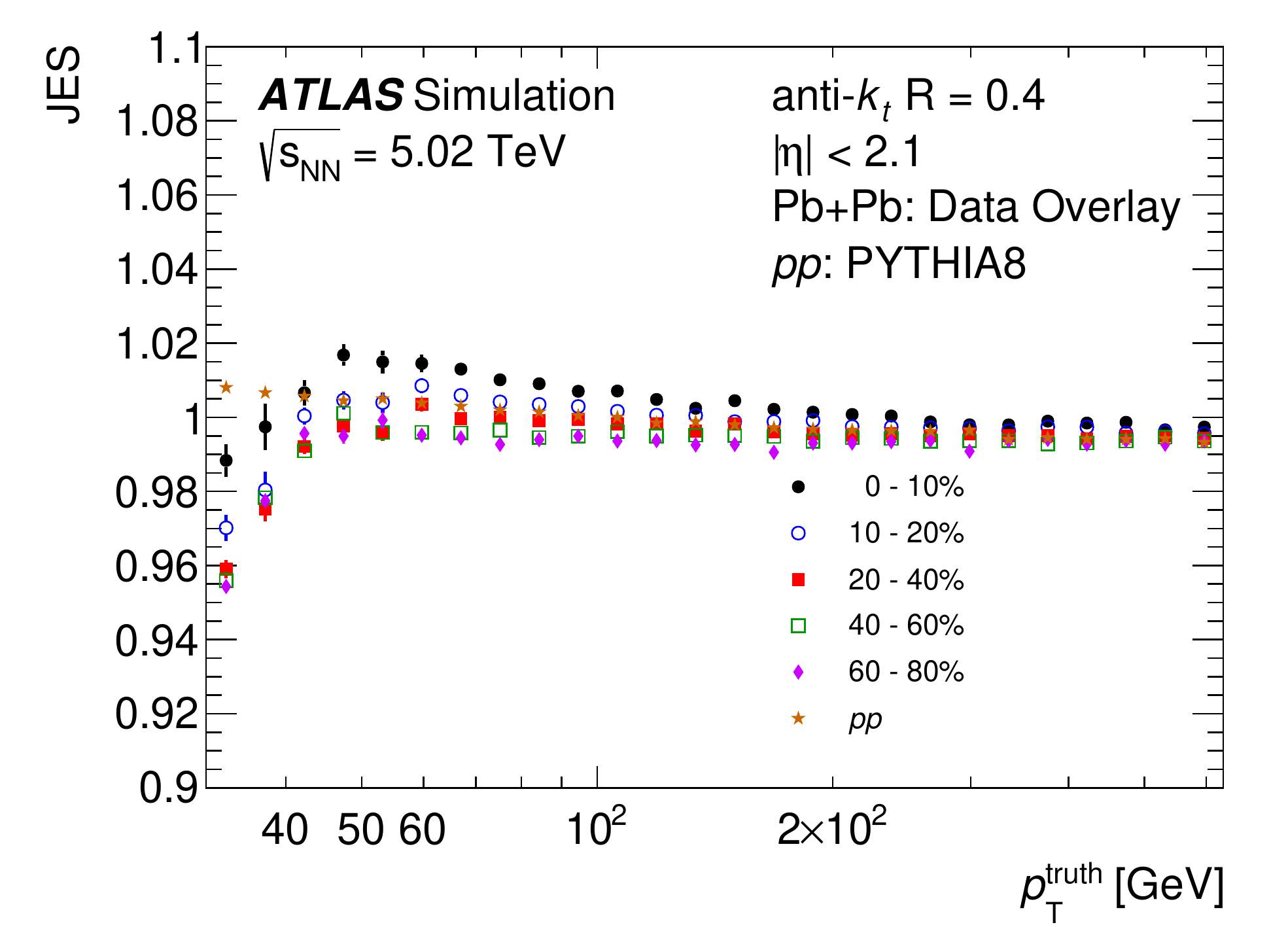}
}
\subfloat{
\centering
\includegraphics[width=0.49\textwidth]{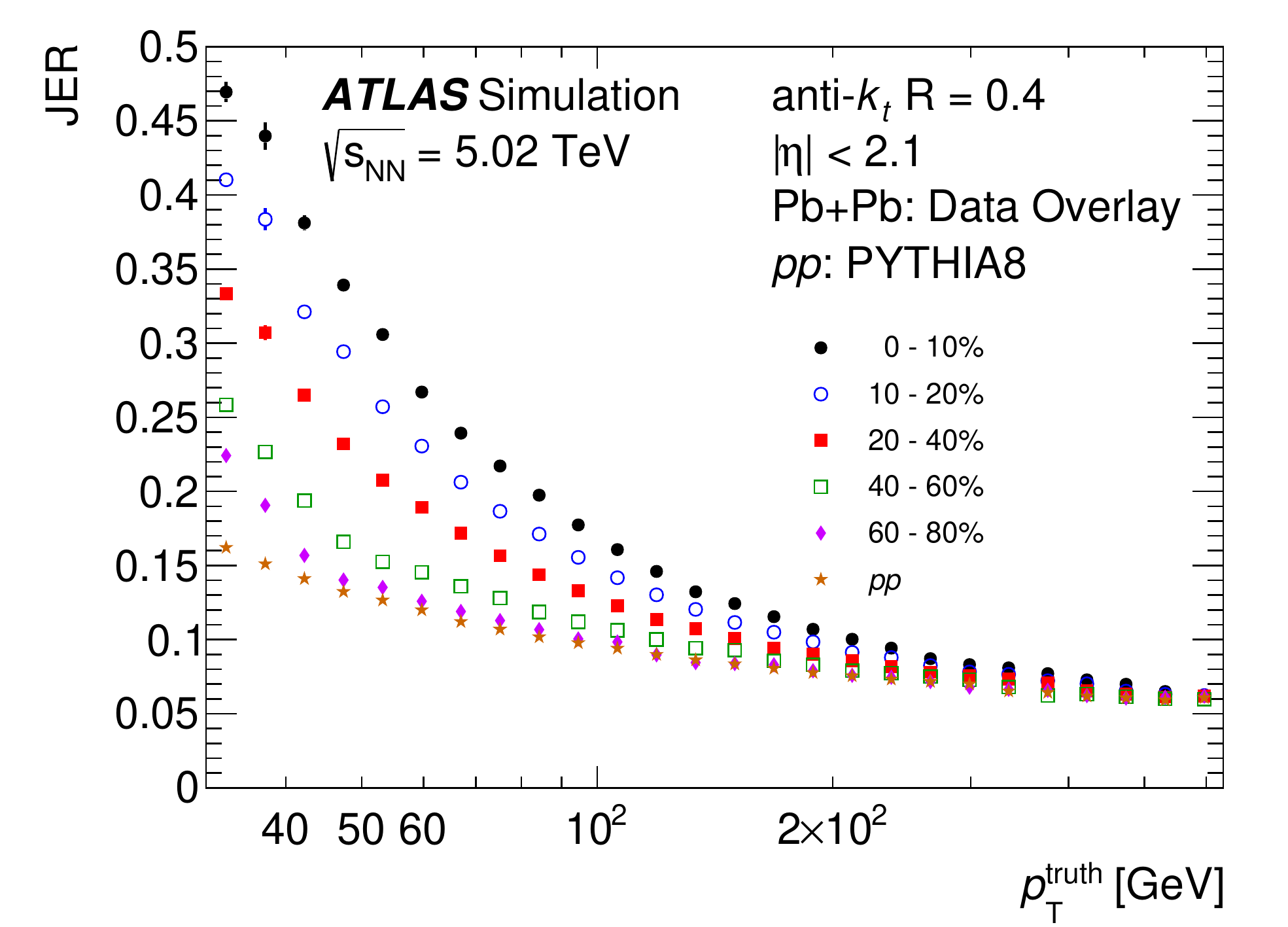}
}
\hfill
 
\caption{\label{fig:jesjer} The jet energy scale, JES, (left) and jet energy resolution, JER, (right) as a function of \pttruth\ for centrality selections in \pbpb\ and in \pp collisions.}
\end{figure}

\begin{figure}[t]
\centering
\subfloat{
\centering
\includegraphics[width=0.45\textwidth]{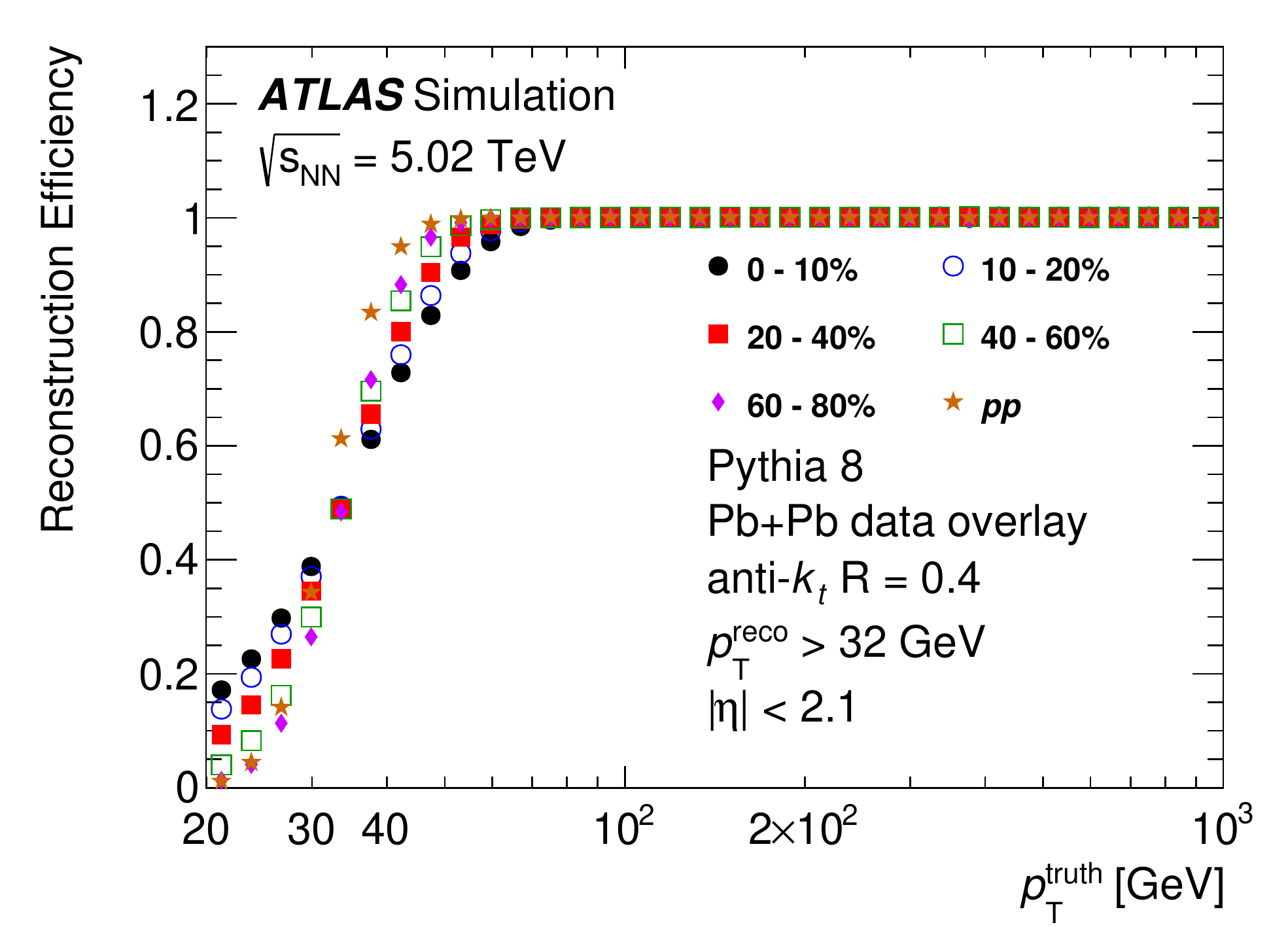}
}    \hfill
 
\caption{\label{fig:recoeff} The efficiency of reconstructing a jet with
$\ptjetreco > 32$~\GeV\ as a function of \ptjettruth\
for centrality selections in \pbpb\ collisions and \pp\ collisions. The statistical uncertainties shown are smaller than the point markers.}
\end{figure}

``Truth''-level jets are defined in the MC sample before detector simulation by applying the \antikt\
algorithm with $R = 0.4$
to stable particles with a proper lifetime greater than 30~ps, but excluding muons and neutrinos, which do not
leave significant energy deposits in the calorimeter.
After the detector simulation the truth jets are matched to the nearest reconstructed jet within $\dR <$~0.3.
The performance of the jet reconstruction is characterized by the JES and jet energy resolution (JER), which correspond to the mean and variance of the $\ptjetreco/\ptjettruth$ distribution,
where \ptjetreco\ is the reconstructed jet \pt\ and \ptjettruth\ is
the \pt\ of the matched truth-level jet. The JES and JER as a function of \pttruth\ can be seen in
Figure~\ref{fig:jesjer}.
The decrease in the JES for jets with $\ptjettruth < 50$~\GeV\ in \pbpb\ collisions stems
from jets for which there is an over subtraction of the underlying event. This is caused by the corresponding
region of the detector not containing a ``seed'' jet above 25~\GeV\ resulting in the jets energy contributing to the UE determination;
this is corrected for in the unfolding procedure discussed below.
The efficiency of reconstructing a jet with $\pt > 32$~\GeV, as evaluated from the probability of a
truth-jet matching to a reconstructed jet with $\pt > 32$~\GeV\ in the MC simulation,
can be seen as a function of \pttruth\ in Figure~\ref{fig:recoeff}.
 
\section{Data analysis}
 
The analysis and dijet selection used here closely follow those in Ref.~\cite{ATLAS:2017xfa}.
In each data event, the reconstructed leading dijet is constructed from the two highest-\ptjetreco\
jets in the event
with reconstructed $\ptjetrecoa > 79$~\GeV\ and $\ptjetrecob > 32$~\GeV,
and both jets are required to have $|y| < 2.1$.
The inclusion of leading jets down to $\ptjetrecoa > 79$~\GeV\ provides an underflow region for the unfolding
to enable inflow and outflow of jets from the measurement region.
These dijets are required to be back-to-back with $|\Delta\phi| > 7\pi/8$.
Events in which the leading dijets do not meet these criteria are discarded.
For dijets matching the selection criteria,
two-dimensional $(\ptjetrecoa,\ptjetrecob)$ distributions are constructed symmetrically across $\ptjetrecoa = \ptjetrecob$.
The distributions are
symmetrized to account for the possibility of swapping the leading and subleading jet
definition due to the finite JER.

\begin{figure}[t]
\centering
\subfloat{
\centering
\includegraphics[width=0.45\textwidth]{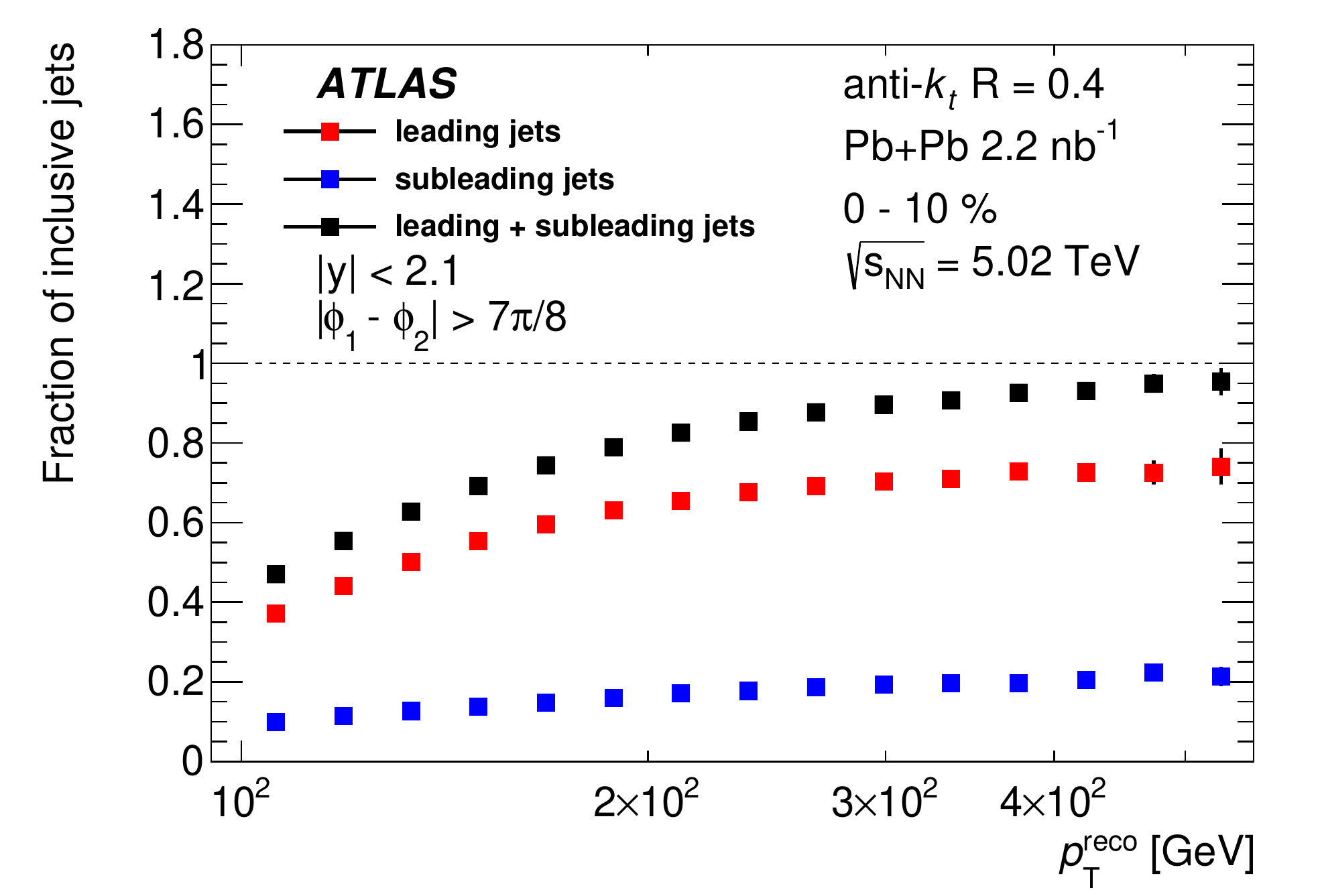}
}
\subfloat{
\centering
\includegraphics[width=0.45\textwidth]{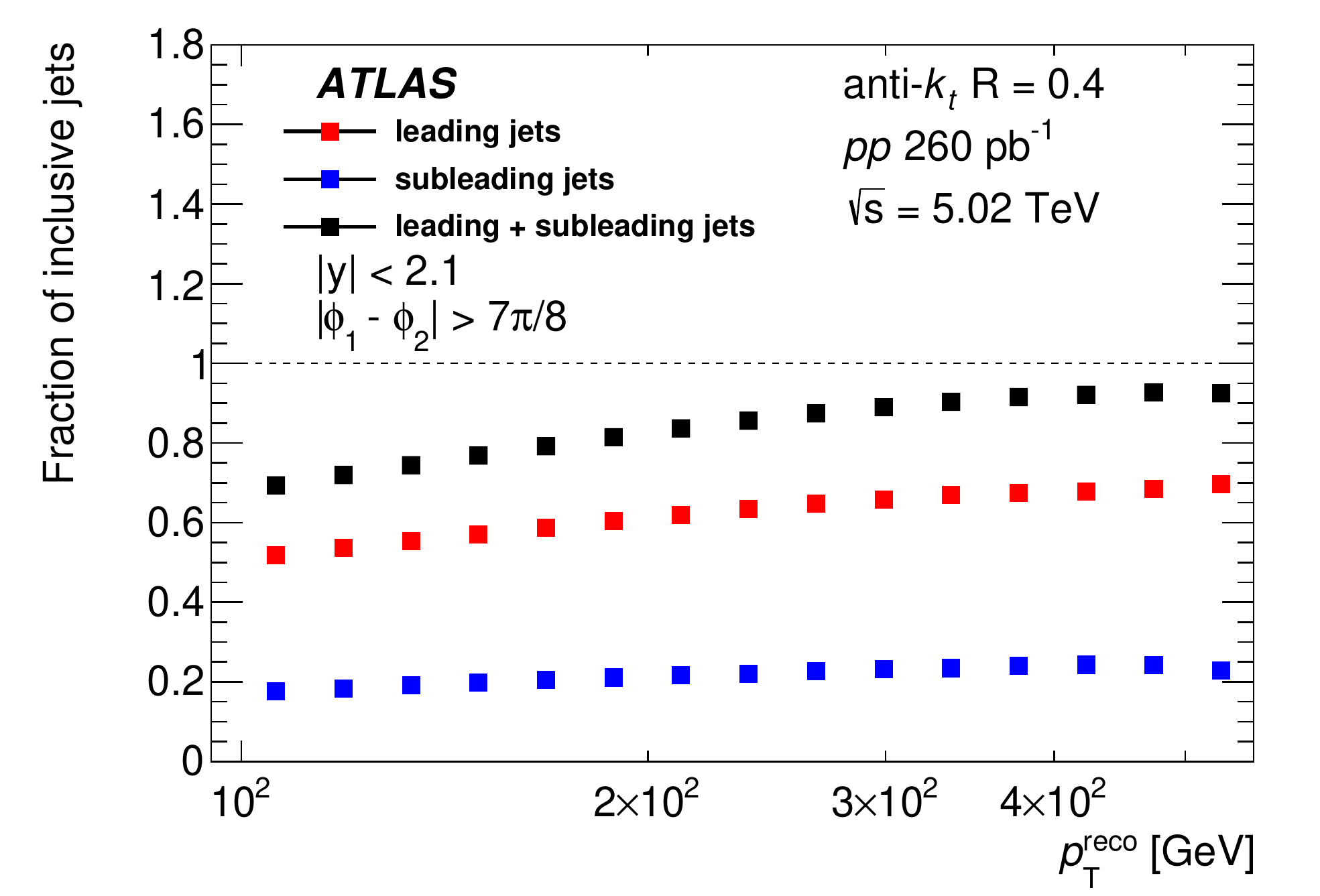}
}
\hfill
 
\caption{\label{fig:breakdown} Measured fractions of the inclusive jets which are
part of the leading dijet (black), the
leading jet of the dijet (red),
or subleading jet of the dijet (blue), all
as a function of \ptjetreco\ for 0--10\% centrality \pbpb\ collisions (left)
and \pp\ collisions (right).}
\end{figure}

The analysis procedure selects at most one dijet per event and therefore has
the potential to produce a biased subsample of the inclusive jets.
Figure~\ref{fig:breakdown} shows the fraction of the inclusive jets which are also included in the leading dijets used
in this analysis.
This quantity is shown as a function of \ptjetreco\, for both central \pbpb\ and \pp\ collisions.
The fraction is
47\% (69\%) at 100~\GeV\ and rises to more than 82\% (83\%) at 200~\GeV\ in central \pbpb\ collisions
(\pp\ collisions).
Due to the steep falloff of the jet yields with increasing \pt\ and the finite JER, jets reconstructed at a given \pt\ typically arise from
jets at a lower truth \pt, and thus the fractions shown here at the reconstructed-\pt scale
underestimate the true fraction of inclusive jets included at a given \pt.
Nevertheless, the large value, and weak \pT\ dependence, of the fraction of inclusive jets that are part of
the leading dijet imply that the dijet selection in this analysis does not produce a particularly biased sample.
The fractions are further broken down by leading and subleading jets. The relative fractions of leading and subleading jets at a given \pt\ are driven largely by the steeply-falling \pt\ spectrum convolved with the shape of the \xJ\ distribution.

\begin{figure}[t]
\centering
\subfloat{
\centering
\includegraphics[width=0.45\textwidth]{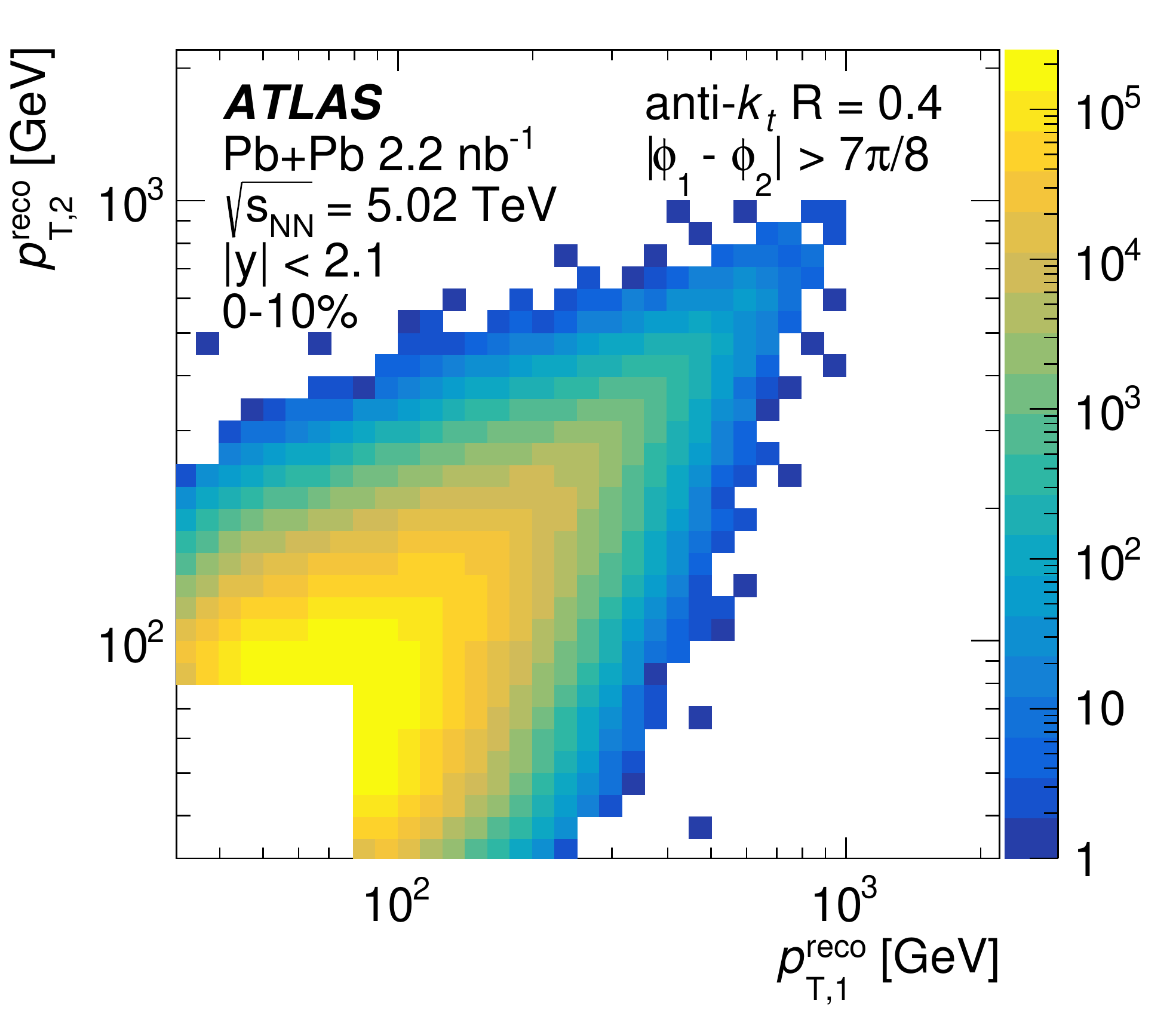}
}
\subfloat{
\centering
\includegraphics[width=0.45\textwidth]{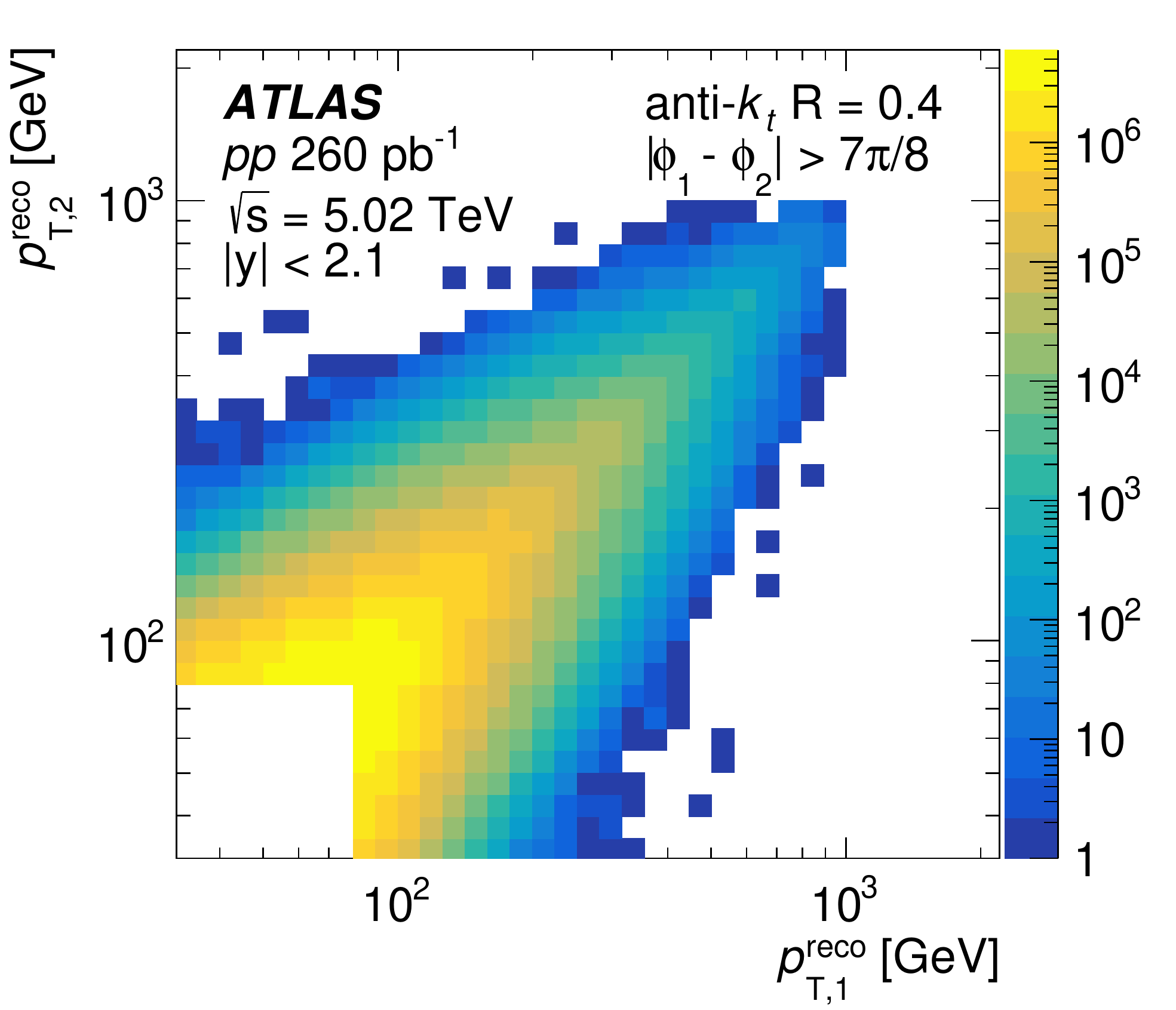}
}
\hfill
 
\caption{\label{fig:rawpt1pt2_2d}
The background-subtracted and symmetrized $(\ptjetrecoa,\ptjetrecob)$ distributions, measured in data, in 0--10\% centrality \pbpb\ collisions (left) and
\pp\ collisions (right).}
\end{figure}

The measured $(\ptjetrecoa,\ptjetrecob)$ distributions are a combination of the dijet signal and pairs of uncorrelated jets.
Since the underlying-event subtraction accounts for azimuthal correlations in the particle production due to hydrodynamic flow, the contribution from
uncorrelated dijets is independent of the $\Delta\phi$ of the jets;
therefore, a $\Delta\phi$ sideband method is used to remove these pairs as a function of
$(\ptjetrecoa,\ptjetrecob)$. The symmetrized two-dimensional
$(\ptjetrecoa,\ptjetrecob)$ distribution of background combinatoric dijets is determined using
dijets with $ 1 < |\Delta\phi| < 1.4$ which, after normalizing to the $\Delta\phi$
window of the signal band, is subtracted from the dijet yields.
This effect is strongest for 0--10\% centrality \pbpb\ events at low \ptjetrecoa,
where combinatoric dijets constitute 13\% of the dijets for $79 < \ptjetrecoa < 100$~\GeV,
and drops off rapidly both with increasing \ptjetrecoa\ and in more peripheral events.
Because of how the leading dijet is defined, the presence of residual combinatoric dijets
in the sample results in an inefficiency for genuine jet pairs,
where one of the jets might be superseded by an uncorrelated third jet.
This effect is corrected for using the measured inclusive jet spectrum from minimum-bias
events to determine the efficiency loss as a function of the measured jet \pt\ following
the method discussed in Ref.~\cite{ATLAS:2017xfa}. The resulting combinatoric-subtracted and
efficiency-corrected symmetrized two-dimensional $(\ptjetrecoa,\ptjetrecob)$
distributions can be seen in Figure~\ref{fig:rawpt1pt2_2d} for both central \pbpb\ and \pp\ collisions. The ability to remove
uncorrelated jets and extract the truth-level leading-dijet \xJ\ distributions in an unbiased way was confirmed by a closure test
performed using the data overlay sample.

In order to correct for the effects of the JES and JER,
the measured $(\ptjetrecoa,\ptjetrecob)$ distributions are unfolded using the
iterative Bayesian unfolding procedure~\cite{DAgostini:1994fjx} as implemented in the
RooUnfold~\cite{Adye:2011gm} software package. A two-dimensional unfolding is used
in order to account for bin migrations of both the leading and the subleading jet \pt,
as well as to account for possible swapping of the leading and subleading jet.
A separate response matrix is generated for \pp\ collisions as well as for each centrality selection
in \pbpb\ collisions.
The response matrix used in the unfolding contains the relationship between
$(\ptjettrutha,\ptjettruthb)$ and $(\ptjetrecoa,\ptjetrecob)$.
It is populated by identifying the leading and subleading truth-level jets in the MC sample
which are matched to the corresponding reconstructed jets within $\Delta R < 0.3$.
In order to account for migration from lower jet \ptjetreco,
the response matrices are populated with truth-level jets down to   \ptjettrutha\ of 20~\GeV\ and
\ptjettruthb\ of 10~\GeV.  As with the reconstructed data, truth dijets are required to have $|\Delta\phi_{\mathrm{truth}}| > 7\pi/8$,
with each jet being within $|y_{\mathrm{truth}}| < 2.1$.
The two selected reconstructed jets from the MC simulations are required to
meet the same selection criteria as applied to dijets measured in data.
Truth dijets that do not match to a reconstructed dijet meeting the selection criteria
are accounted for by using an efficiency correction in the unfolding.
Similarly to the construction of the data distributions, the response matrix is populated
symmetrically in \pta\ and \ptb.
To construct a Bayesian prior which is similar to the measured distributions,
the response matrices are reweighted along the \ptjettrutha\
and \ptjettruthb\ axes by the ratio of the two-dimensional reconstructed yields in data to those from simulation.
The number of iterations used in the unfolding is tuned separately for each centrality in \pbpb\
collisions and for \pp\ collisions.
The number of iterations in each case was
selected to optimize the balance between the accuracy of the final unfolded yield,
and the increased statistical uncertainty which results from a larger number of iterations.
Six iterations were used for the 0--10\% centrality \pbpb\ events, while four iterations were used for all other centralities as well as for \pp\ collisions.
 
In order to evaluate the statistical uncertainties of the data and the
MC simulation, 1000 unfoldings were performed by varying each bin in either the
data distribution or the response matrix independently according to its
statistical uncertainty while preserving the symmetrization across
\pta\ = \ptb. The RMS of the resulting unfolded distributions in each
$(\pta,\ptb)$ bin was taken as the contribution to the
statistical uncertainty on the corresponding unfolded $(\pta,\ptb)$ bin. The data and response matrix statistical
uncertainty components were combined in quadrature in order to obtain
the total statistical uncertainty of the unfolded distributions.  Due
to the large number of bins in the two-dimensional unfolding, in some
cases the uncertainties due to the finite MC sample sizes are
comparable to those in the data.
 
To extract measurements of the dijet momentum balance observable, \xj,
the unfolded two-dimensional $(\pta,\ptb)$ distributions are first
reflected about $\pta = \ptb$ in order to restore the
leading/subleading hierarchy.  Then, following the procedure
discussed in Ref.~\cite{ATLAS:2017xfa}, the two-dimensional
distributions are projected in slices of \pta\ into bins of \xJ.  The upper edge of the $i^{\mathrm{th}}$
\xJ\ bin boundary is defined by $((p_{\mathrm{T},N}/p_{\mathrm{T,0}})^{1/N})^{i-N}$,
where $N$ is the number of
logarithmic \pt\ bins (40), and $p_{\mathrm{T,0}}$ and $p_{\mathrm{T},N}$ are the minimum (10~\GeV) and maximum (1~\TeV) bin edges covered
by the binning, respectively.  This enables the mapping of the logarithmic \pta\ and
\ptb\ bins to \xJ\ in such a way that each reported $(\pta, \ptb)$ bin
is fully contained within two adjacent \xJ\ bins.  In this analysis,
half of the yield in each $(\pta, \ptb)$ bin is apportioned to each of
the two adjacent \xJ\ bins. An exception to this are the bins along the diagonal
\pta\ = \ptb, which contribute solely to the highest \xJ\ bin.
 
By projecting the resulting distributions over selections of
\pta\ while normalizing by the number of dijets
($N_{\textrm{pair}}$) and accounting for the \xj\ bin widths,
measurements of the dijet yield-normalized \xJ\ distributions are
extracted.  The absolutely-normalized \xJ\ distributions are
extracted by instead normalizing the \xJ\ distributions by the number
of events and either the \TAA\ in \pbpb\ collisions or the integrated luminosity in
\pp\ collisions, as described by Exp.~(\ref{eq:absxjpbpb}).
 
In order to quantify the energy losses of the leading and subleading jets
within a dijet, the nuclear modification factor, \RAApair, is measured
for each of these selections. The \RAApair\ compares the per-event
differential yield of jets at a given \pt\ in \pbpb\ collisions with the
cross-section in \pp\ collisions scaled by the \TAA, as described in
Eqs.~(\ref{eq:raa_lead}) and~(\ref{eq:raa_sublead}).

\section{Systematic uncertainties}

Systematic uncertainties for this measurement
are attributed to three categories of sources, those which arise from the analysis and unfolding procedure, those which stem from uncertainties in the jet energy resolution (JER) and scale (JES), and those connected with global normalization. For each uncertainty component in the first two categories, the entire analysis procedure, including unfolding, is repeated accounting for the modification to the analysis procedure or the response matrix and the result is compared with the nominal one. The third category applies to the absolutely-normalized \xJ\ distributions,
\RAAlead,  and \RAAsublead; it contains
the uncertainty in the determination of the mean nuclear thickness function, \avgTAA,
and the \pp\ luminosity.  These uncertainties are independent of the jet
transverse momentum and are noted on the figures.

The systematic uncertainty in the JES has four parts. First, a centrality-independent baseline component is determined from in situ
studies of the calorimeter response to jets reconstructed with the procedure used in 13~\TeV\ \pp\ collisions~\cite{ATLAS:2017bje}.
A second, centrality-independent component accounts for the relative energy scale difference between the jet reconstruction procedures
used in this analysis and those in 13~\TeV\ \pp\ collisions.  This
is evaluated using the cross-calibration procedure described in Ref.~\cite{ATLAS:2015twa}. Potential inaccuracies in the \pythiaeight\ MC sample's description
of the relative abundances of jets initiated by quarks and gluons and of the calorimetric response to quark and gluon jets
are accounted for by the third, centrality independent, component -- based on evaluating these same quantities
using the \herwig\ MC sample. The fourth, centrality-dependent, component accounts for modifications of the parton shower due to quenching~\cite{ATLAS:2018bvp}, which is not
modeled in the simulations. The modifications to the parton shower can impact the detector response to jets in \pbpb\ collisions resulting in a small disagreement in the JES
between data and simulations. The extent of this disagreement, and corresponding uncertainty contribution is evaluated by the method used
in Ref.~\cite{ATLAS:2015twa}
for 2015 and 2011 data, which compares the jet \pt\ measured in the calorimeter with the sum of the transverse momenta of charged particles within the jet, in both the data and MC samples.
This uncertainty is determined as a function of event centrality and was found to be independent of jet \pT\ and $\eta$.
The selected charged-particle tracks have $\pt > 4$~\GeV\  in order to exclude particles from the UE.
The sum of the charged-particle transverse momenta provides a data-driven estimate of the centrality dependence of the JES arising from the observed centrality-dependent modification
of the jet fragmentation due to jet quenching
in \pbpb\ collisions~\cite{ATLAS:2018bvp}.
The size of this centrality-dependent uncertainty in the JES reaches 1.2\% in the most central
collisions and the value is applied independent of \xJ.
For each individual component, the JES in the MC simulation was modified as a function of \pt\ and $\eta$ by
one standard deviation, and the response matrix was recomputed.

The uncertainty due to the JER is evaluated by repeating the unfolding procedure with modified response matrices, where an additional contribution is added to the
resolution of the reconstructed \pT\ in the MC sample using a Gaussian smearing procedure.
The smearing factor is evaluated using an in situ technique in 13~\TeV\ \pp\ data that involves studies of dijet
\pT\ balance~\cite{ATLAS:2020cli}.
Further, an uncertainty is included to account for differences between the tower-based jet reconstruction and the jet reconstruction used in
analyses of 13~\TeV\ \pp\ data, as well as differences in calibration procedures.
The modifications to the response are propagated through the unfolding and the resulting uncertainty is symmetrized.

Two sources of systematic uncertainty were included to account for uncertainties in the removal
of the combinatoric background. The first contribution stems from the combinatoric subtraction method, and was determined by extracting the
two-dimensional $(\pta, \ptb)$ distribution of combinatoric jets from an
alternative sideband of 1.1 $ < |\Delta\phi|  < $ 1.5 as was done in Ref.~\cite{ATLAS:2017xfa}.
The second
contribution stems from the sensitivity of the analysis to the efficiency correction for combinatoric jets, and was evaluated by repeating the
analysis without the inclusion of this efficiency correction. The deviation from the nominal result is taken as the uncertainty contribution.

Additional sources of systematic uncertainty which account for the
unfolding procedure were considered.
The sensitivity to the Bayesian prior was evaluated by modifying the weights applied when producing the response matrix in a centrality-dependent manner in order to
enclose the data $(\ptjetrecoa,\ptjetrecob)$
distributions between the corresponding MC distributions based on the
nominal and alternative priors. There is a sensitivity to the minimum
\ptjet\ in the analysis at small \xJ\ and small \pta\
due to the efficiency correction made as part of the unfolding.
The sensitivity of the
result to this effect is evaluated by varying the
minimum reconstructed \ptjet, motivated by the magnitude of the JER, from 32 to 39~\GeV\ for both data and simulation.
This results in a significant contribution to the systematic uncertainties at low \xJ\ in low \pta\ bins. For each of these contributions the deviation of the unfolded result from the nominal is symmetrized and taken as a contribution to the systematic uncertainties.
A closure test was performed using the MC samples by
splitting them into two equal parts and using one part to produce a response matrix and the other in place of  the data sample in the unfolding.
The deviation between the unfolded result and the underlying truth-level distribution was taken as an estimate of this uncertainty. The uncertainty from each of these contributions is taken independently and symmetrized.
 
\begin{figure}[t]
\centering
\subfloat{
\centering
\includegraphics[width=0.42\textwidth]{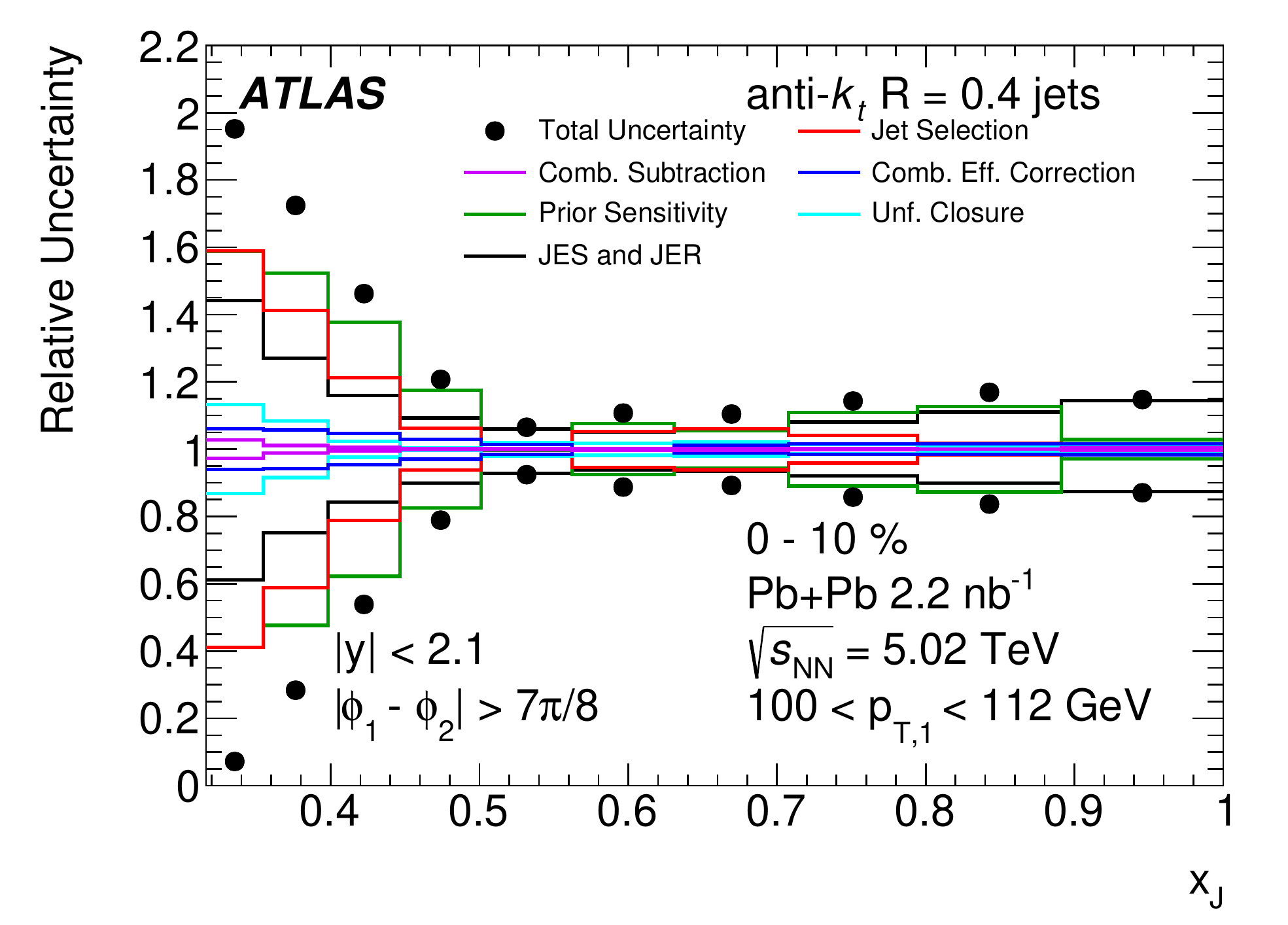}
}
\subfloat{
\centering
\includegraphics[width=0.42\textwidth]{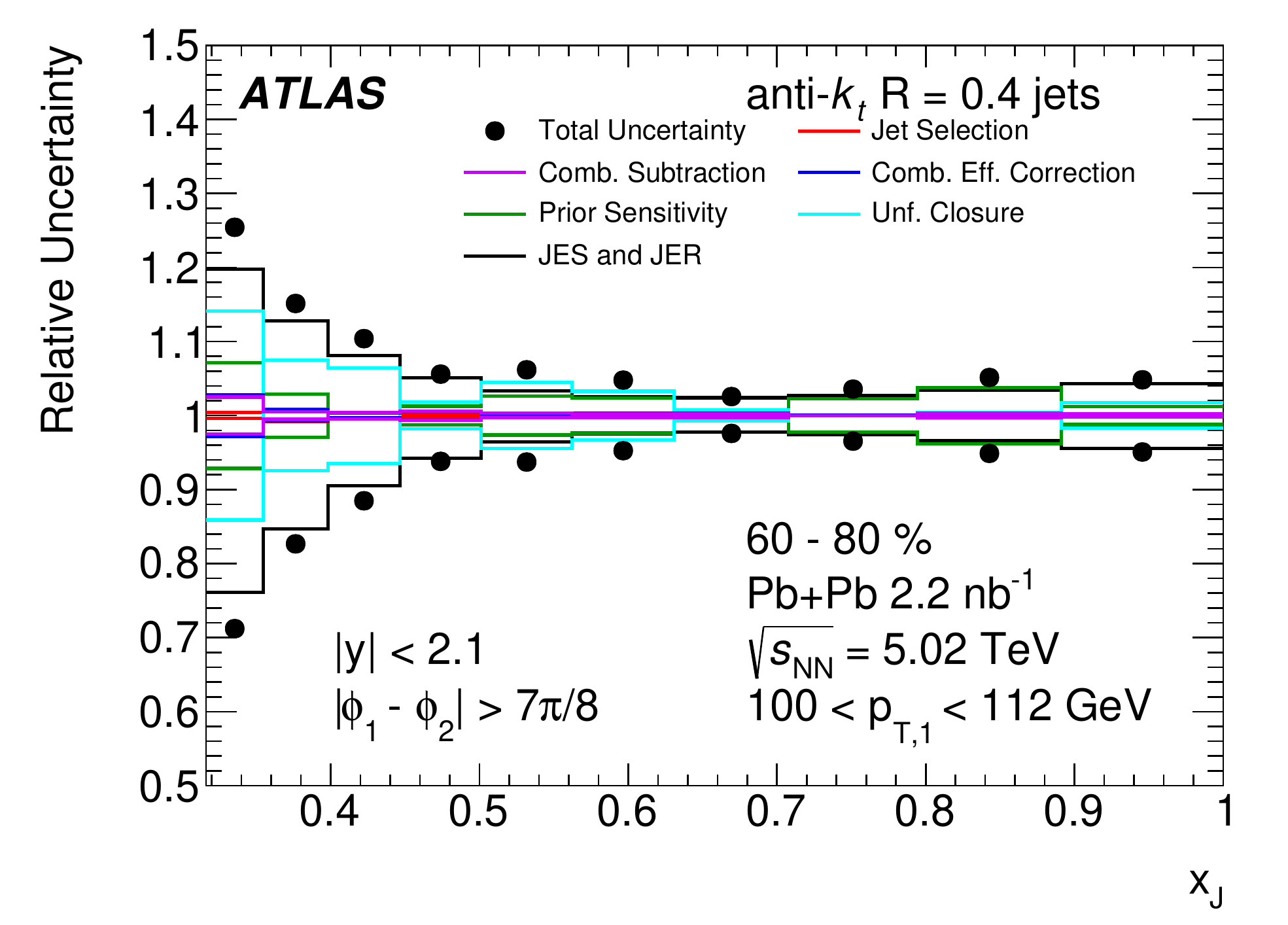}
}
\hfill
\subfloat{
\centering
\includegraphics[width=0.42\textwidth]{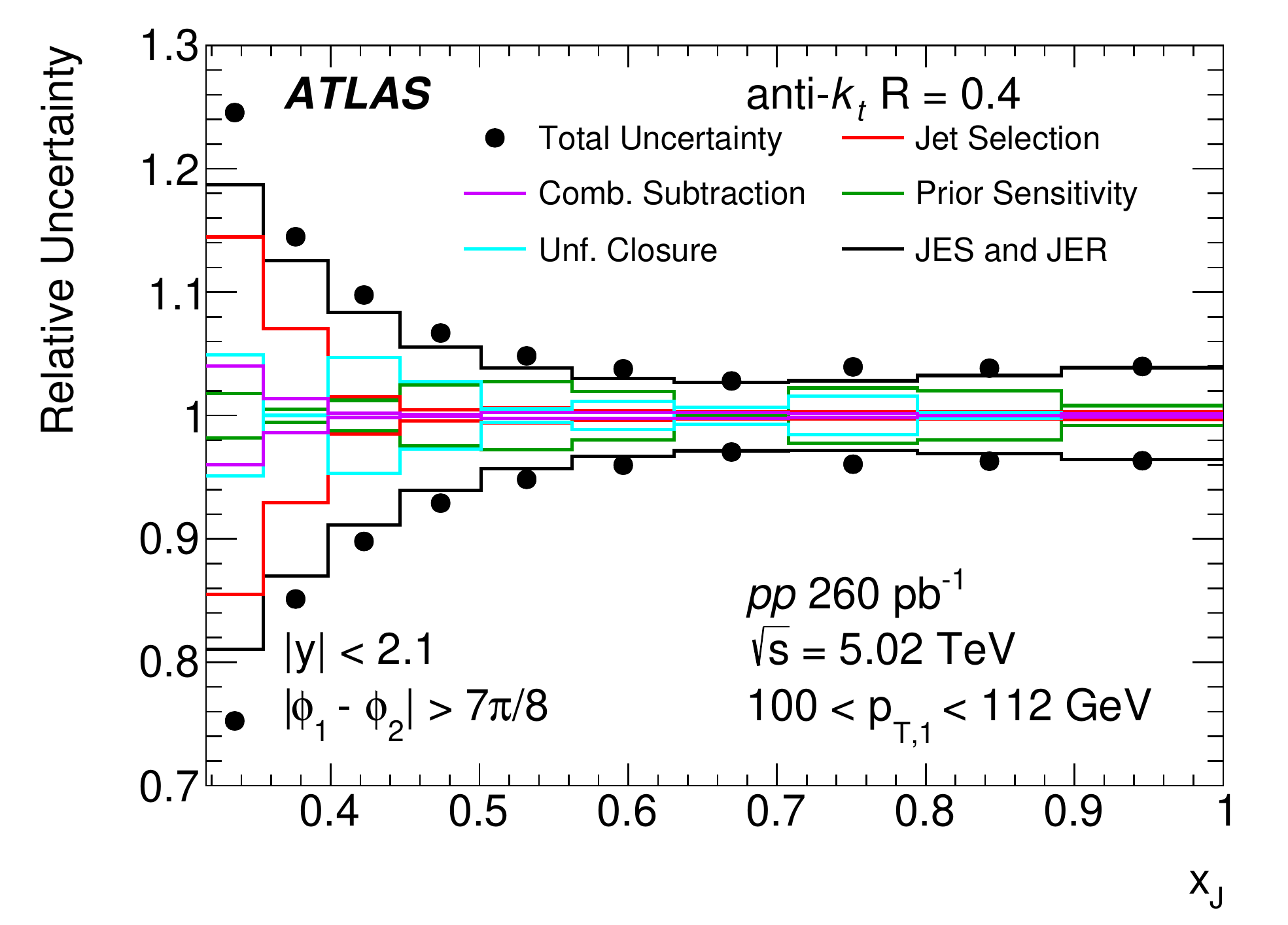}
}
\hfill
 
\caption{\label{fig:systematic_xj} The systematic uncertainty breakdown of the \xj\ distributions in central (left) and peripheral (right) \pbpb\ collisions as well as \pp\ collisions (bottom).
The panels have different vertical axis ranges.}
\end{figure}

The magnitude of the systematic uncertainties in the dijet-yield-normalized \xJ\ distributions can be seen in Figure~\ref{fig:systematic_xj} for both central and peripheral \pbpb\ collisions. In the central collisions,
the total systematic uncertainties are largely driven  by
the sensitivity of the unfolding to the choice of prior. In contrast, in peripheral collisions the total uncertainties are largely driven by the uncertainty in the JES and JER.
The fractional uncertainties are largest at low \xJ\ in both collision systems; however,
the yield in these \xJ\ regions is small.
 
\begin{figure}[t]
\centering
\subfloat{
\centering
\includegraphics[width=0.45\textwidth]{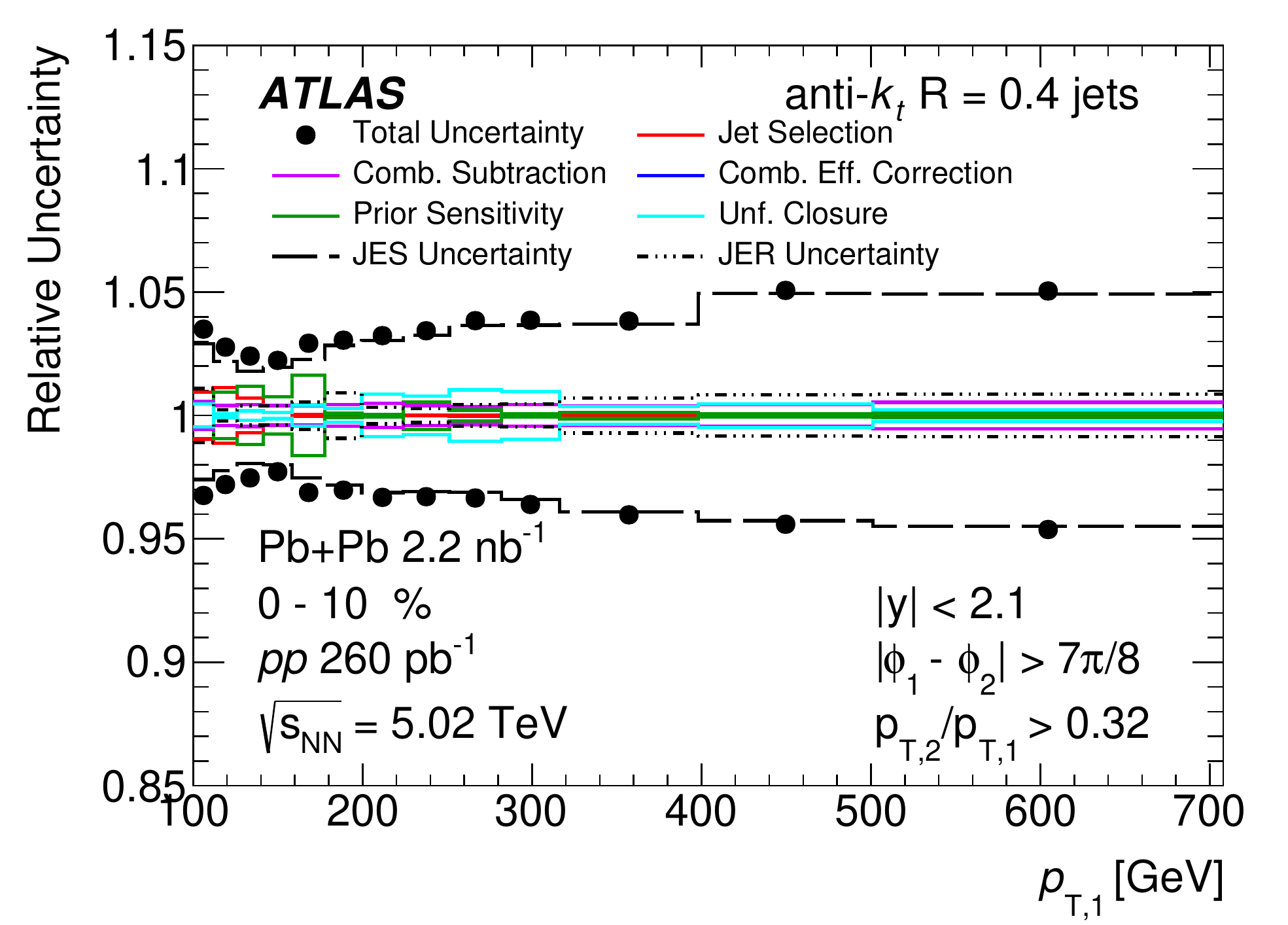}
}
\subfloat{
\centering
\includegraphics[width=0.45\textwidth]{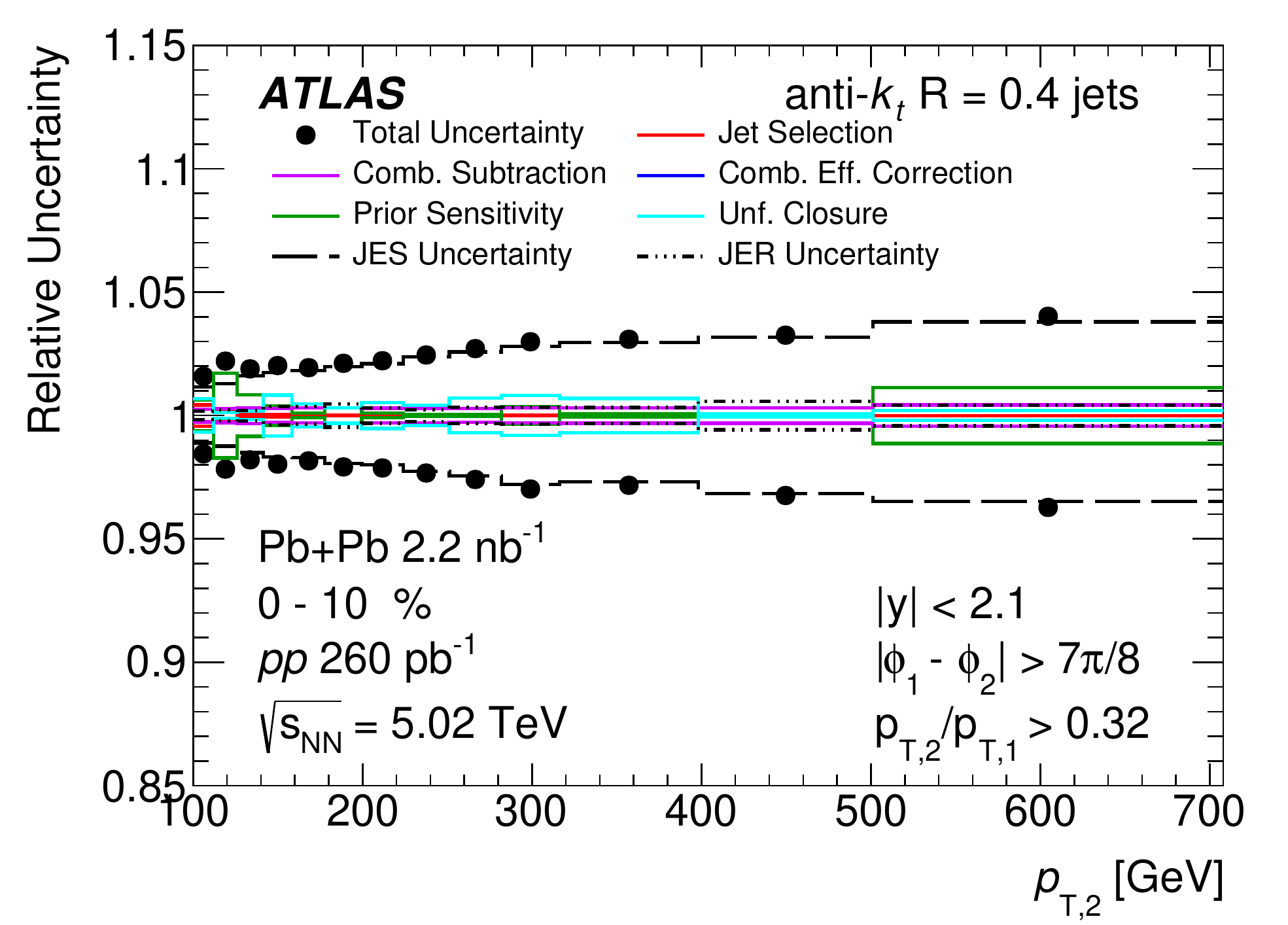}
}
\hfill
 
\caption{\label{fig:systematic_raa}The systematic uncertainty breakdown of the \RAApair\ for leading jets (left) and subleading jets (right). Uncertainty on the \TAA\ and \pp\ luminosity are not shown.}
\end{figure}

The systematic uncertainty contributions are similarly propagated to the calculation of the nuclear modification factor. The centrality-independent components of the JES and JER, as well as the sensitivity of the unfolding to the minimum measured \ptjet, are treated as correlated between \pbpb\ and \pp\ collisions.
The remainder of the contributions to the systematic uncertainty, including the centrality-dependent component of the JES, are treated as uncorrelated between \pbpb\ and \pp.
The resulting uncertainties in \RAAlead\ and \RAAsublead\ are shown for 0--10\% centrality \pbpb\ collisions in Figure~\ref{fig:systematic_raa}; these uncertainties are dominated by the JES uncertainty.
In the ratio of \RAAsublead\ to \RAAlead, each source of systematic uncertainty is treated as fully correlated between \RAAsublead\ and \RAAlead, including the global systematic uncertainties.

\section{Results}
\subsection{Dijet-yield-normalized \xJ\ distributions}
\begin{figure}[t]
\centering
\subfloat{
\centering
\includegraphics[width=0.48\textwidth]{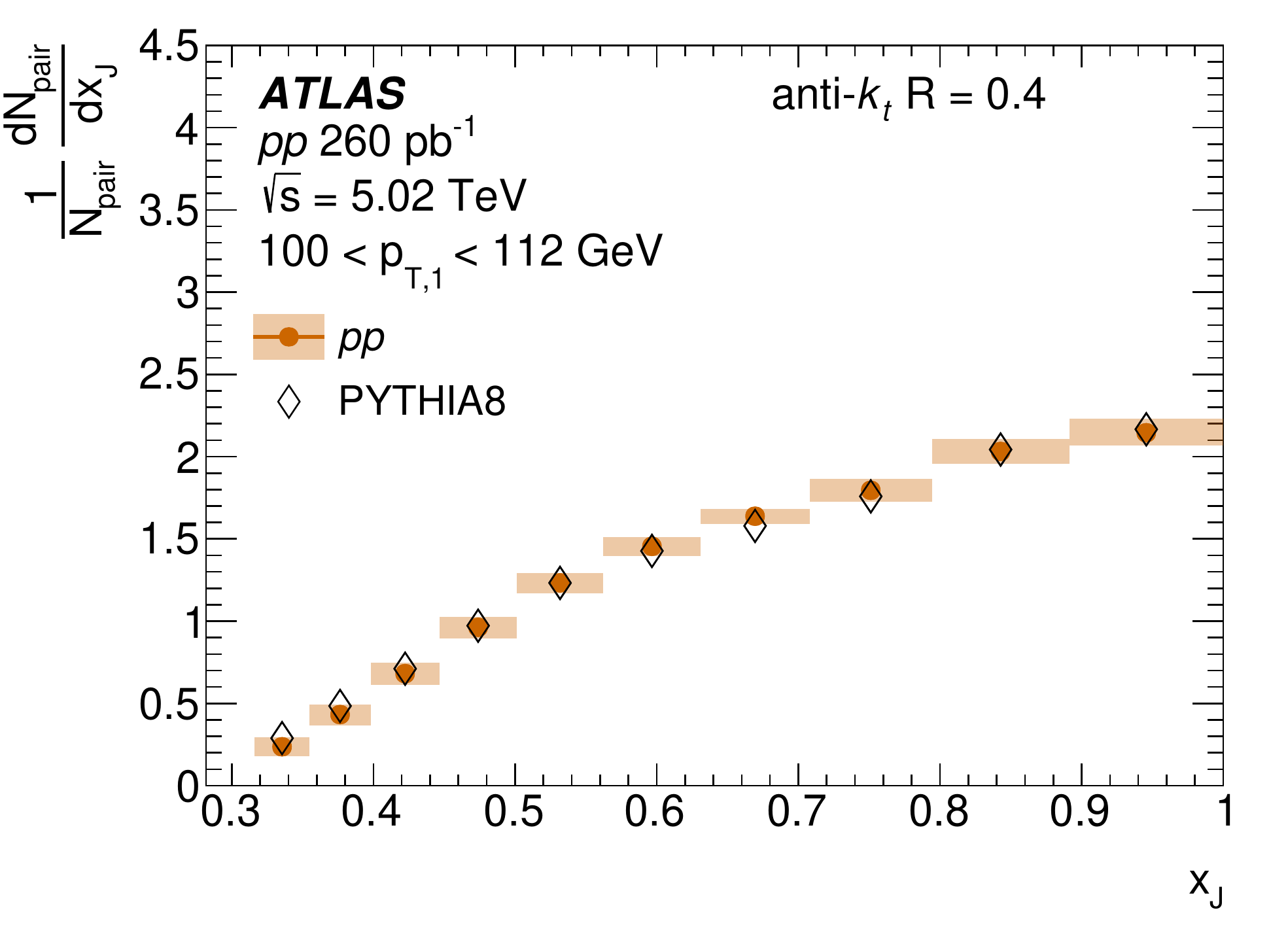}
}
\hfill
\subfloat{
\centering
\includegraphics[width=0.48\textwidth]{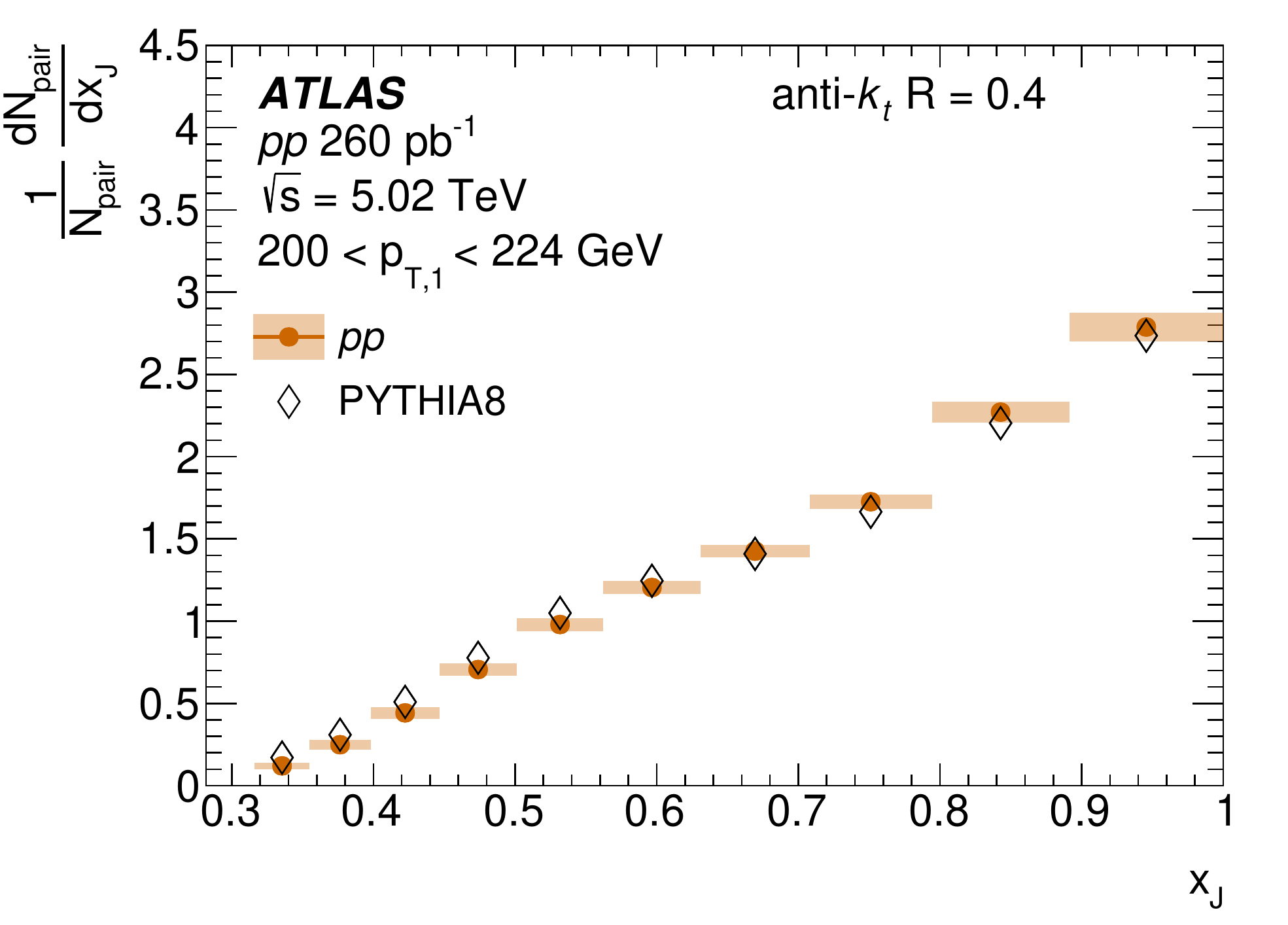}
}
\hfill
\subfloat{
\centering
\includegraphics[width=0.48\textwidth]{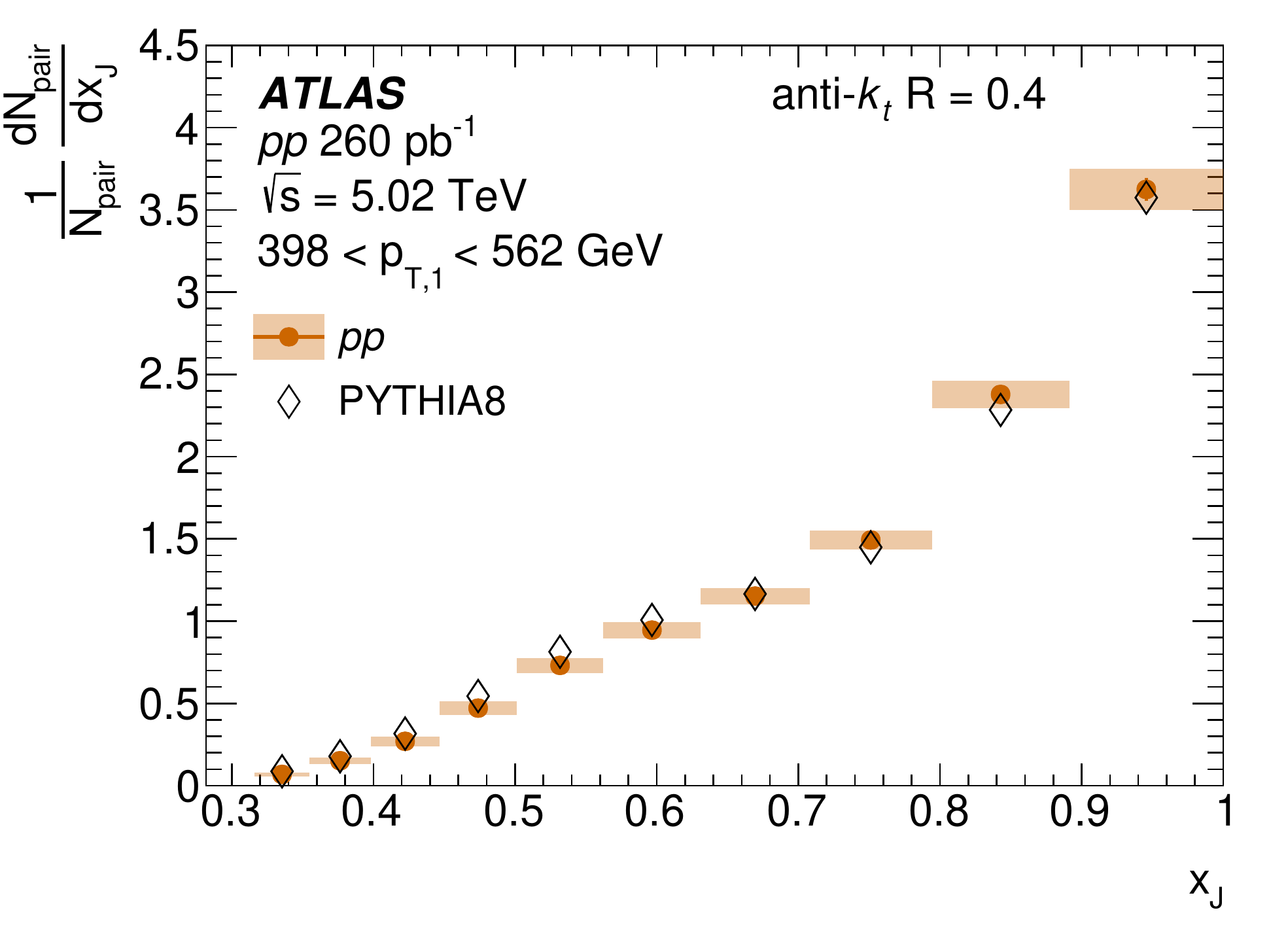}
}
\hfill
\caption{\label{fig:ppresult}The dijet-yield-normalized \dNdxJ\ distributions in \pp\ collisions for three
\pta\ selections from 100~\GeV\ to 562~\GeV\
overlaid with the corresponding distributions from \pythiaeight. The statistical uncertainties
are shown as error bars (which are smaller than the markers in many cases), and the
systematic uncertainties are shown as boxes.}
\end{figure}

The dijet-yield-normalized \xJ\ distributions in \pp\ collisions are
shown for three selections of \pta\ from 100 to 562~\GeV\ in Figure~\ref{fig:ppresult}.
The measured distributions are compared with those from the
\pythiaeight\ generator~\cite{SJOSTRAND2015159} using the A14
tune~\cite{ATL-PHYS-PUB-2014-021} in combination with the \textsc{NNPDF23lo}
PDFs~\cite{Ball:2012cx}. The \pythiaeight\ sample is able to describe
the data over the full \xJ\ range considered.
 
\begin{figure}[t]
\centering
\subfloat{
\centering
\includegraphics[width=0.33\textwidth]{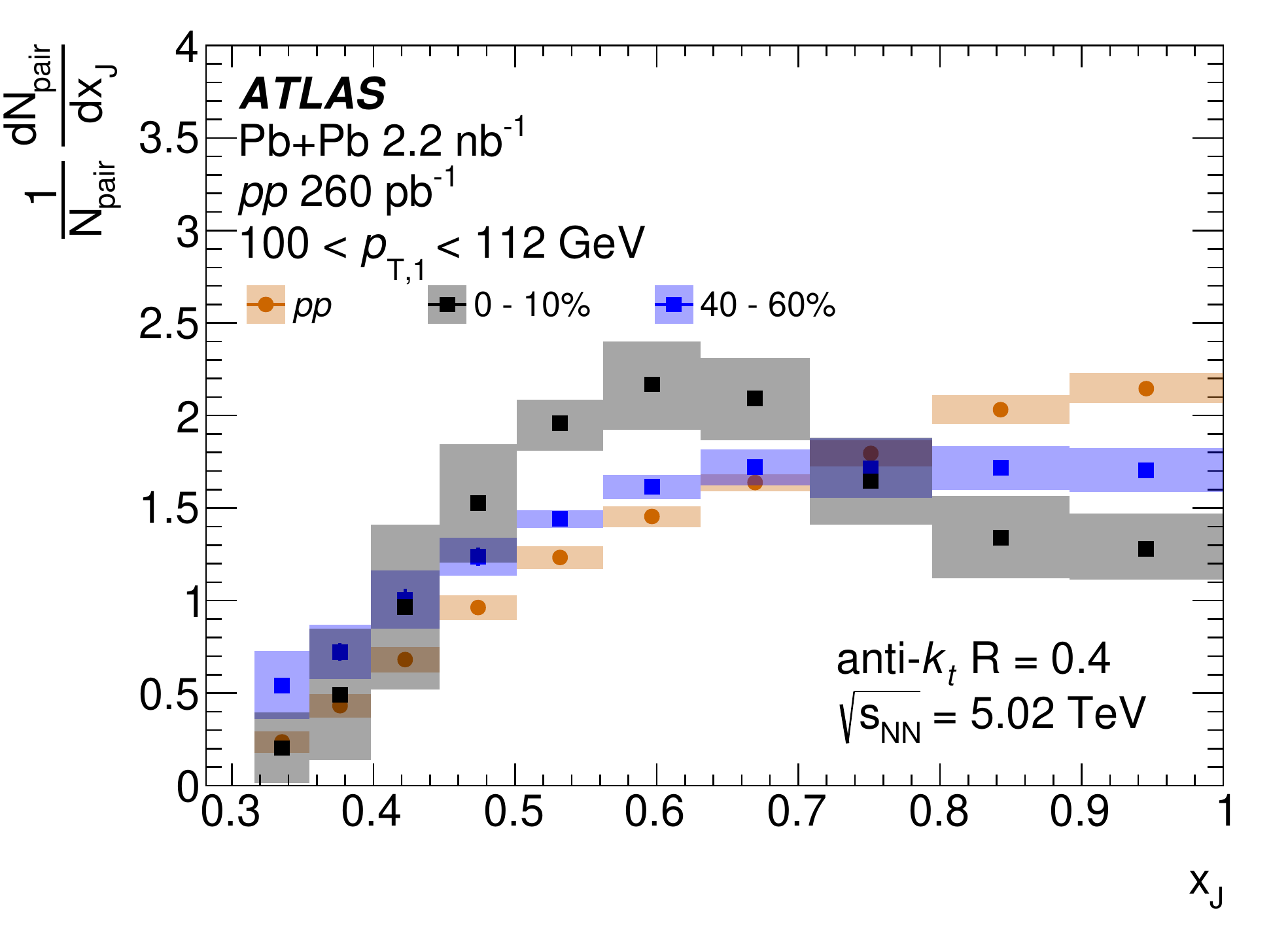}
}
\subfloat{
\centering
\includegraphics[width=0.33\textwidth]{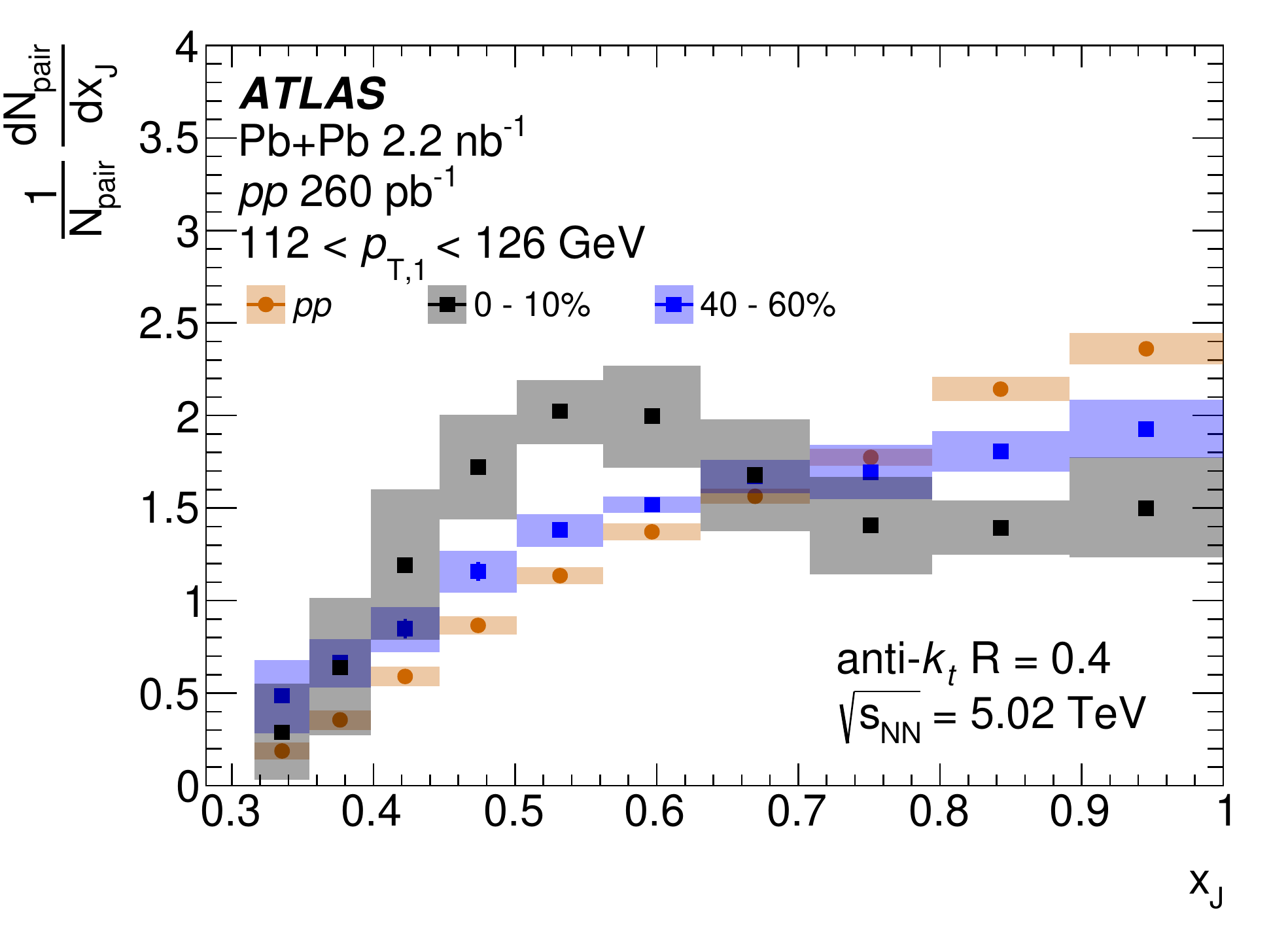}
}
\subfloat{
\centering
\includegraphics[width=0.33\textwidth]{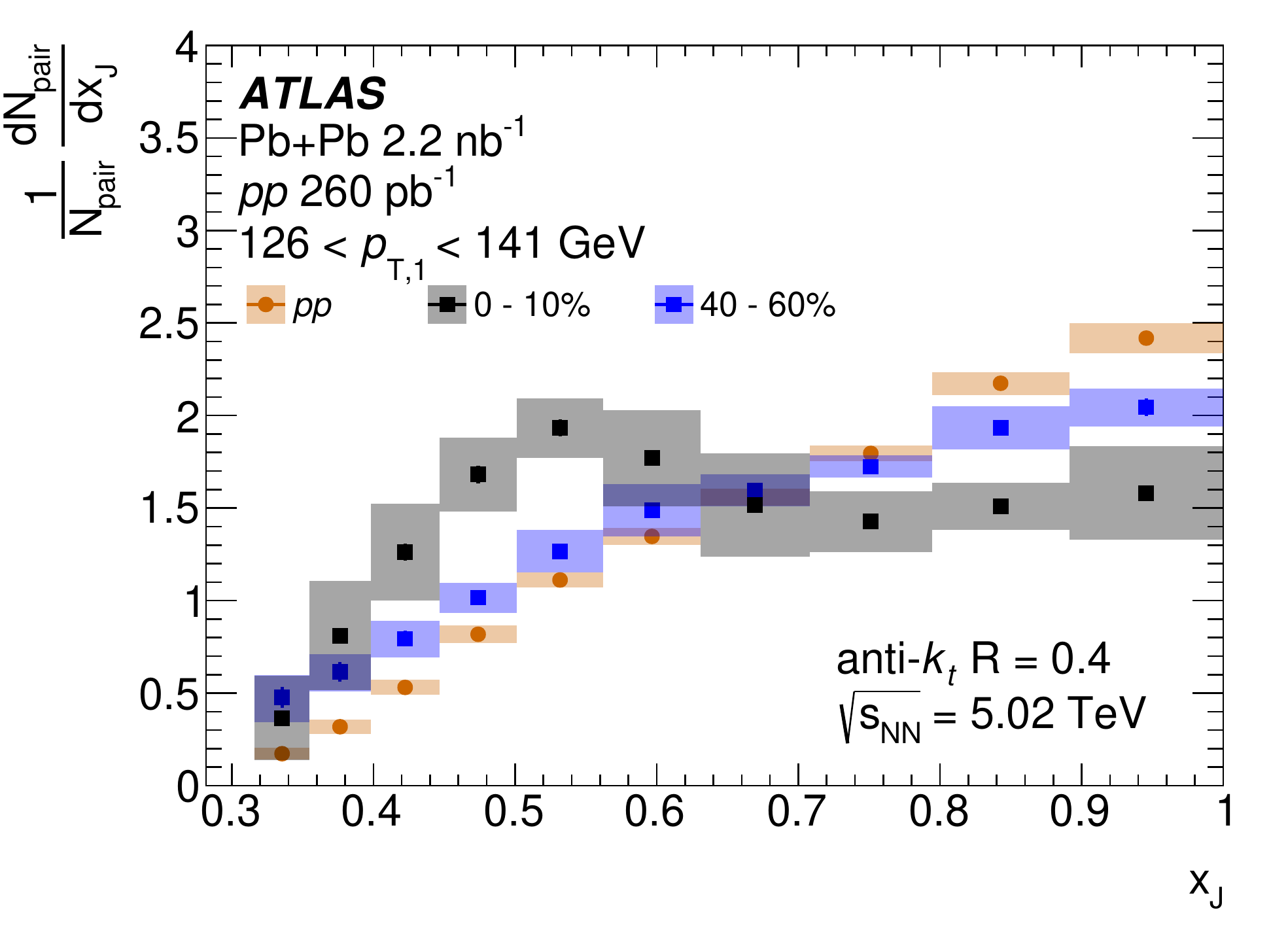}
}
\hfill

\subfloat{
\centering
\includegraphics[width=0.33\textwidth]{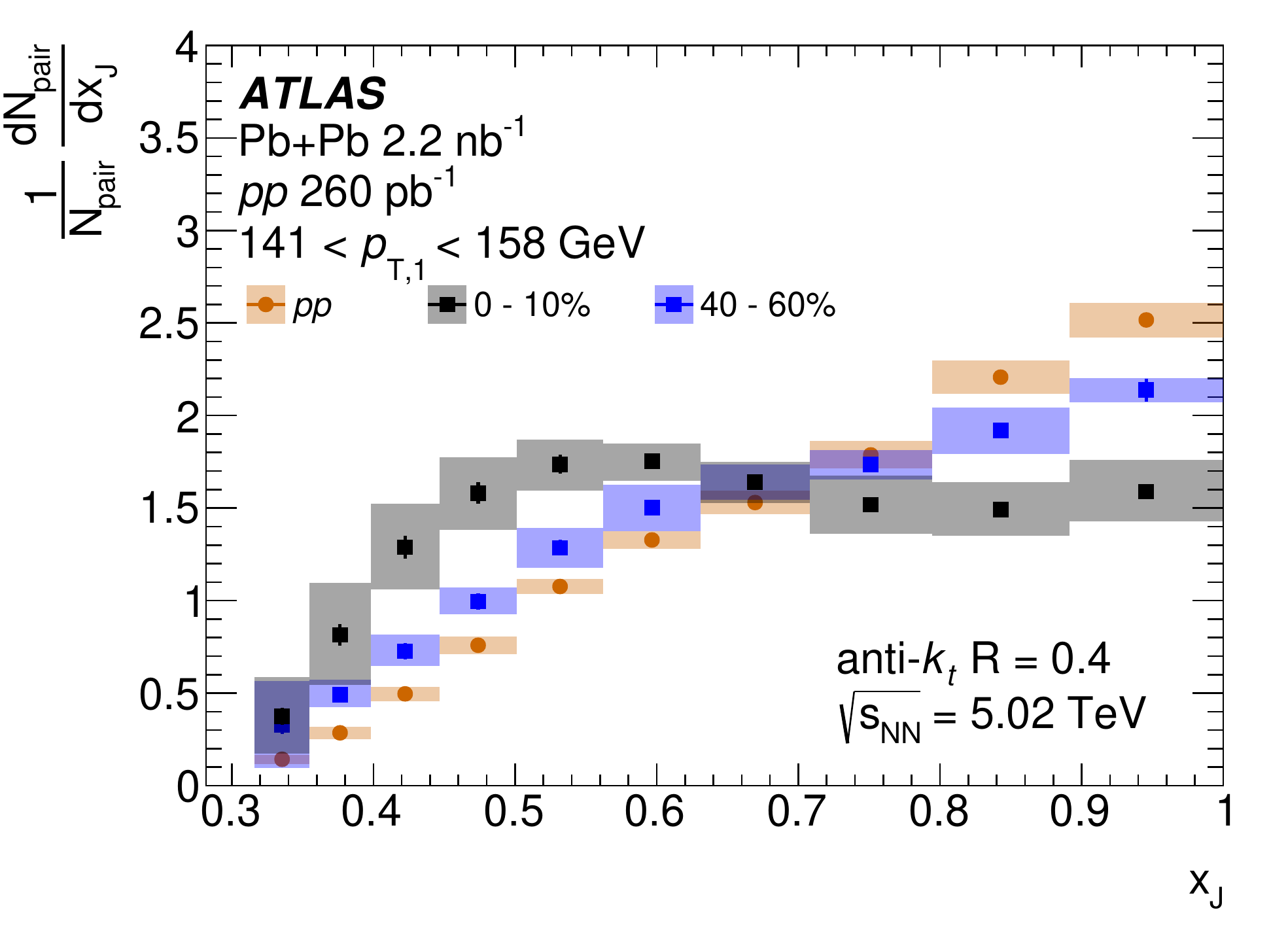}
}
\subfloat{
\centering
\includegraphics[width=0.33\textwidth]{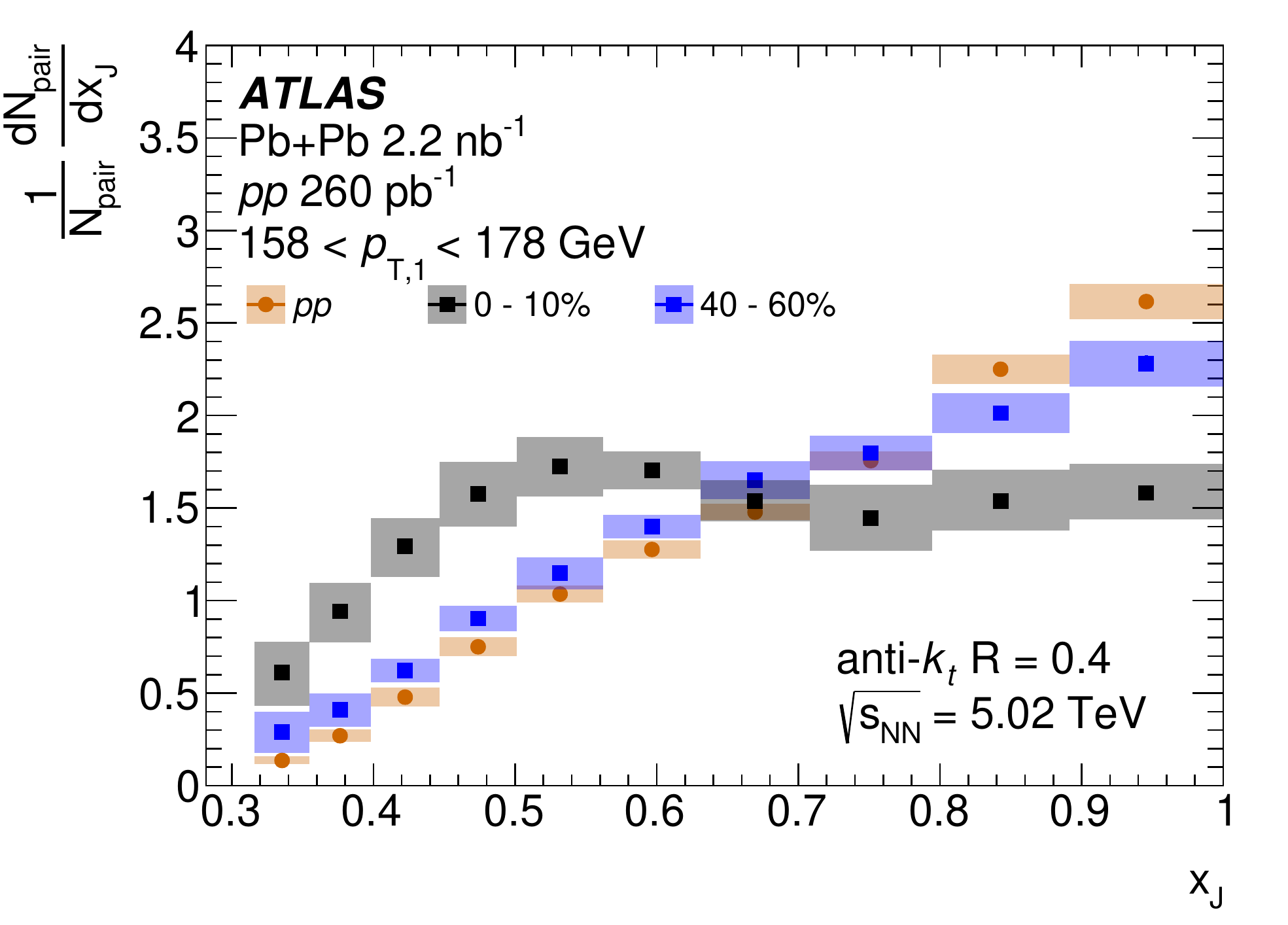}
}
\subfloat{
\centering
\includegraphics[width=0.33\textwidth]{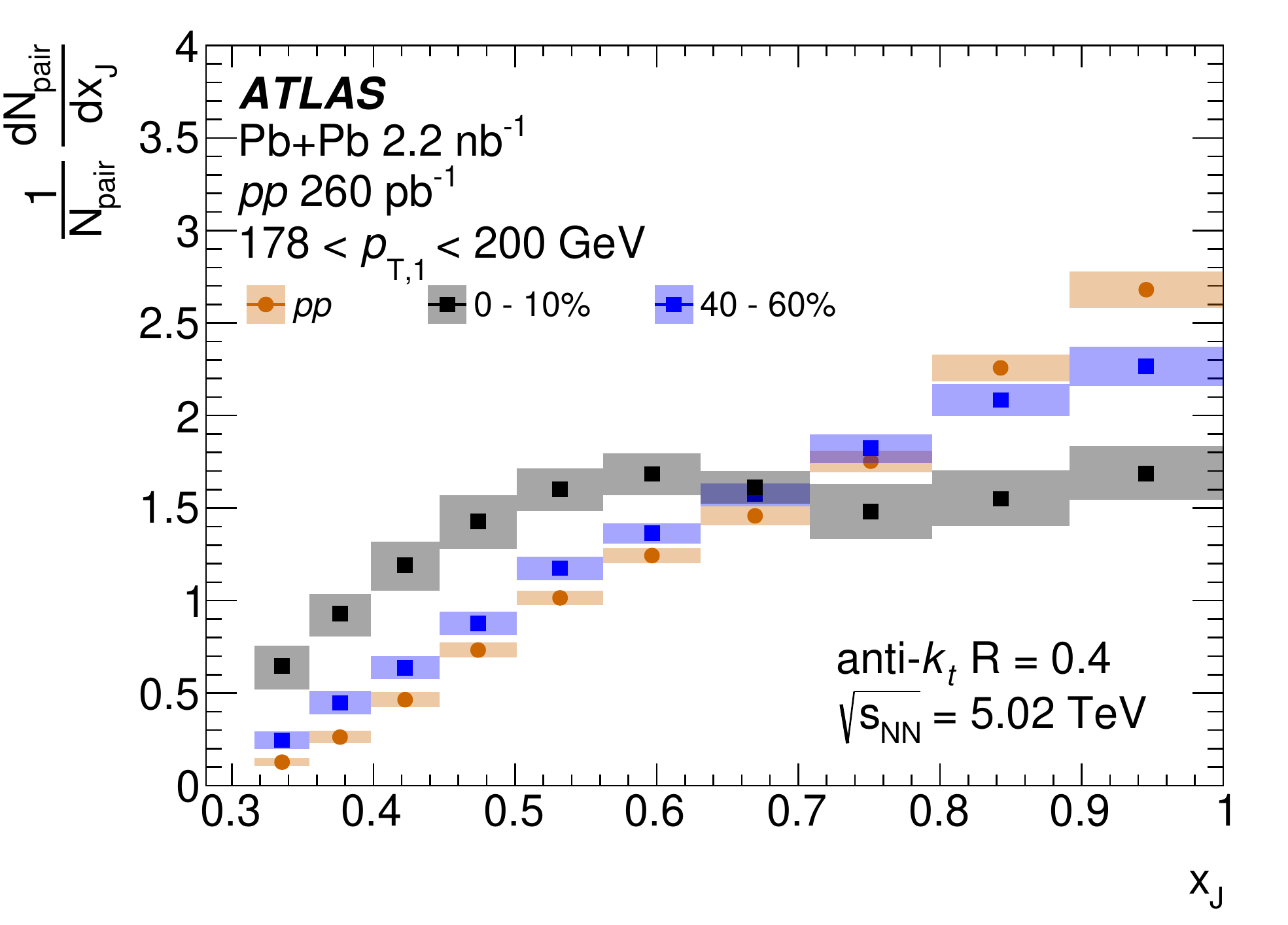}
}
\hfill
\subfloat{
\centering
\includegraphics[width=0.33\textwidth]{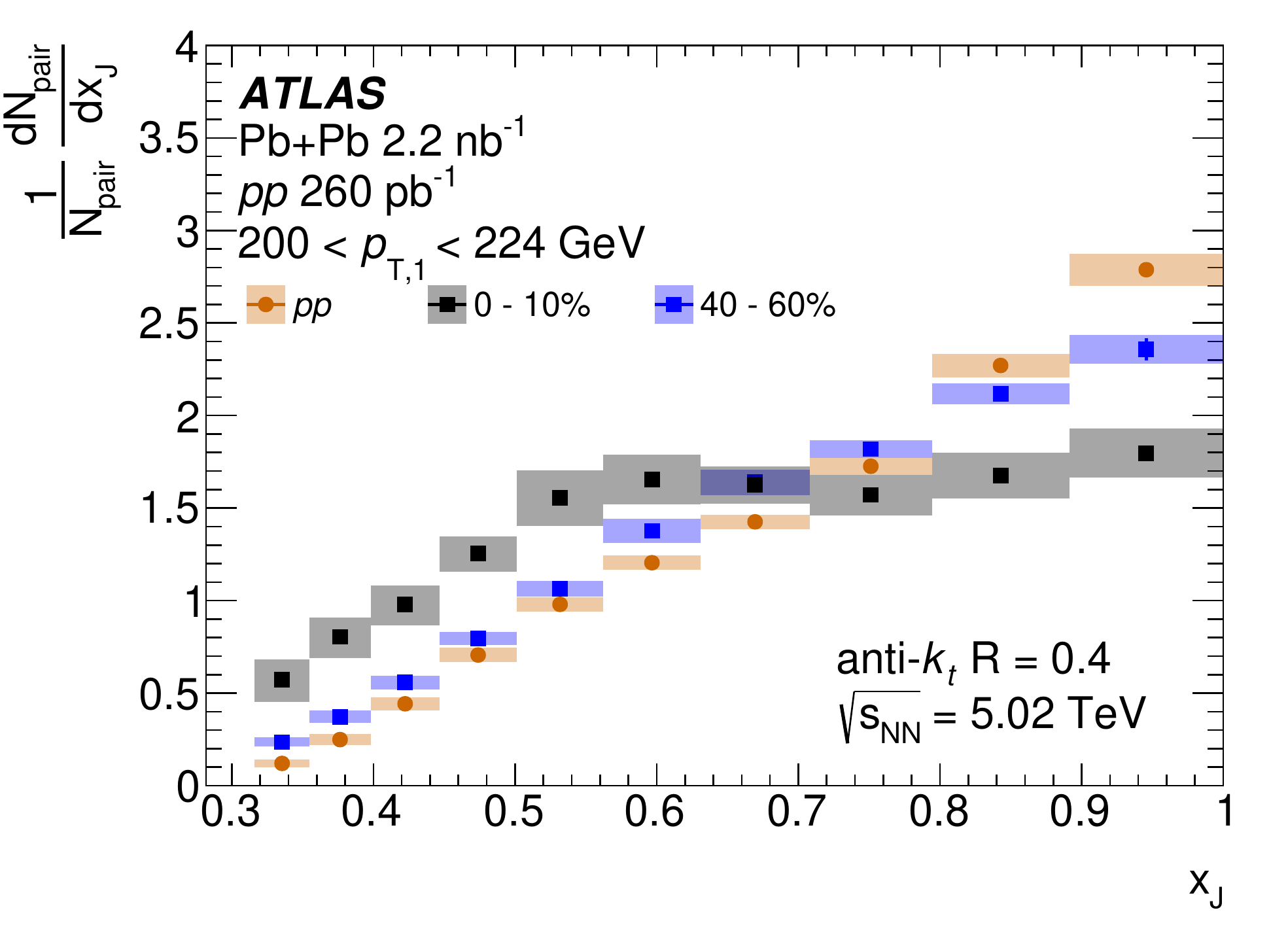}
}
\subfloat{
\centering
\includegraphics[width=0.33\textwidth]{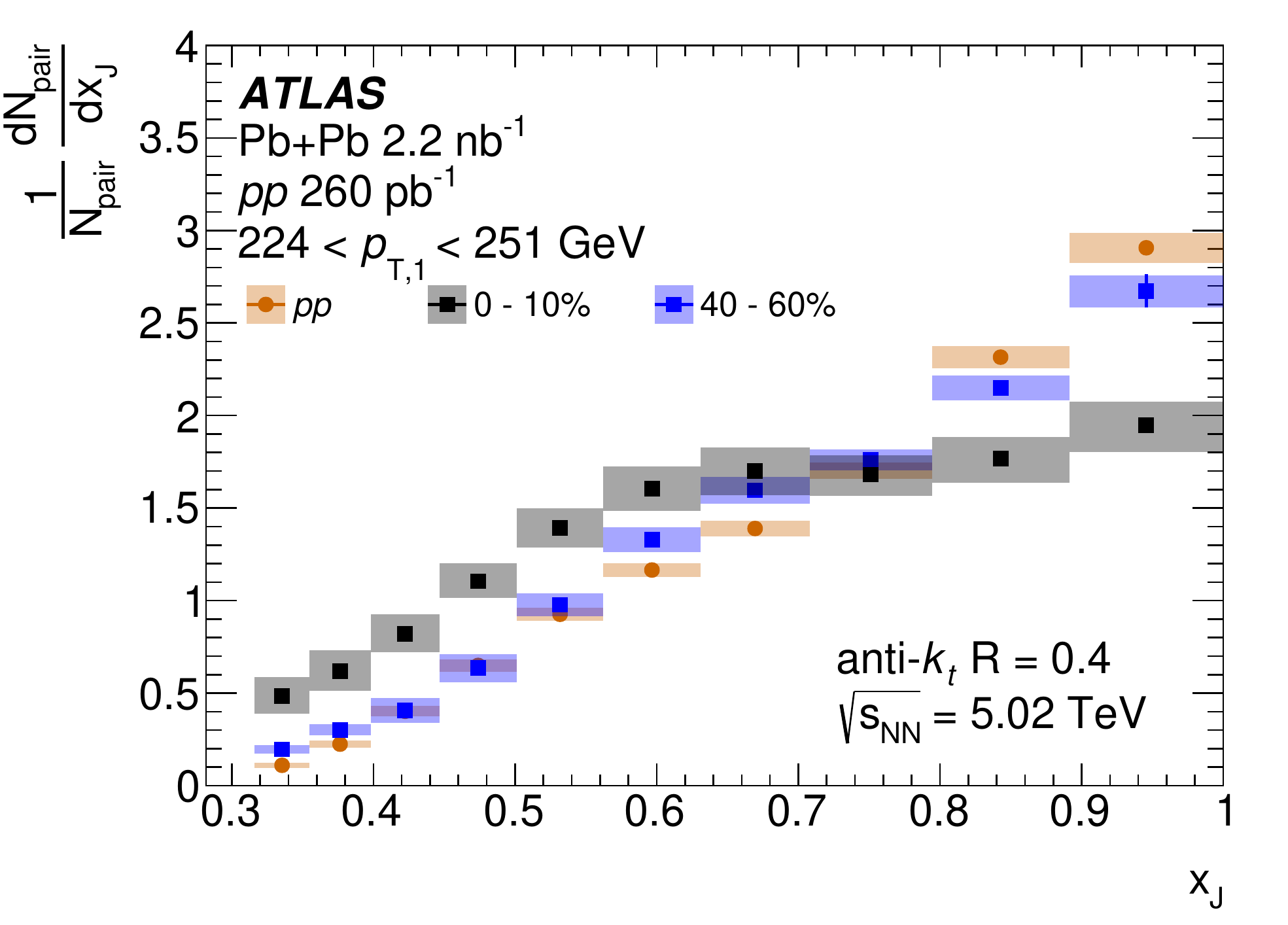}
}
\subfloat{
\centering
\includegraphics[width=0.33\textwidth]{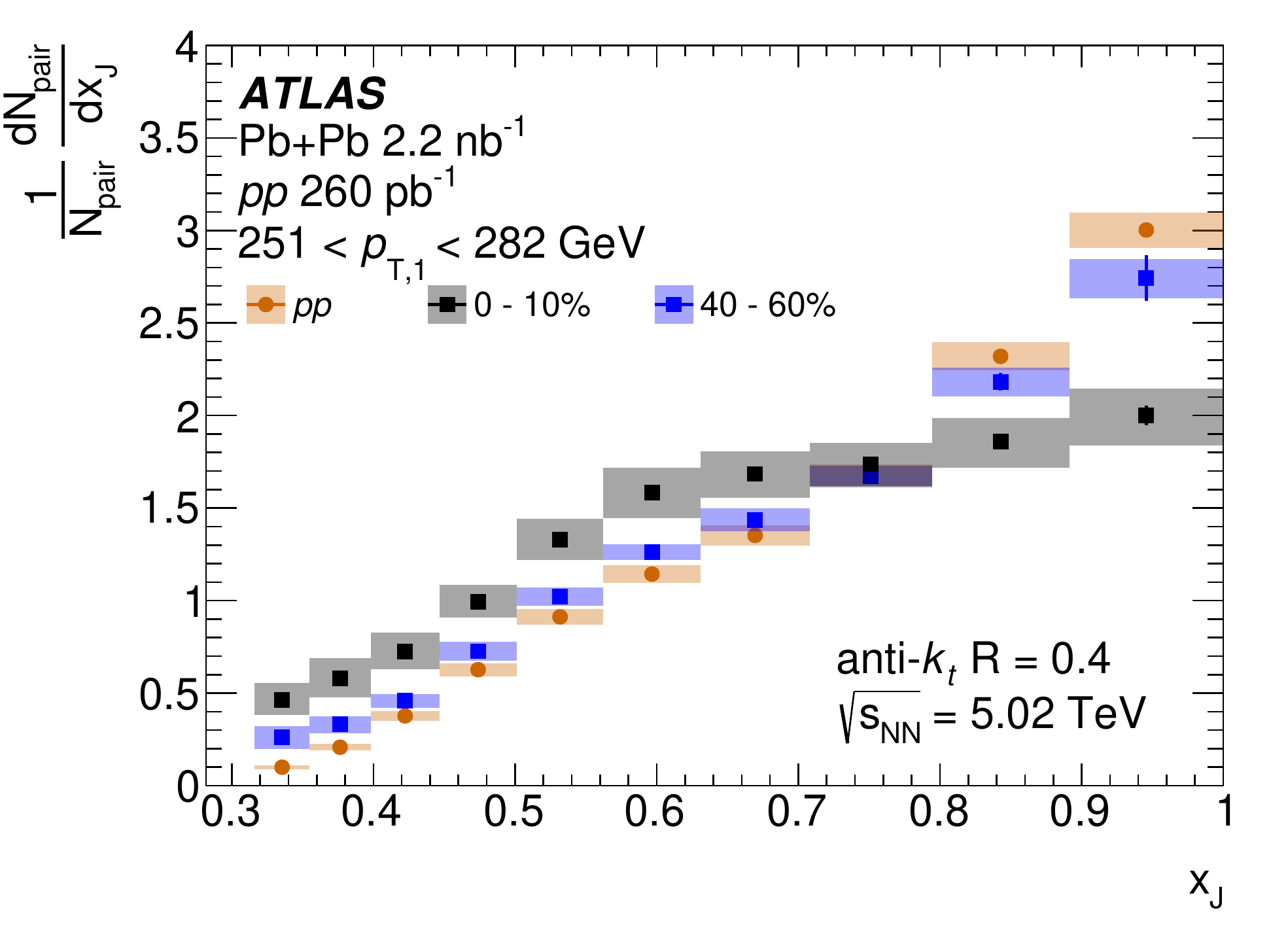}
}
\hfill
\subfloat{
\centering
\includegraphics[width=0.33\textwidth]{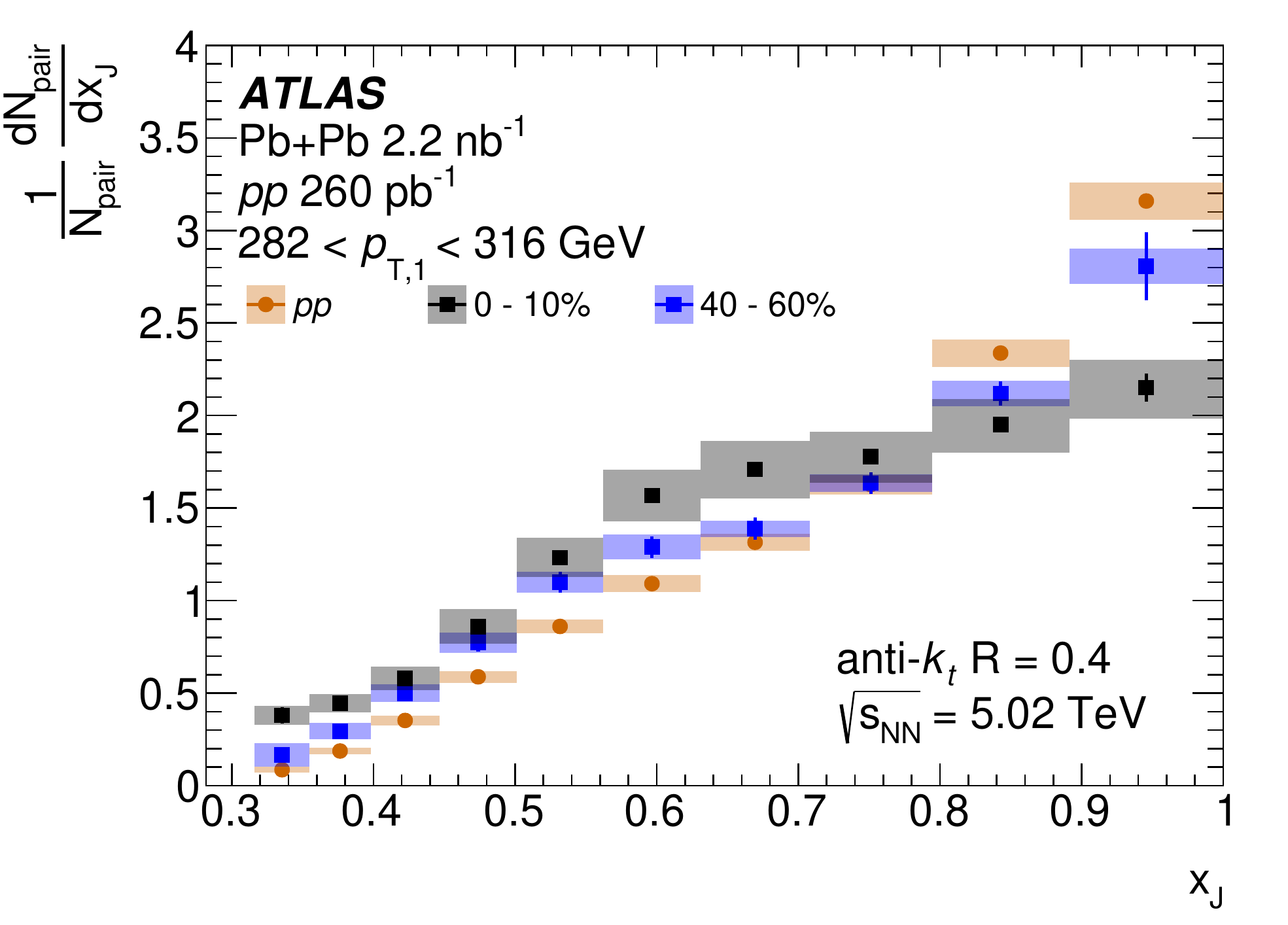}
}
\subfloat{
\centering
\includegraphics[width=0.33\textwidth]{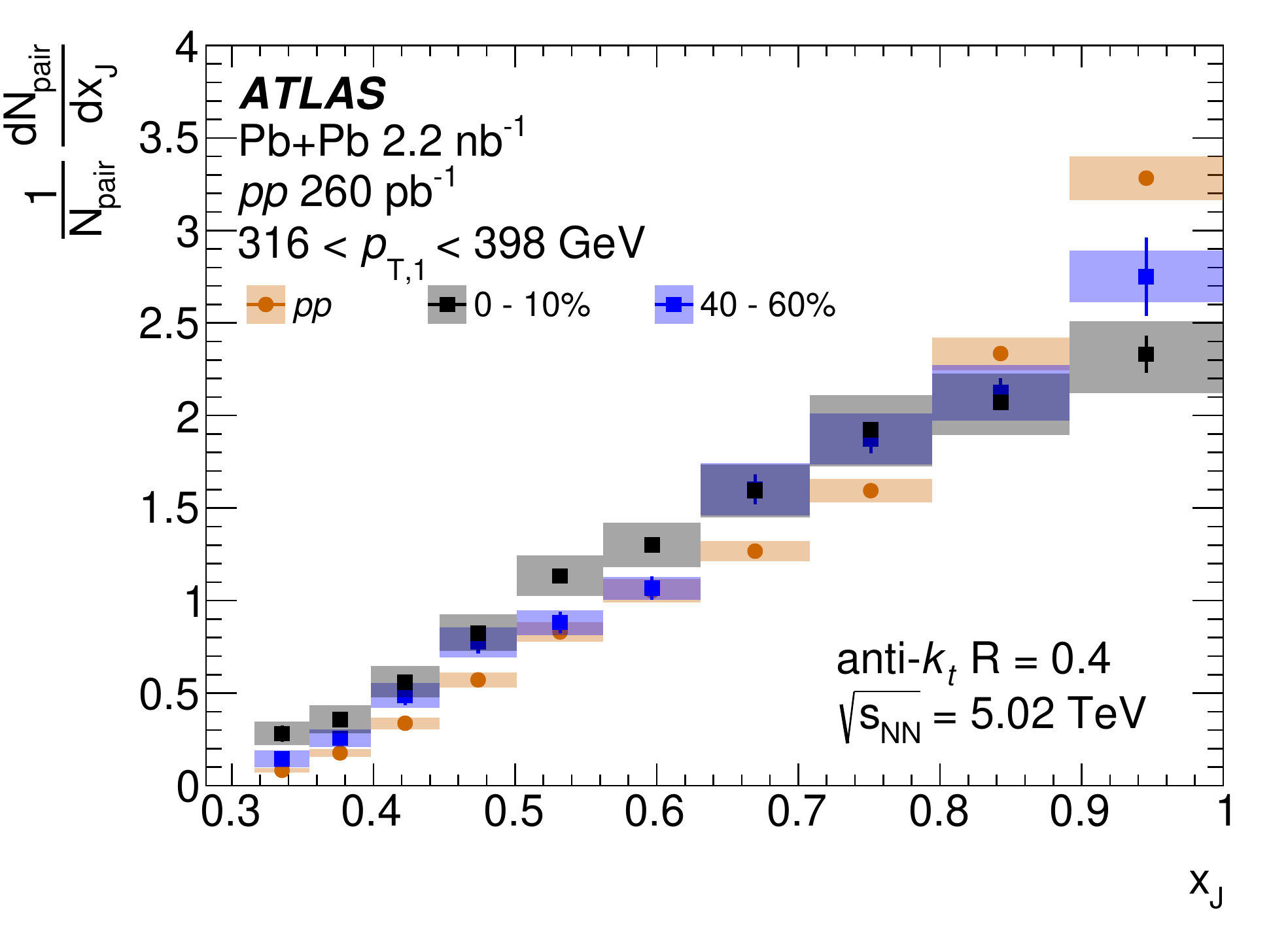}
}
\subfloat{
\centering
\includegraphics[width=0.33\textwidth]{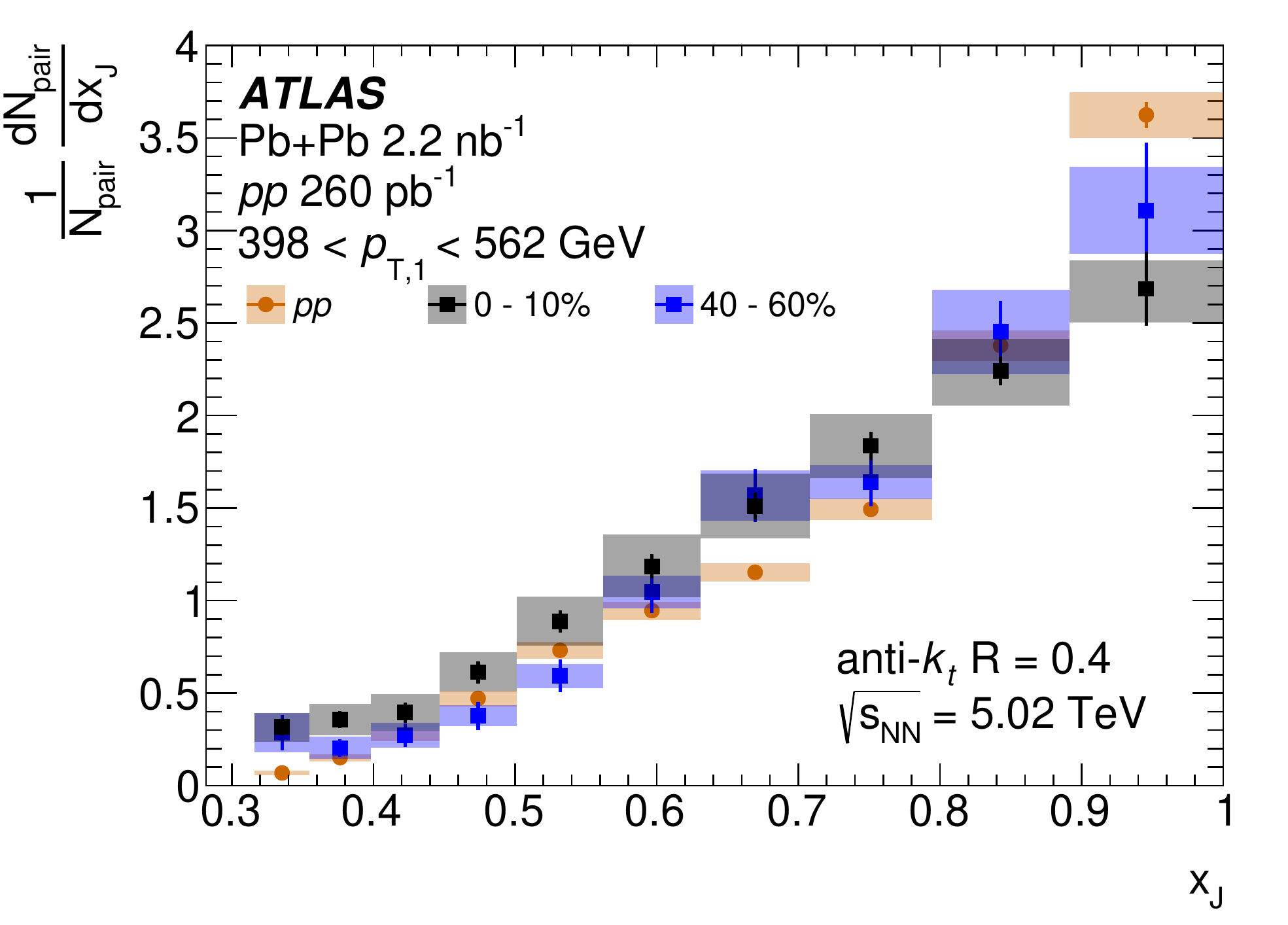}
}
\hfill
\caption{\label{fig:result} The dijet-yield-normalized \dNdxJ\ distributions in 0--10\% and 40--60\% centrality
\pbpb\ collisions and \pp\ collisions for
twelve
\pta\ selections from 100~\GeV\ to 562~\GeV.
Statistical uncertainties are shown as
error bars (which are smaller than the markers in many cases)
and systematic uncertainties are shown as boxes.}
\end{figure}

Figure~\ref{fig:result} shows the measured dijet-yield-normalized
\xJ\ distributions in \pp\ collisions and as a function of centrality in \pbpb\ collisions
for twelve \pta\ selections between 100 and 562~\GeV. In central
collisions, at $100 < \pta < 112$~\GeV\ a broad maximum is observed in
the distributions at \xJ\ around 0.6. With increasing \pta\ as well as
with decreasing centrality, this feature becomes less pronounced.
Significant modification from \pp\ collisions is observed up to $398
< \pta < 562$~\GeV\ in the full centrality range; the
\xj\ distributions have a reduced fraction of balanced jets and an
increased fraction of imbalanced jets.
 
\begin{figure}[t]
\centering
\centering
\includegraphics[width=0.45\textwidth]{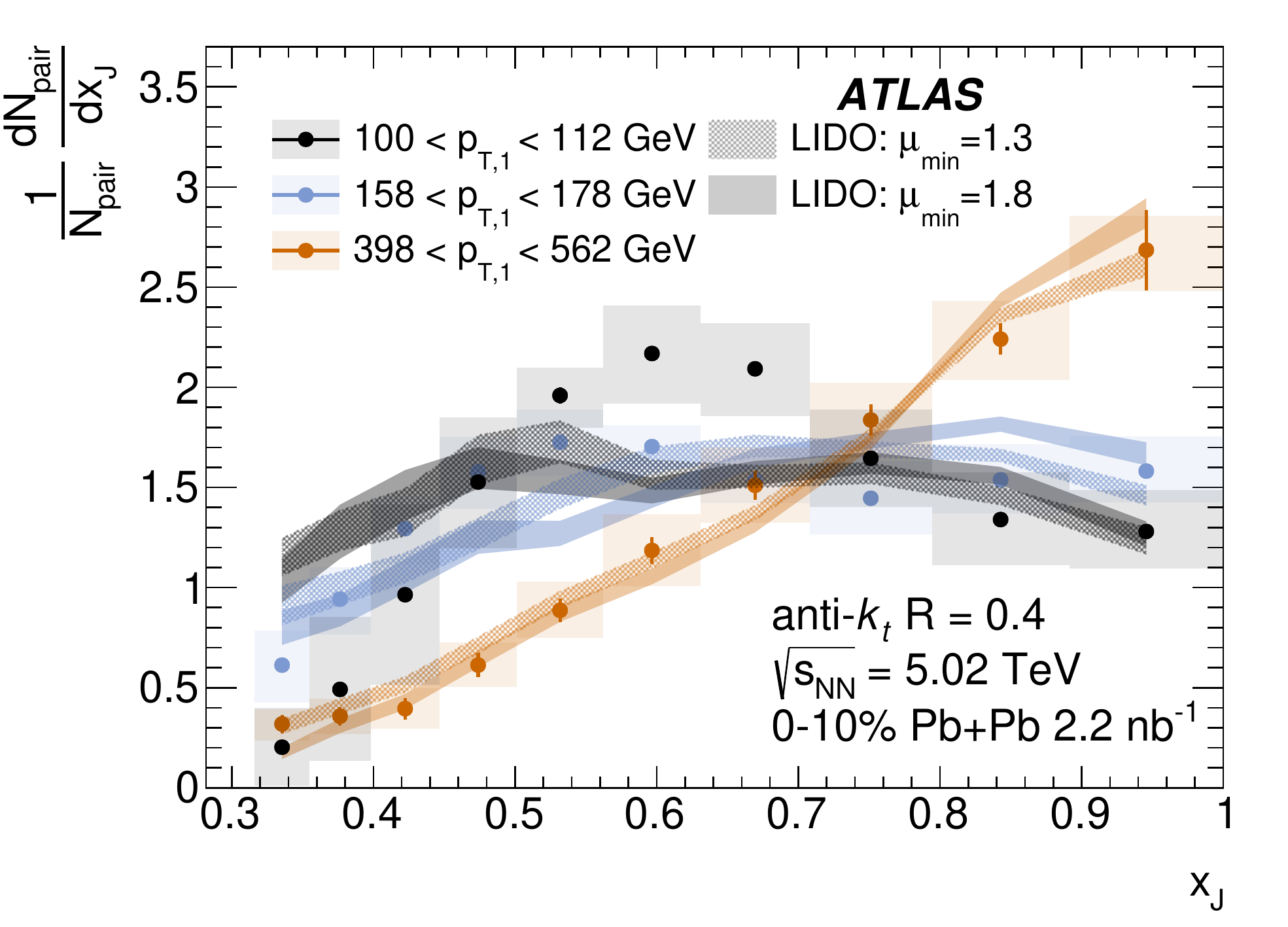}
\includegraphics[width=0.45\textwidth]{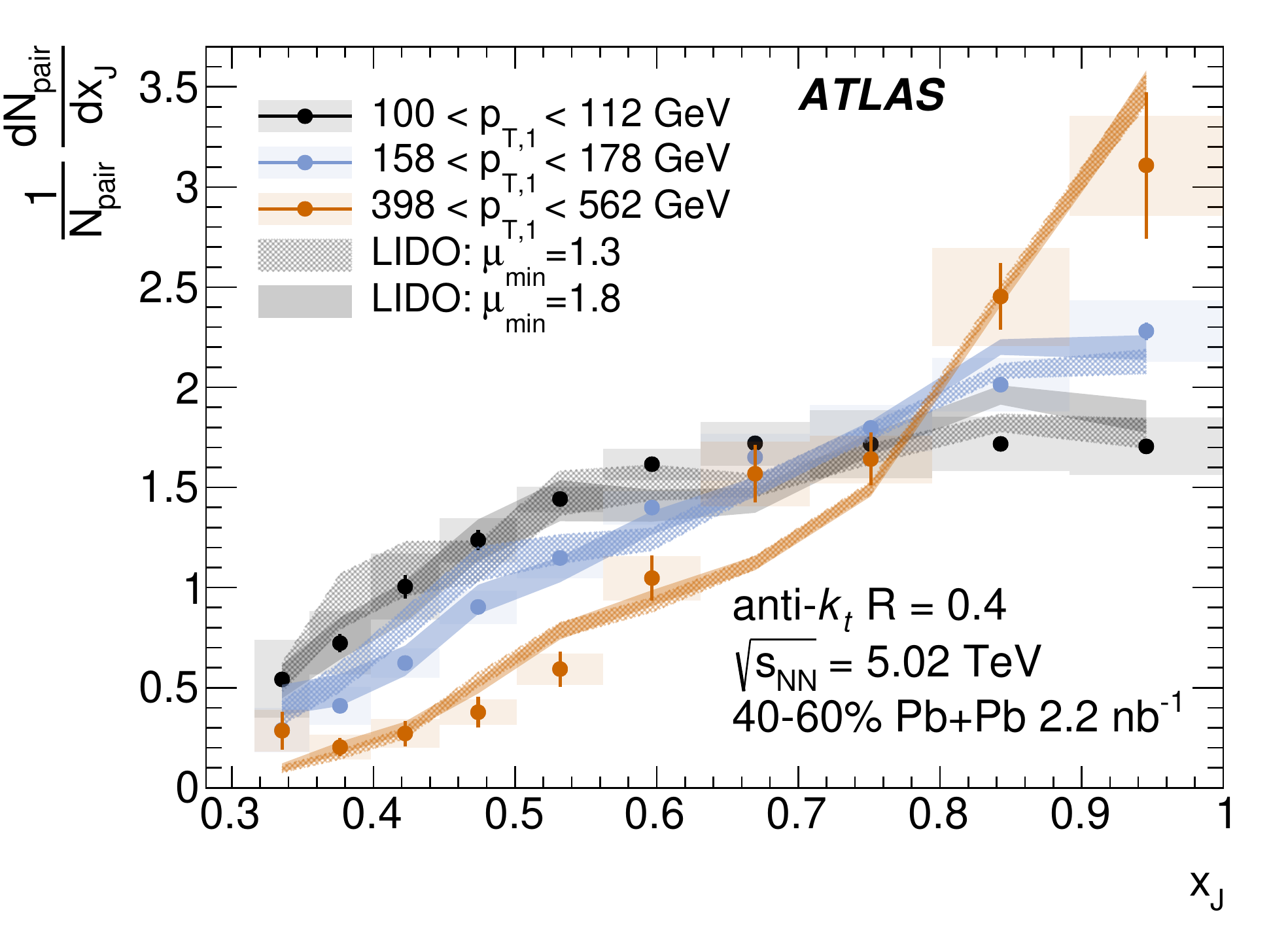}
\caption{\label{fig:theorycomp} Comparison of the dijet-yield-normalized \xJ\ distributions in 0--10\% (left) and 40--60\% (right) centrality \pbpb\ collisions with predictions from LIDO~\cite{Ke:2020clc} for two values of the jet--medium coupling cutoff parameter, $\mu_\text{min} = 1.3$, and $\mu_\text{min} = 1.8$.
Statistical uncertainties are shown as
error bars (which are smaller than the markers in many cases)
and systematic uncertainties are shown as boxes.
}
\end{figure}

Figure~\ref{fig:theorycomp} compares the measured distributions in  \pbpb\
collisions with the predictions from
the LIDO model~\cite{Ke:2020clc}. LIDO is a transport model including both elastic jet--medium collisions and medium-induced radiative processes, with event-by-event fluctuations implemented for both processes,
as well as a simple model for the response of the medium. LIDO uses \pythiaeight~\cite{SJOSTRAND2015159}
with the A14 tune~\cite{ATL-PHYS-PUB-2014-021} and the CTEQL1 PDF~\cite{Pumplin:2002vw} to initialize their energy-loss calculation.
The calculations are performed using two high likelihood values for the jet--medium coupling cutoff parameter: $ \mu_{\mathrm{min}} = 1.3$ and $\mu_{\mathrm{min}} = 1.8$. This parameter is related to the strength of the coupling between the jet and the medium, where a smaller $\mu_{\mathrm{min}}$ value corresponds to a larger coupling, and was constrained using inclusive jet and hadron suppression measurements~\cite{Ke:2020clc}. Only minimal sensitivity to the jet--medium coupling cutoff parameter is observed in the dijet-yield-normalized \xJ\ distributions, with $\mu_{\mathrm{min}} = 1.3$ and $\mu_{\mathrm{min}} = 1.8$ demonstrating equal ability to reproduce the measured results. The model reproduces the behavior observed at high \pta\ and in 40--60\% centrality \pbpb\ collisions while not reproducing the relative enhancement of intermediate \xJ\ observed at low \pta\ in 0--10\% centrality \pbpb\ collisions.

\begin{figure}[t]
\centering
\centering
\includegraphics[width=0.6\textwidth]{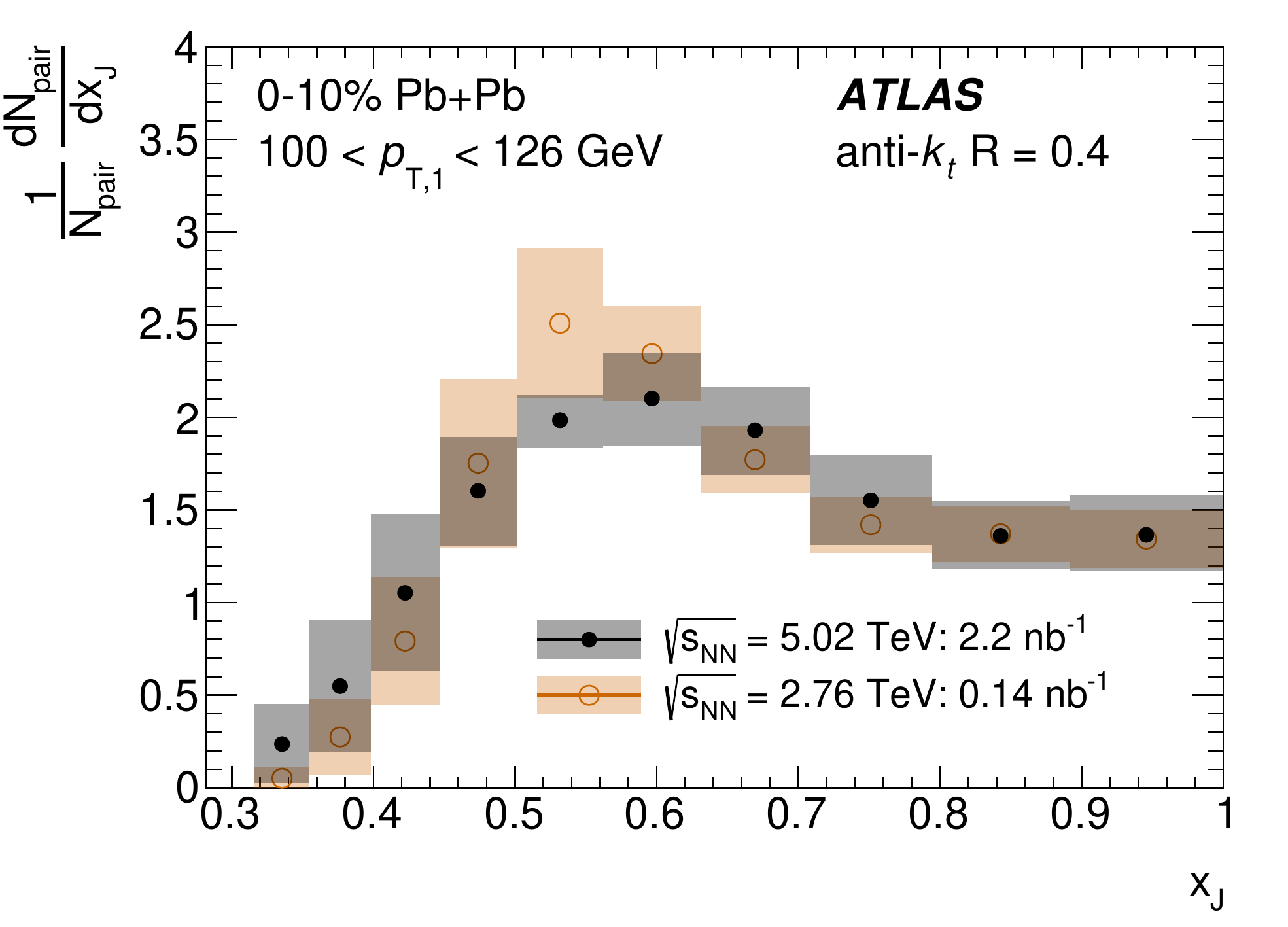}
\includegraphics[width=0.6\textwidth]{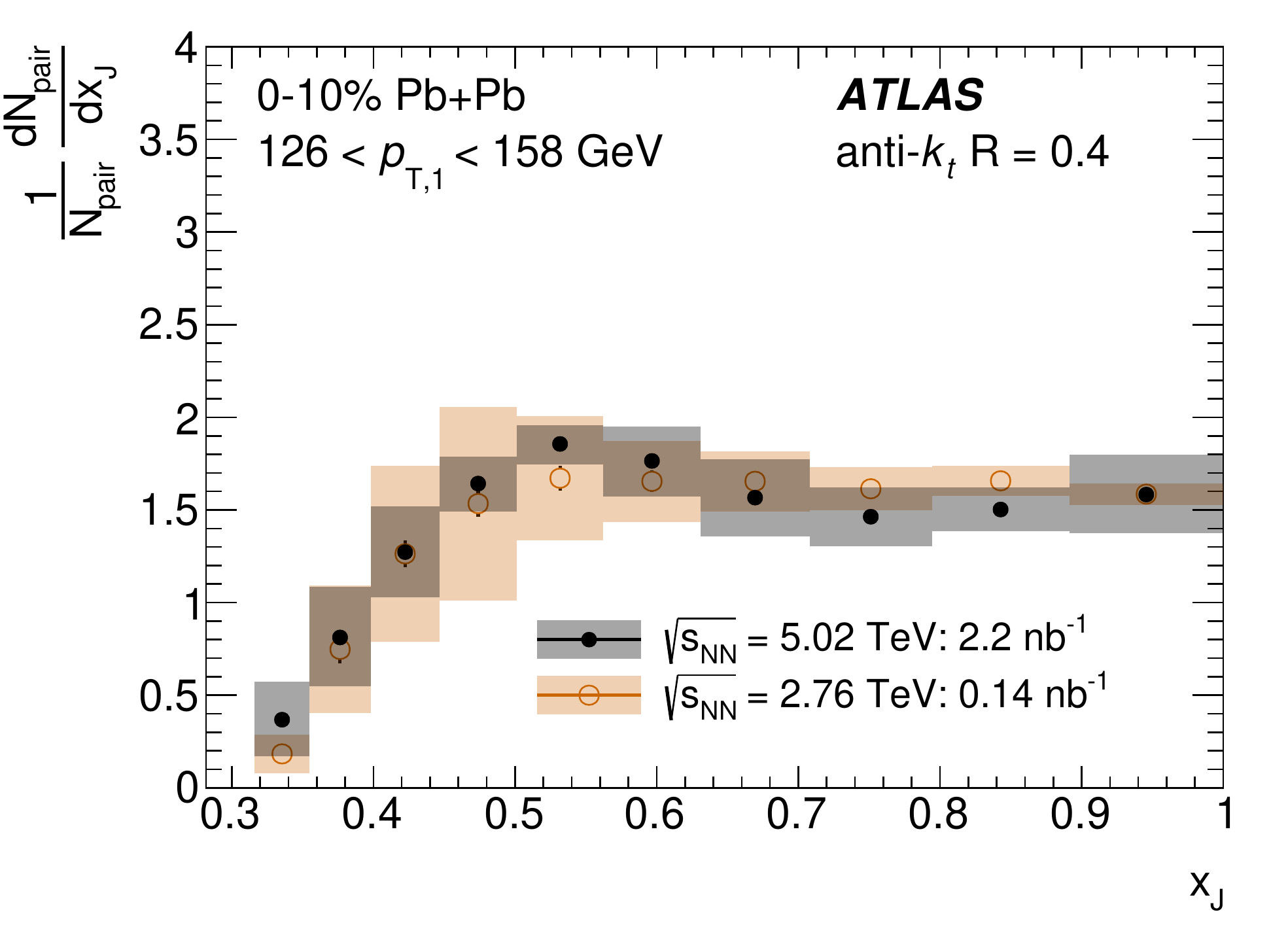}
\includegraphics[width=0.6\textwidth]{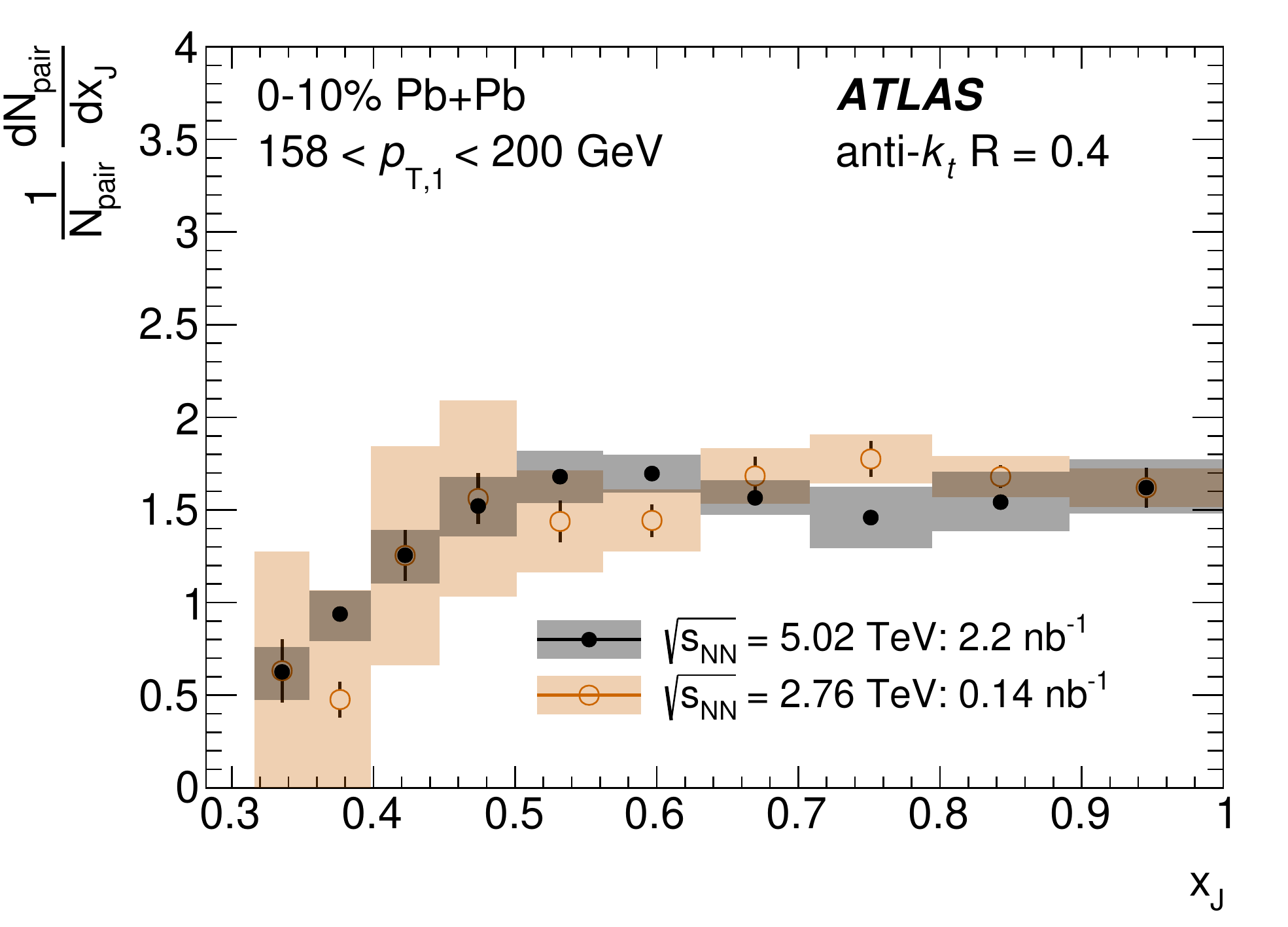}
\caption{\label{fig:central_pbpb_comparison} Comparison of the measured \xJ\
distribution at 5.02~\TeV\ in 0--10\% centrality collisions for
100~$<\pta<$~126~\GeV\ (top), 126~$<\pta<$~158~\GeV\ (middle), and
158~$<\pta<$~200~\GeV\ (bottom) with those at $2.76$~\TeV\ from Ref.~\cite{ATLAS:2017xfa}.
Statistical uncertainties are shown as
error bars (which are smaller than the markers in many cases)
and systematic uncertainties are shown as boxes.}
 
\end{figure}

Figure~\ref{fig:central_pbpb_comparison} shows a direct comparison
between the dijet-yield-normalized \xJ\ distributions in 0--10\% centrality
\pbpb\ collisions at 5.02~\TeV\  for $100 < \pta < 200 $~\GeV\  and
results at 2.76~\TeV\ from Ref.~\cite{ATLAS:2017xfa} for the same
collision centrality and \pta\ selections.
Although these two measurements are performed at different
collision energies, similar levels of quenching are
expected, based measurements of inclusive jets~\cite{ATLAS:2014ipv,ATLAS:2018gwx}.
The  \xj\ distributions are observed to be consistent within systematic uncertainties
at the two collision energies.
For 100~$<\pta<$~126~\GeV, a peak is observed at intermediate \xJ\ for both collision
energies.

\subsection{Absolutely normalized \xJ\ distributions}

\begin{figure}[t]
\centering
\centering
\includegraphics[width=0.6\textwidth]{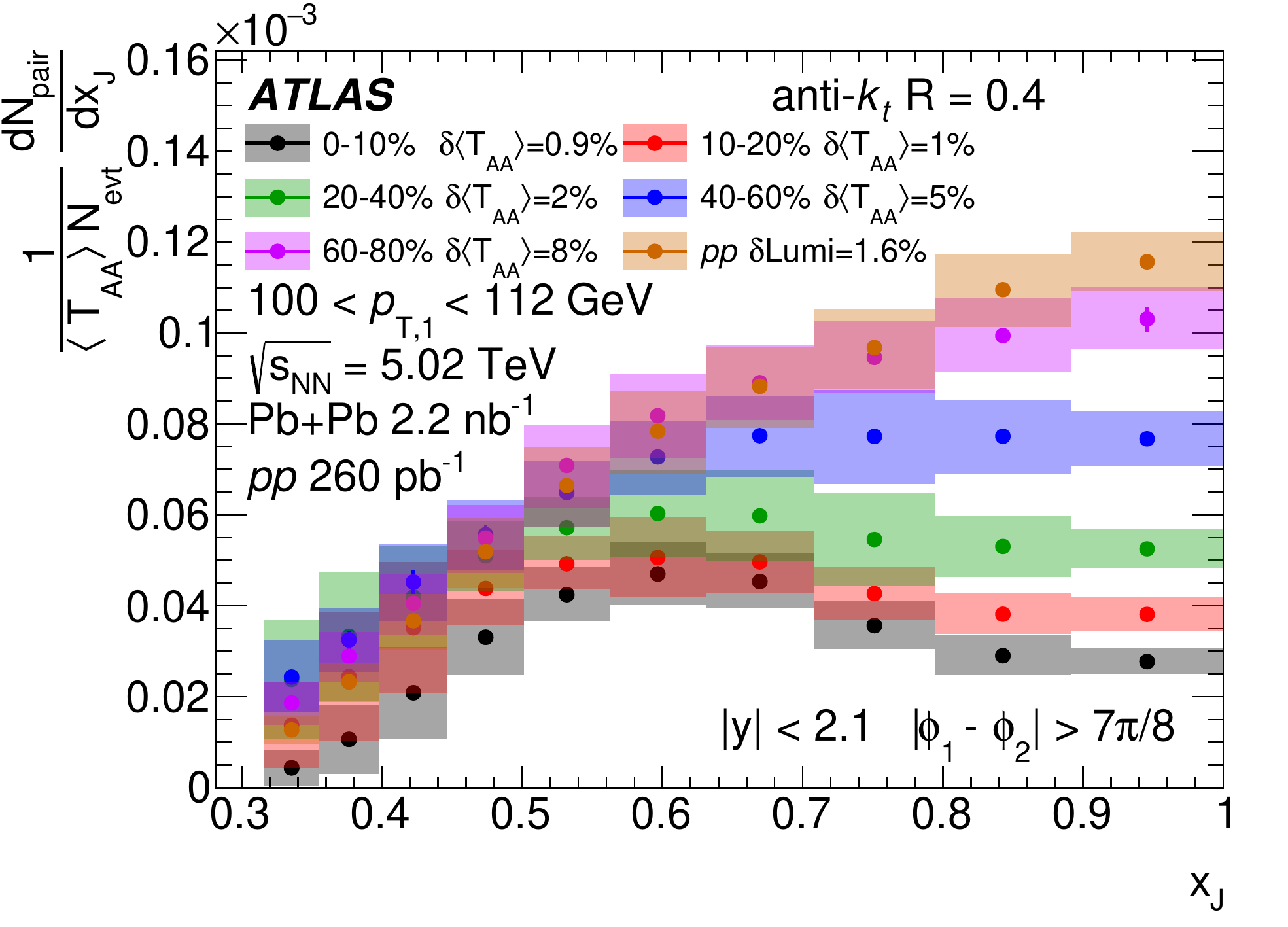}
\hfill
\includegraphics[width=0.6\textwidth]{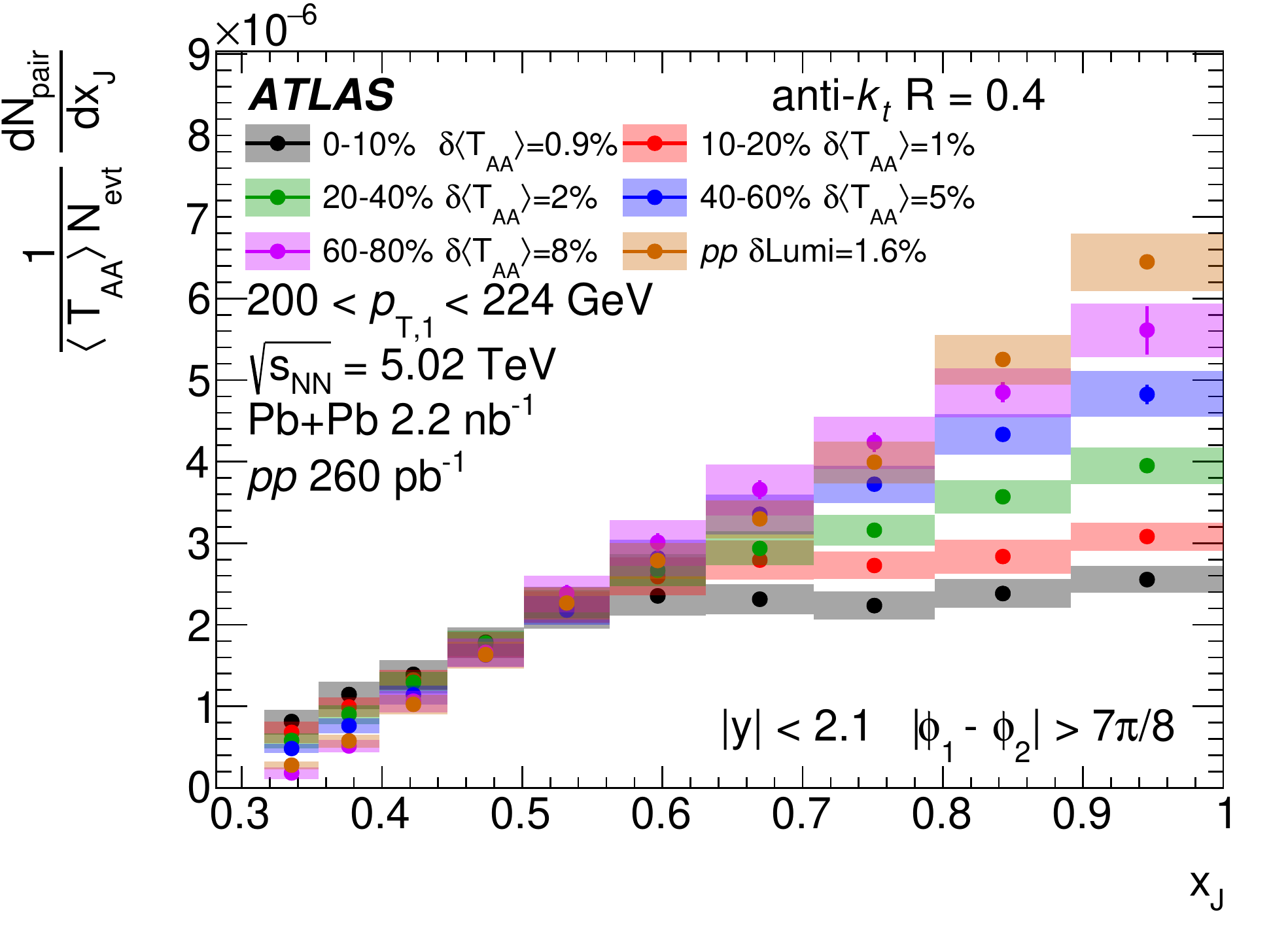}
\hfill
\includegraphics[width=0.6\textwidth]{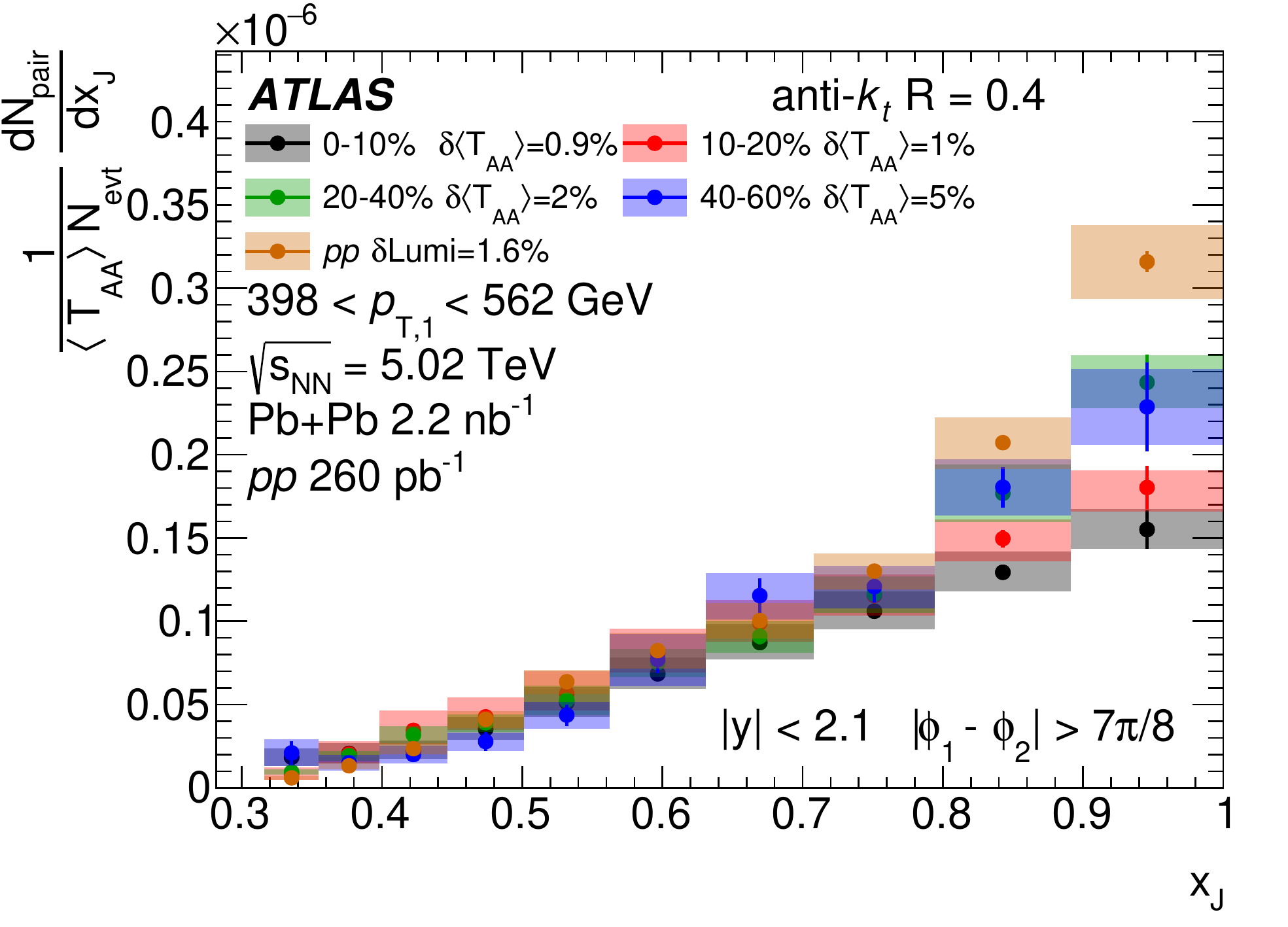}
\hfill

\caption{\label{fig:per_event_xJ} The absolutely normalized \xJ\ distributions for $100 < \pta < 112$~\GeV\ (top), $200 < \pta < 224$~\GeV\ (center), and $398 < \pta < 562$~\GeV\ (bottom) in \pp\ collisions and for five centrality selections in \pbpb\ collisions. Statistical uncertainties are shown as error bars (which are smaller than the markers in many cases) and systematic uncertainties are shown as boxes. The global normalization uncertainties due to the \TAA\ for each centrality as well as the \pp\ luminosity are listed in the legends.}
 
\end{figure}

The absolutely normalized \xJ\ distributions, as calculated with Exps.~(\ref{eq:absxjpbpb}) and~(\ref{eq:absxjpp})
for \pbpb\ and \pp\ collisions respectively,
allow a direct comparison of the absolute dijet yields between \pbpb\ and \pp\
collisions as a function of \xJ.
Figure~\ref{fig:per_event_xJ} shows these distributions for \pp\ and \pbpb\ collisions
as a function of centrality for three \pta\ selections: 100--112~\GeV, 200--224~\GeV, and
398--562~\GeV. Balanced dijets are suppressed in central \pbpb\ collisions relative to asymmetric dijets,
leading to the peak structure observed in the dijet-yield-normalized \xJ\ distributions at low \pta.
The depletion of balanced dijets persists over all \pta\ and centrality selections measured here, with
the magnitude of the depletion decreasing with increasing \pta\ and towards more
peripheral collisions.
When going from central \pbpb\ collisions to \pp\ collisions for 100~$ < \pta\ <$~112~\GeV,
the suppression of dijets is smaller for $\xJ < 0.6$ than for higher \xJ.  For higher \pta, there is
a smaller difference between \pp\ and \pbpb\ collisions at low \xJ.

\begin{figure}[t]
\centering
\centering
\includegraphics[width=0.6\textwidth]{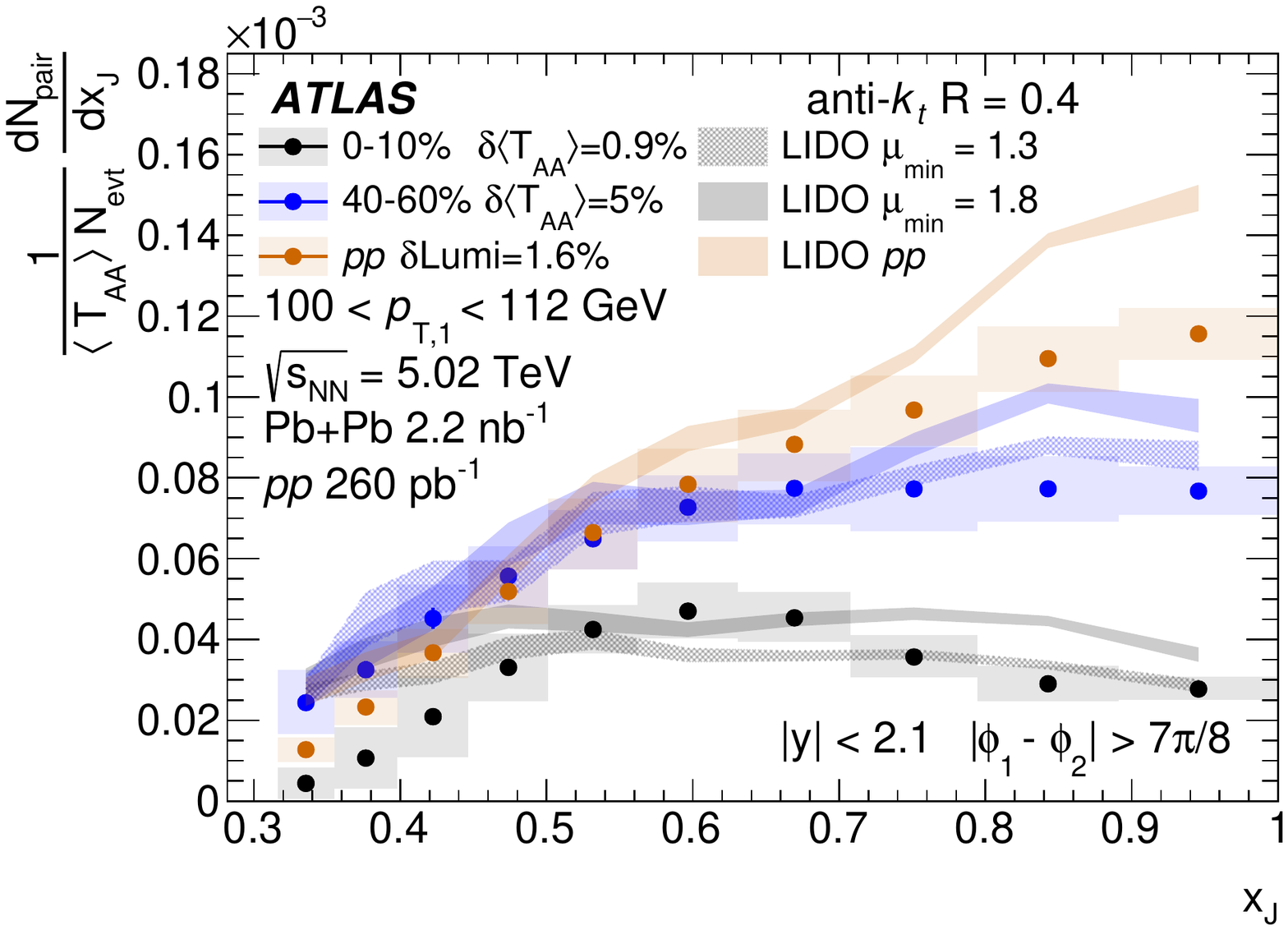}
\hfill
\includegraphics[width=0.6\textwidth]{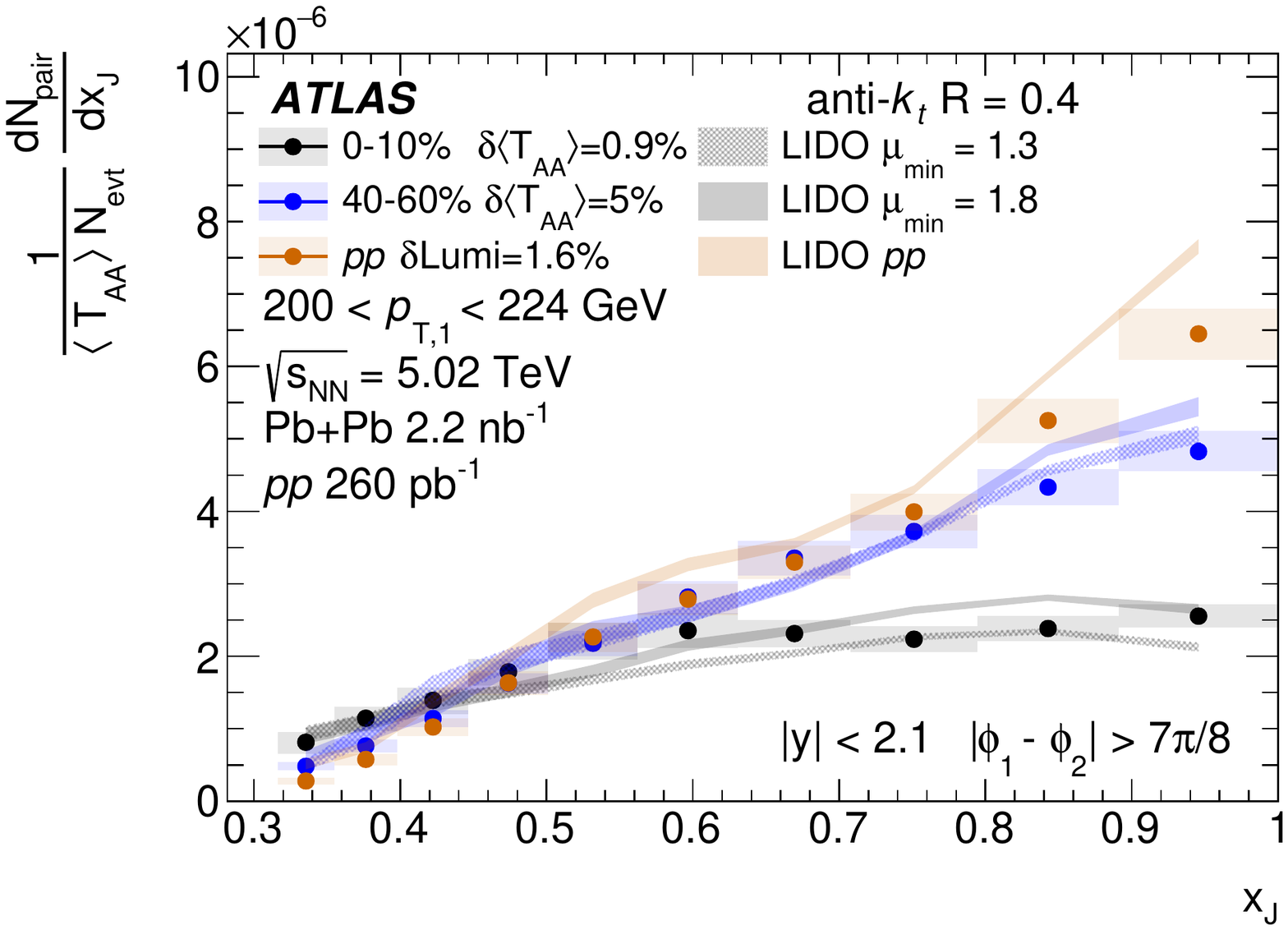}
\hfill
\includegraphics[width=0.6\textwidth]{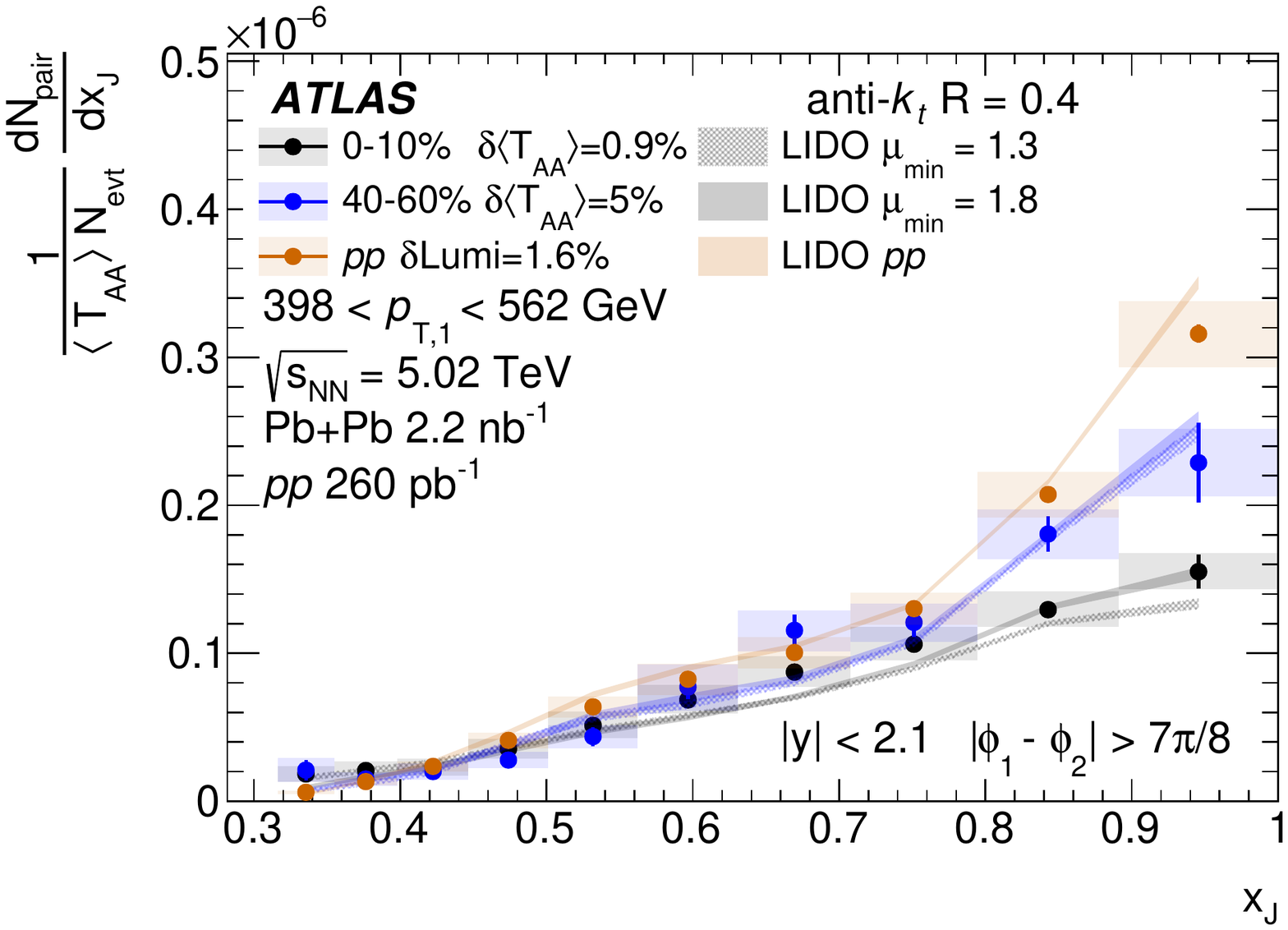}
\hfill

\caption{\label{fig:per_event_xJ_lido} The absolutely normalized \xJ\ distributions for $100 < \pta < 112$~\GeV\ (top), $200 < \pta < 224$~\GeV\ (center), and $398 < \pta < 562$~\GeV\ (bottom)
in \pp\ collisions and for two centrality selections in \pbpb\ collisions.
For \pbpb\ collisions, the measurements are compared with predictions from
LIDO~\cite{Ke:2020clc} for two values of the jet--medium coupling cutoff parameter, $\mu_\text{min} = 1.3$ and $\mu_\text{min} = 1.8$, and the \pp\ collision measurements are compared with the LIDO
\pp\ collision calculation.}

\end{figure}

Figure~\ref{fig:per_event_xJ_lido} shows the absolutely normalized \xJ\ distributions
compared with the same LIDO calculations~\cite{Ke:2020clc} as in the previous section for
three \pta\ selections from 100 to 562~\GeV.  The LIDO \pp\ calculation
is also shown.  For \pp\ collisions, LIDO uses \pythiaeight~\cite{SJOSTRAND2015159}
with the A14 tune~\cite{ATL-PHYS-PUB-2014-021} and the CTEQL1 PDF~\cite{Pumplin:2002vw}.
The LIDO model overestimates the \pp\ collision results at large \xJ; the
size of the discrepancy between the model and the data decreases with increasing \pta.  This
trend persists, but with a smaller difference in 40--60\%
centrality \pbpb\ collisions.  In 0--10\% centrality \pbpb\ collisions,
the LIDO model's prediction is flatter than the data in the lowest \pta\ range, but agrees
well at higher \pta.

\subsection{\RAApair\ of leading and subleading jets}

\begin{figure}[ht]
\centering
\subfloat{
\centering
\includegraphics[width=0.6\textwidth]{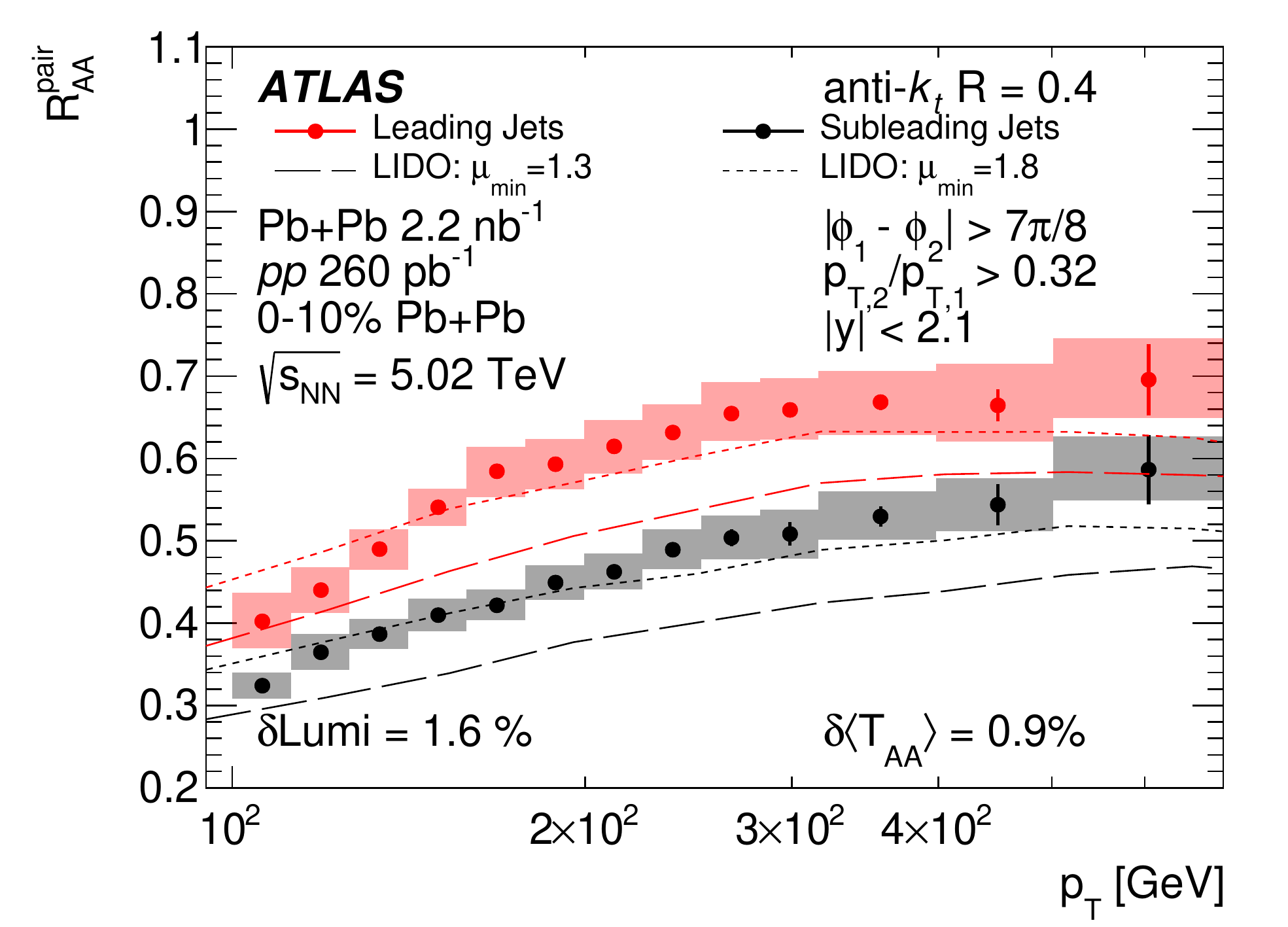}
}
\hfill
\subfloat{
\centering
\includegraphics[width=0.6\textwidth]{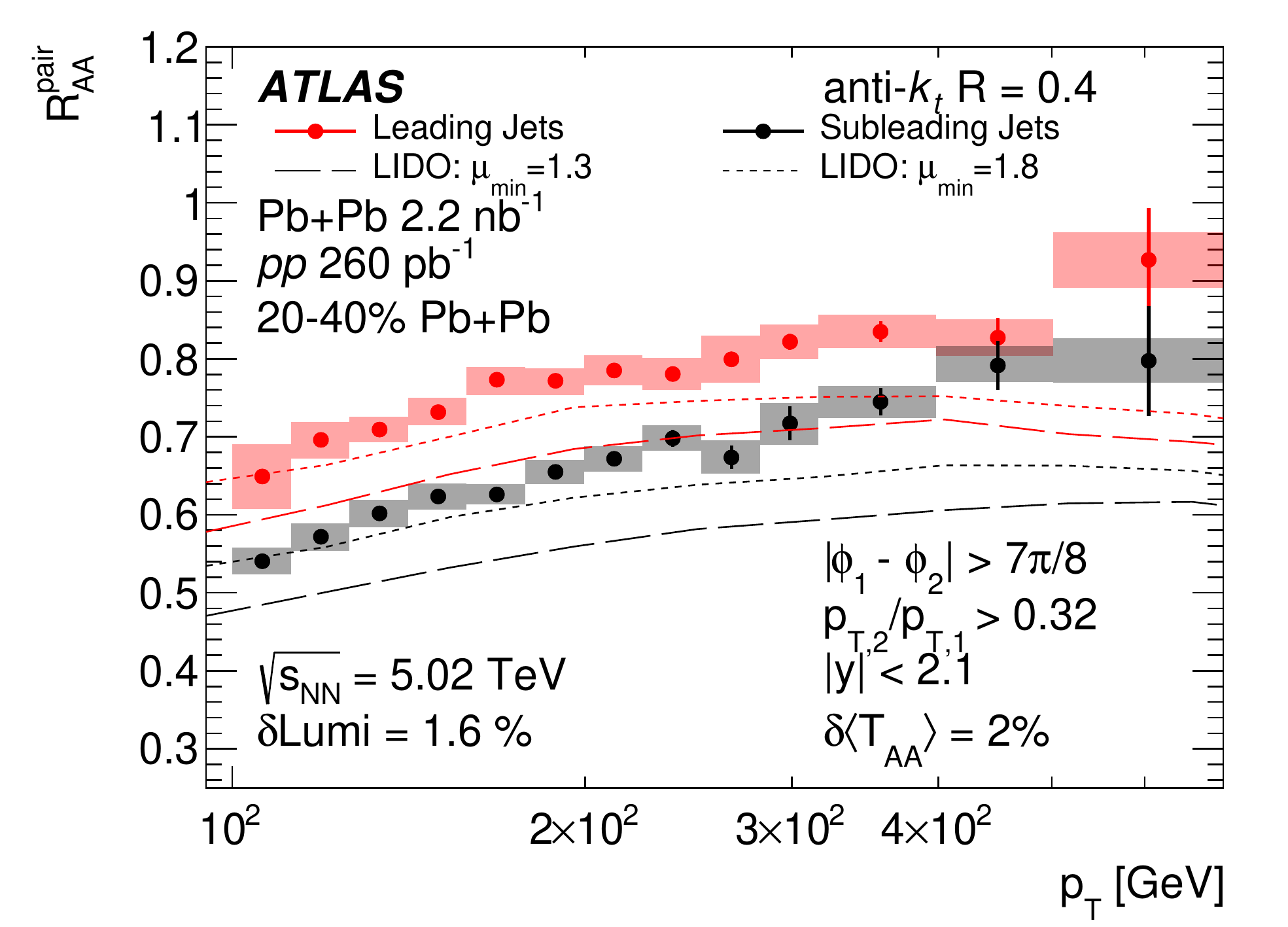}
}
\hfill
 
\subfloat{
\centering
\includegraphics[width=0.6\textwidth]{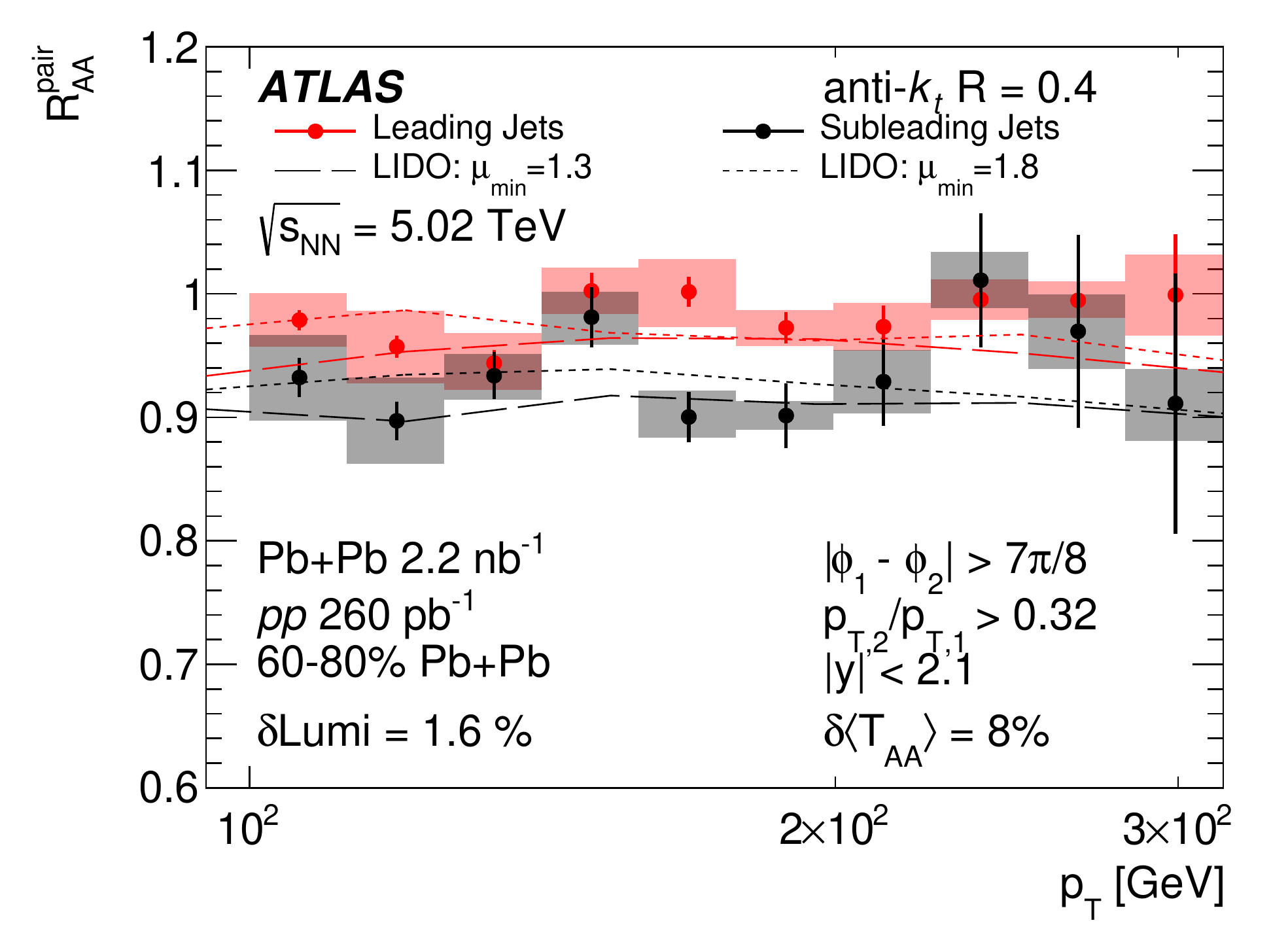}
}
 
\caption{\label{fig:result_leadsublead_raa}The leading and subleading jet \RAApair\ distributions measured for a selection of centrality regions compared with predictions from
LIDO~\cite{Ke:2020clc}.
Statistical uncertainties are shown as
error bars (which are smaller than the markers in many cases)
and systematic uncertainties are shown as boxes.}
\end{figure}

The pair nuclear modification factors projected for leading and
subleading jets, as defined in Eqs.~(\ref{eq:raa_lead})
and~(\ref{eq:raa_sublead}), are shown for three centrality selections
as a function of the jet \pt\ in
Figure~\ref{fig:result_leadsublead_raa}.
Both \RAAlead\ and \RAAsublead\ show significant modification from
a value of one in all but the most peripheral collisions, with a stronger
suppression observed at low \ptjet.
It is observed that the subleading jets are more suppressed than
leading jets, particularly at $\pt >$~200~\GeV\ where nearly all the
inclusive jets (within the rapidity range of this analysis) are
included in the dijet sample.
This behavior is qualitatively reproduced in predictions made by the
LIDO~\cite{Ke:2020clc} model.  In the LIDO model, the jet--medium
coupling cutoff parameter $\mu_\text{min}=1.3$ clearly predicts a larger
suppression for both jets than is observed in the data.  The
calculation with $\mu_\text{min}=1.8$ is closer to the data but does not
describe the trends with \pt\ and centrality.
 
\begin{figure}[ht]
\centering
\includegraphics[width=0.85\textwidth]{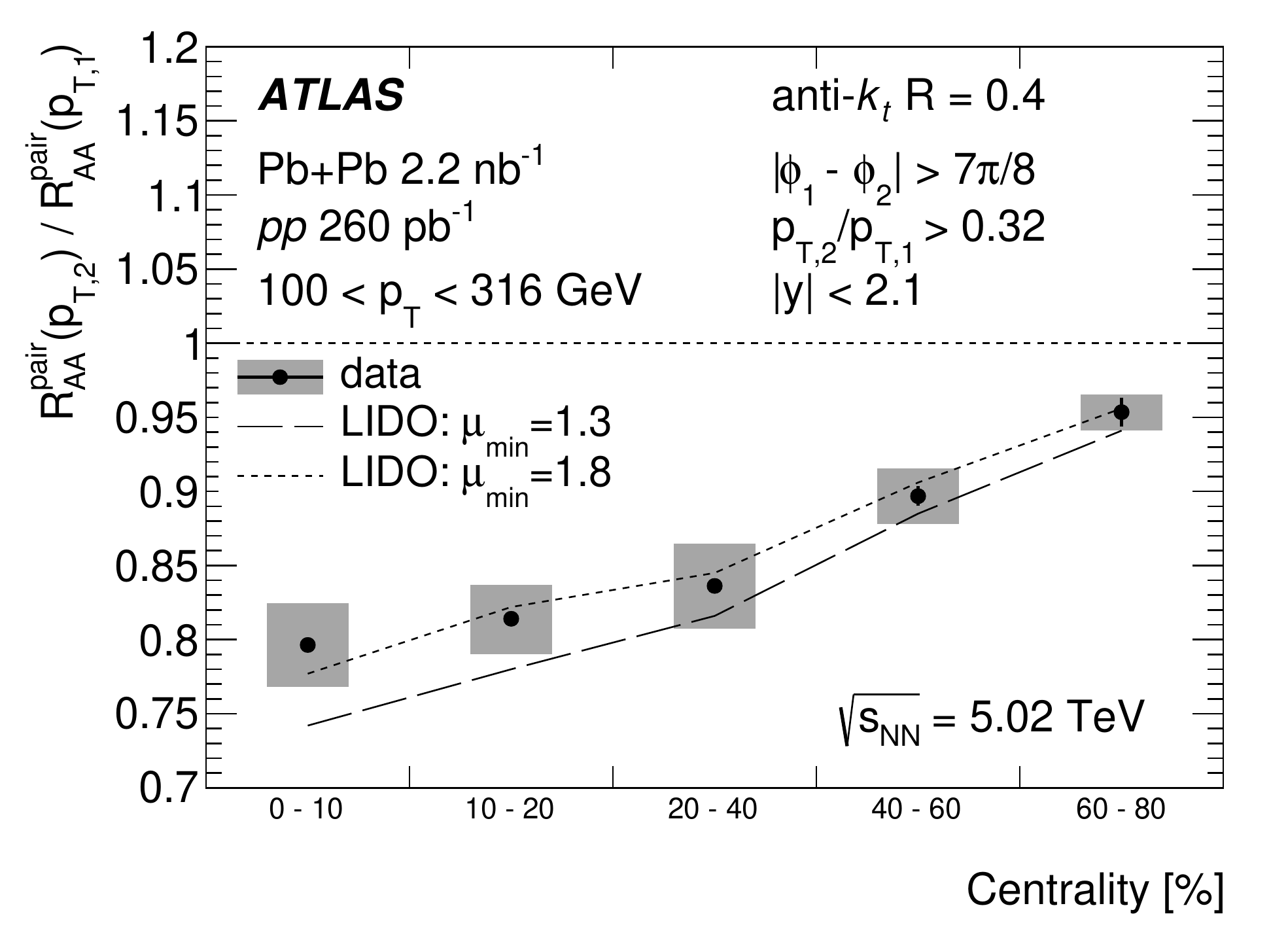}
\caption{\label{fig:raa_doubleratio} The ratio of the subleading-jet to leading-jet nuclear modification factors integrated over jet \pt\ between 100 and 316~\GeV.
Statistical uncertainties are shown as
error bars (which are smaller than the markers in many cases)
and systematic uncertainties are shown as boxes.}
\end{figure}

The different levels of suppression observed for leading and subleading jets is quantified as a function of centrality in the ratio
$\RAAsublead/\RAAlead$ integrated over \ptjet\ between 100 and 316~\GeV\ as shown in Figure \ref{fig:raa_doubleratio},
where \RAAsublead\ is seen to be smaller than \RAAlead\ in all centrality selections. The suppression
of subleading jets relative to leading jets is most significant in 0--10\% centrality \pbpb\ collisions
and decreases smoothly towards peripheral
\pbpb\ collisions. In 0--10\% centrality \pbpb\ events, subleading jets
are approximately 20\% more suppressed than leading jets, with 60--80\% centrality \pbpb\
events seeing only about 5\% more suppression for subleading jets than for leading jets. The
value of  $\RAAsublead/\RAAlead$ in 60--80\% centrality \pbpb\ events is 3$\sigma$ away from one,
providing evidence of jet quenching in peripheral \pbpb\ collisions. In contrast,
in inclusive jet
measurements, due to the larger uncertainties, no significant suppression is observed in this
\pt\ range for the most peripheral \pbpb\ collisions~\cite{ATLAS:2018gwx}.
This is in qualitative agreement with results from CMS~\cite{CMS-HIN-19-013}, which find
that, compared to \pp\ collisions, the subleading jets in dijet events in \pbpb\ collisions have
stronger modification to their structure than leading jets.
 
\clearpage
 
\section{Conclusion}
 
This paper presents a measurement of the unfolded double-differential yields of leading dijets.
The yields are presented as measurements of the dijet \xJ\ distributions and the \RAApair\ projected for leading and subleading jets, using 2.2~\invnb\ of \pbpb\ data and 260~\invpb\ of \pp\ data
collected at $\sqrtsnn = 5.02$~\TeV\ by the ATLAS detector at the LHC.
The measurement of the \xJ\ distributions is performed differentially in \pta\ and centrality, and the \RAA\ of leading and subleading jets is measured differentially in jet \pt\ and centrality. The presented results are unfolded to correct for detector response and inefficiencies.
 
The dijet-yield-normalized \xJ\ distributions, across the full
\pta\ range considered in this analysis, show significant broadening
in central \pbpb\ collisions relative to those in \pp\ collisions.
Up to the highest \pta\ interval considered in this measurement, $398 < \pta <
562$~\GeV, significant modification from the distributions in
\pp\ collisions is observed for central \pbpb\ collisions.  With
decreasing \pta\ the modifications from the distributions in
\pp\ collisions become larger, with significant modifications observed
for all centralities at $ 100 < \pta < 112$~\GeV. In 0--10\% centrality
\pbpb\ collisions, between 100 and 112~\GeV, the distribution is
peaked around \xJ\ of 0.6, indicating that the most probable
configuration for dijets is to be highly imbalanced in \pt.
Measurements of absolutely normalized \xJ\ distributions show that
this comes from the suppression of balanced dijets, as opposed to an
enhancement of imbalanced jets, in comparison with \pp\ collisions.
This suppression of balanced
dijets evolves smoothly as a function of the event centrality, with
peripheral \pbpb\ collisions demonstrating only slight modification
from \pp\ collisions. The measurements of the pair nuclear modification factor, \RAAlead\ and
\RAAsublead, show that subleading jets are more
suppressed than leading jets. This is observed to be largely
independent of jet \pt\ and this behavior is seen in all centralities.
In central collisions, subleading jets are observed to be 20\% more
suppressed than the leading jets.  In peripheral collisions, the
suppression of subleading jets relative to leading jets is reduced,
but the stronger suppression of subleading jets remains significant.
These measurements provide new information about the role of
path-length dependence and fluctuations in jet energy loss which will
help constrain models of parton energy loss in the quark--gluon plasma.


\section*{Acknowledgments}


We thank CERN for the very successful operation of the LHC, as well as the
support staff from our institutions without whom ATLAS could not be
operated efficiently.
 
We acknowledge the support of
ANPCyT, Argentina;
YerPhI, Armenia;
ARC, Australia;
BMWFW and FWF, Austria;
ANAS, Azerbaijan;
CNPq and FAPESP, Brazil;
NSERC, NRC and CFI, Canada;
CERN;
ANID, Chile;
CAS, MOST and NSFC, China;
Minciencias, Colombia;
MEYS CR, Czech Republic;
DNRF and DNSRC, Denmark;
IN2P3-CNRS and CEA-DRF/IRFU, France;
SRNSFG, Georgia;
BMBF, HGF and MPG, Germany;
GSRI, Greece;
RGC and Hong Kong SAR, China;
ISF and Benoziyo Center, Israel;
INFN, Italy;
MEXT and JSPS, Japan;
CNRST, Morocco;
NWO, Netherlands;
RCN, Norway;
MEiN, Poland;
FCT, Portugal;
MNE/IFA, Romania;
MESTD, Serbia;
MSSR, Slovakia;
ARRS and MIZ\v{S}, Slovenia;
DSI/NRF, South Africa;
MICINN, Spain;
SRC and Wallenberg Foundation, Sweden;
SERI, SNSF and Cantons of Bern and Geneva, Switzerland;
MOST, Taiwan;
TENMAK, T\"urkiye;
STFC, United Kingdom;
DOE and NSF, United States of America.
In addition, individual groups and members have received support from
BCKDF, CANARIE, Compute Canada and CRC, Canada;
PRIMUS 21/SCI/017 and UNCE SCI/013, Czech Republic;
COST, ERC, ERDF, Horizon 2020 and Marie Sk{\l}odowska-Curie Actions, European Union;
Investissements d'Avenir Labex, Investissements d'Avenir Idex and ANR, France;
DFG and AvH Foundation, Germany;
Herakleitos, Thales and Aristeia programmes co-financed by EU-ESF and the Greek NSRF, Greece;
BSF-NSF and MINERVA, Israel;
Norwegian Financial Mechanism 2014-2021, Norway;
NCN and NAWA, Poland;
La Caixa Banking Foundation, CERCA Programme Generalitat de Catalunya and PROMETEO and GenT Programmes Generalitat Valenciana, Spain;
G\"{o}ran Gustafssons Stiftelse, Sweden;
The Royal Society and Leverhulme Trust, United Kingdom.
 
The crucial computing support from all WLCG partners is acknowledged gratefully, in particular from CERN, the ATLAS Tier-1 facilities at TRIUMF (Canada), NDGF (Denmark, Norway, Sweden), CC-IN2P3 (France), KIT/GridKA (Germany), INFN-CNAF (Italy), NL-T1 (Netherlands), PIC (Spain), ASGC (Taiwan), RAL (UK) and BNL (USA), the Tier-2 facilities worldwide and large non-WLCG resource providers. Major contributors of computing resources are listed in Ref.~\cite{ATL-SOFT-PUB-2021-003}.


\clearpage
\FloatBarrier
\appendix
\part*{Appendix}
\addcontentsline{toc}{part}{Appendix}


For completeness, the data including the full range of \pta\ intervals and centrality selections in \pbpb\ collisions are included
here.
\begin{figure}[t]
\centering
\subfloat{
\centering
\includegraphics[width=0.33\textwidth]{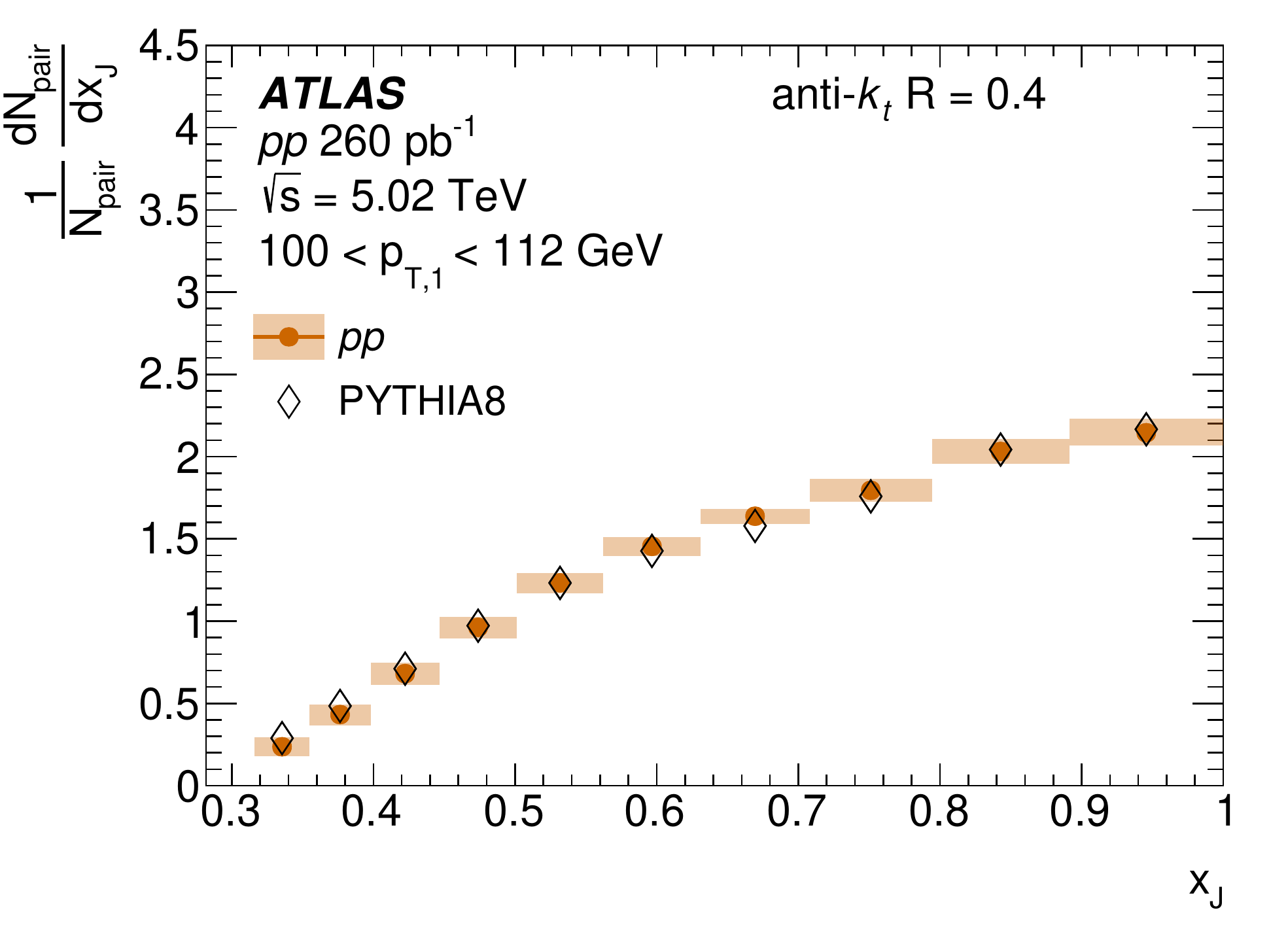}
}
\subfloat{
\centering
\includegraphics[width=0.33\textwidth]{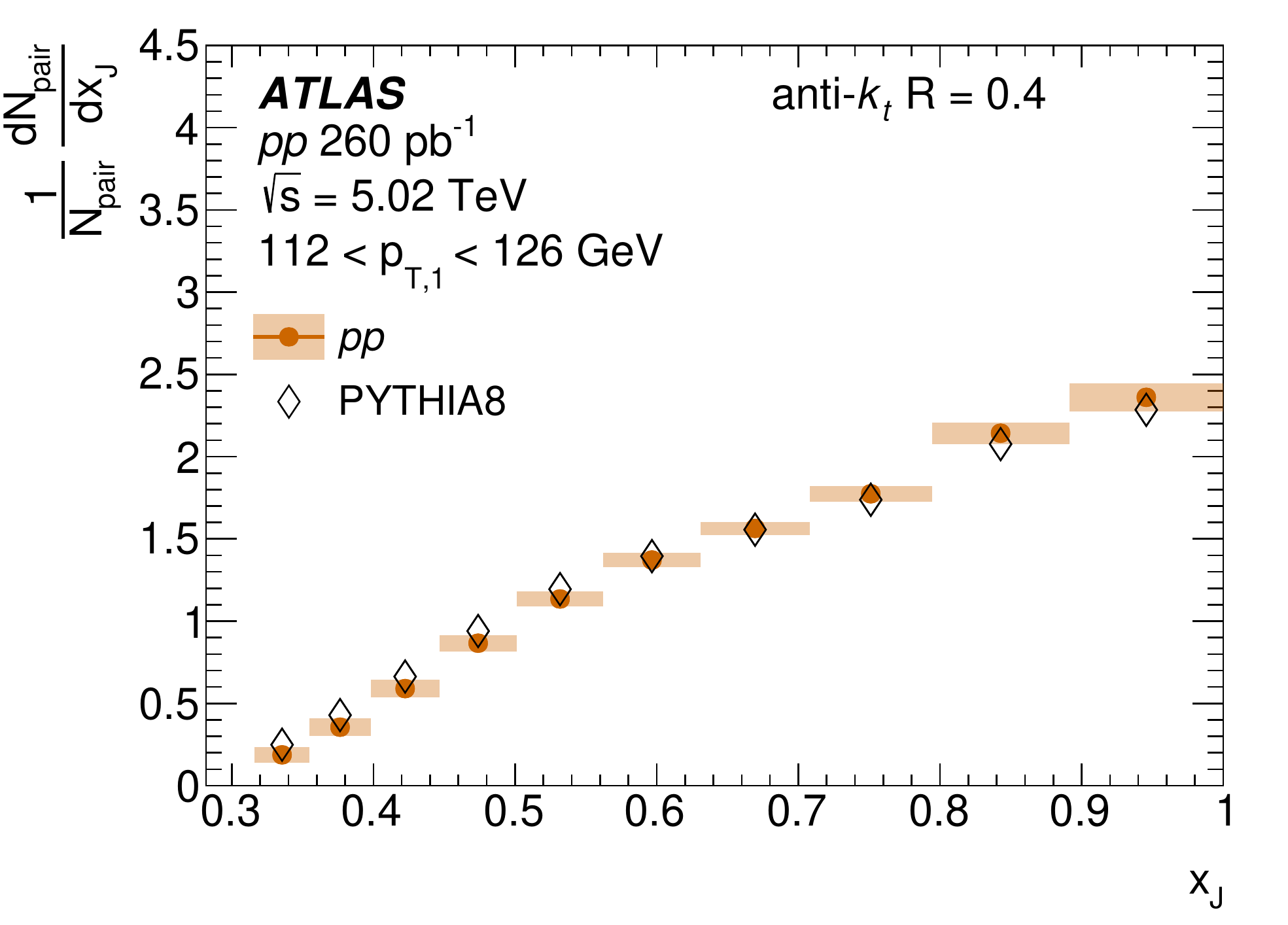}
}
\subfloat{
\centering
\includegraphics[width=0.33\textwidth]{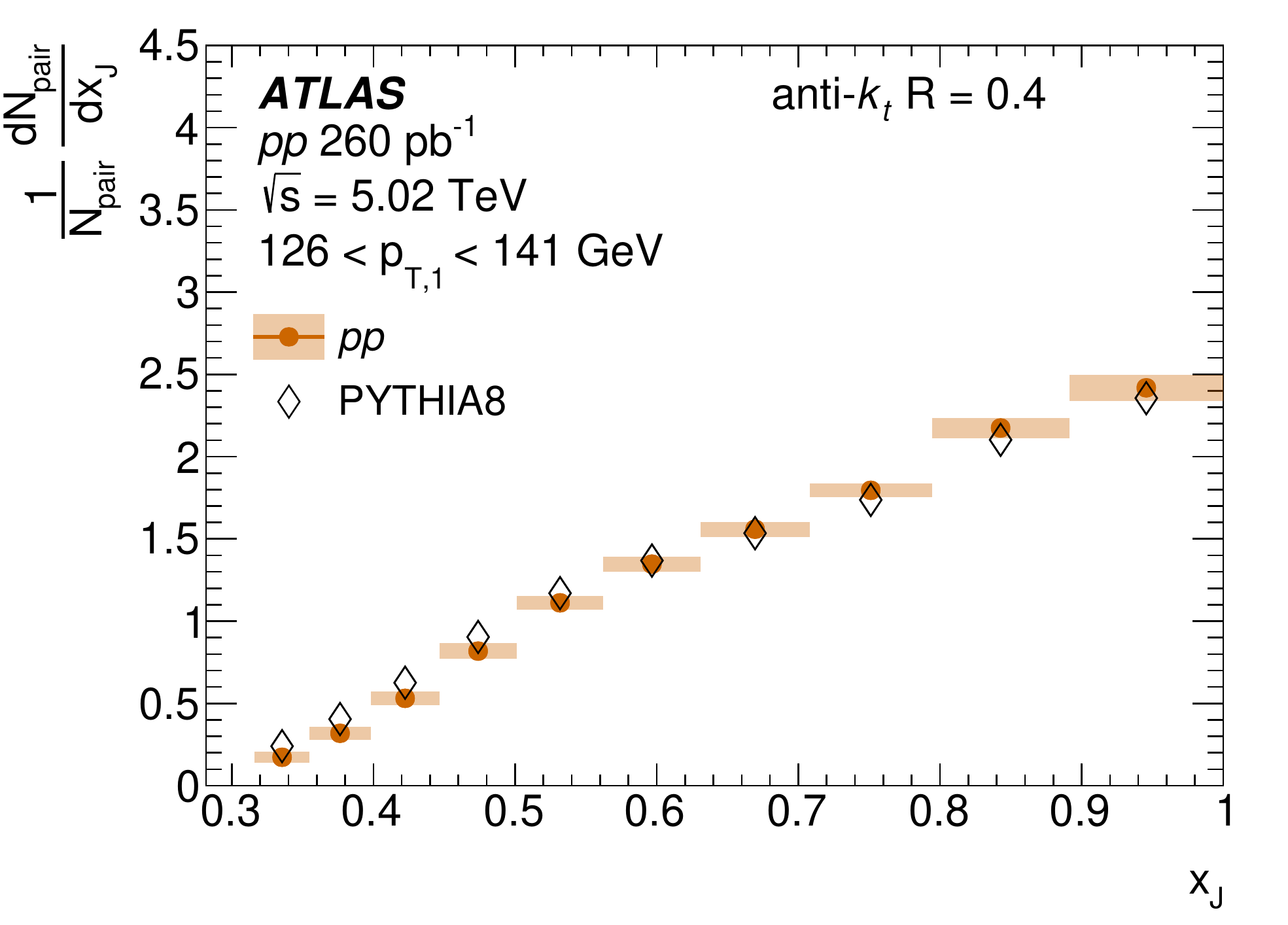}
}
\hfill

\subfloat{
\centering
\includegraphics[width=0.33\textwidth]{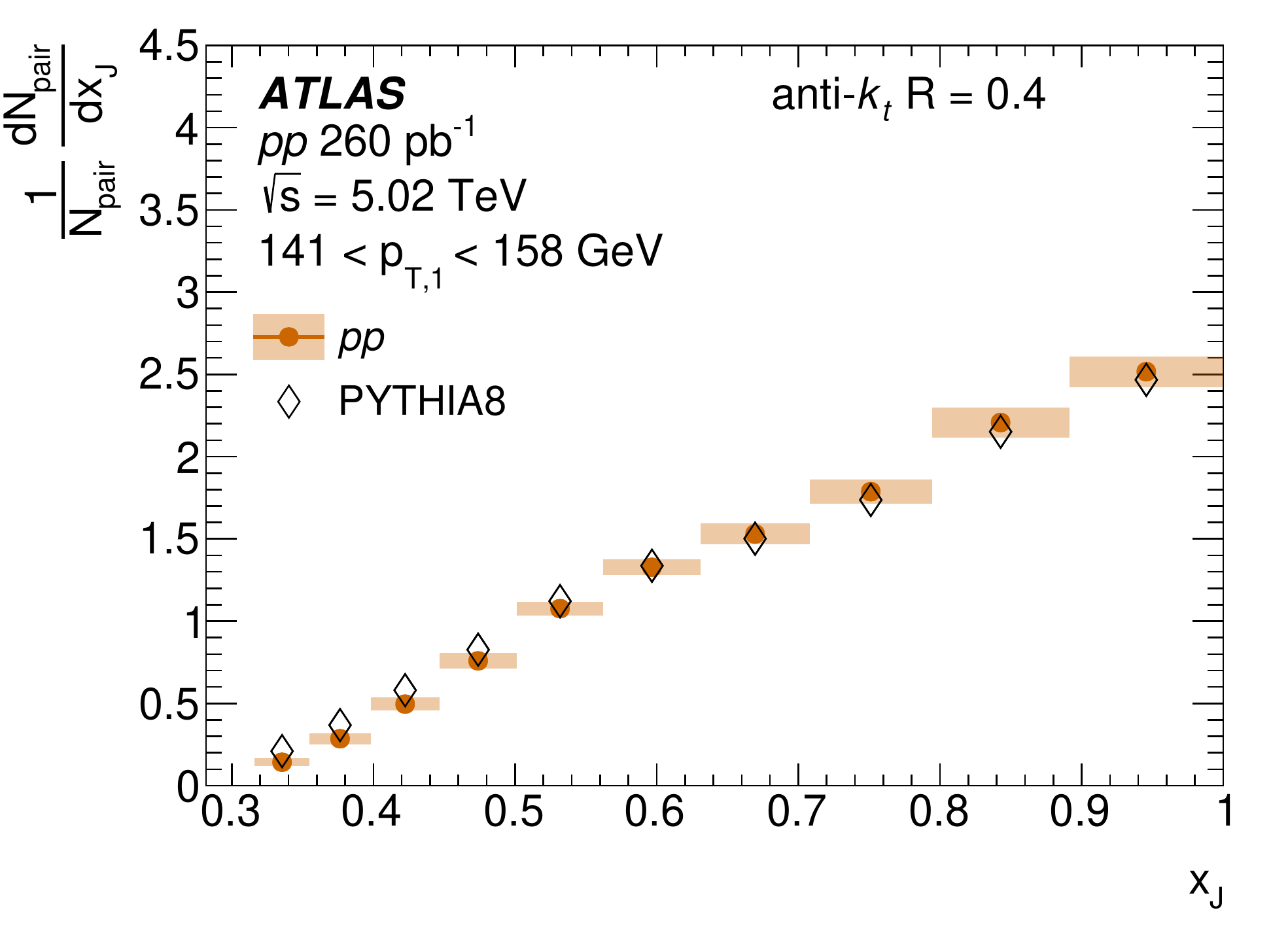}
}
\subfloat{
\centering
\includegraphics[width=0.33\textwidth]{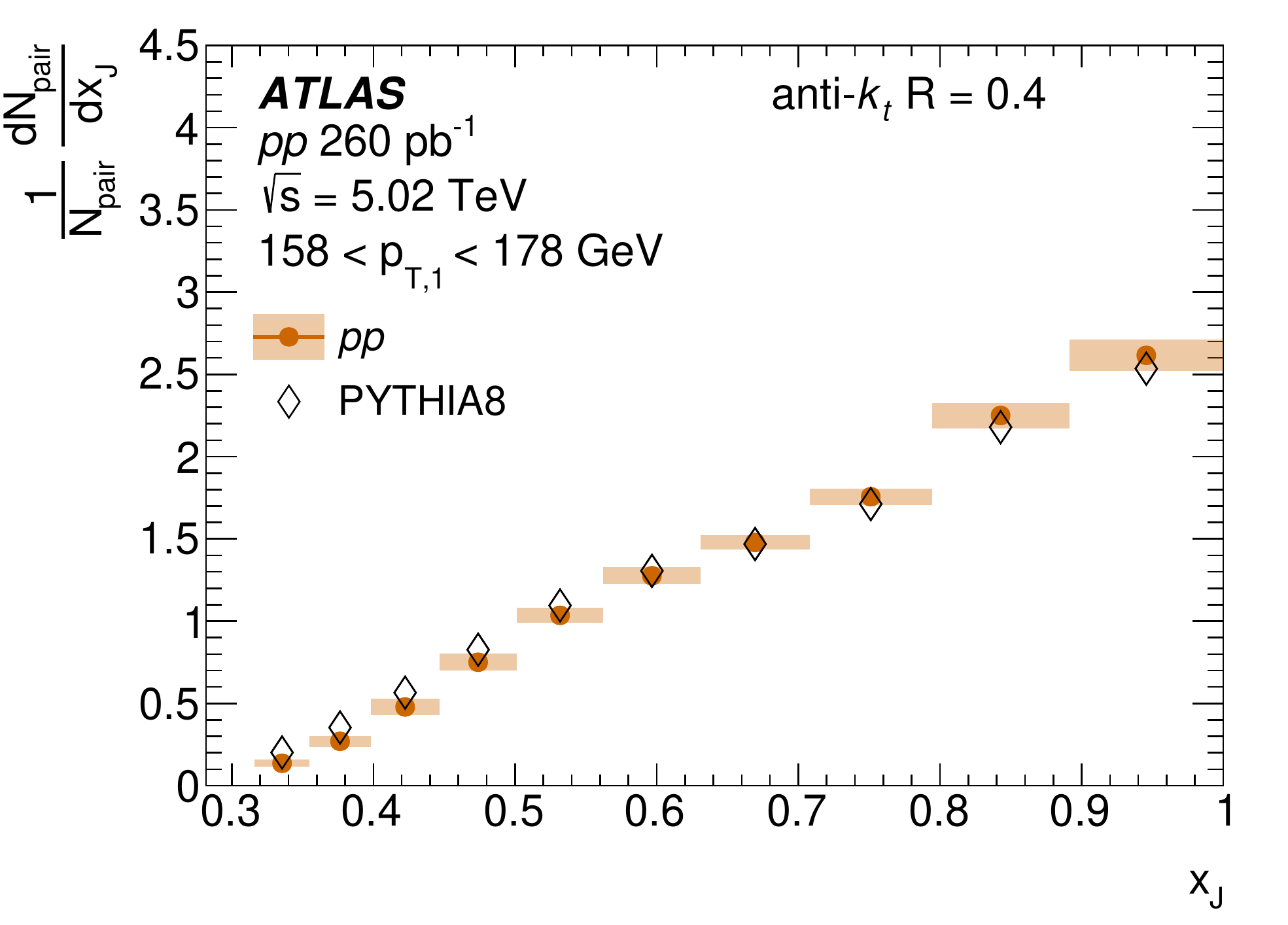}
}
\subfloat{
\centering
\includegraphics[width=0.33\textwidth]{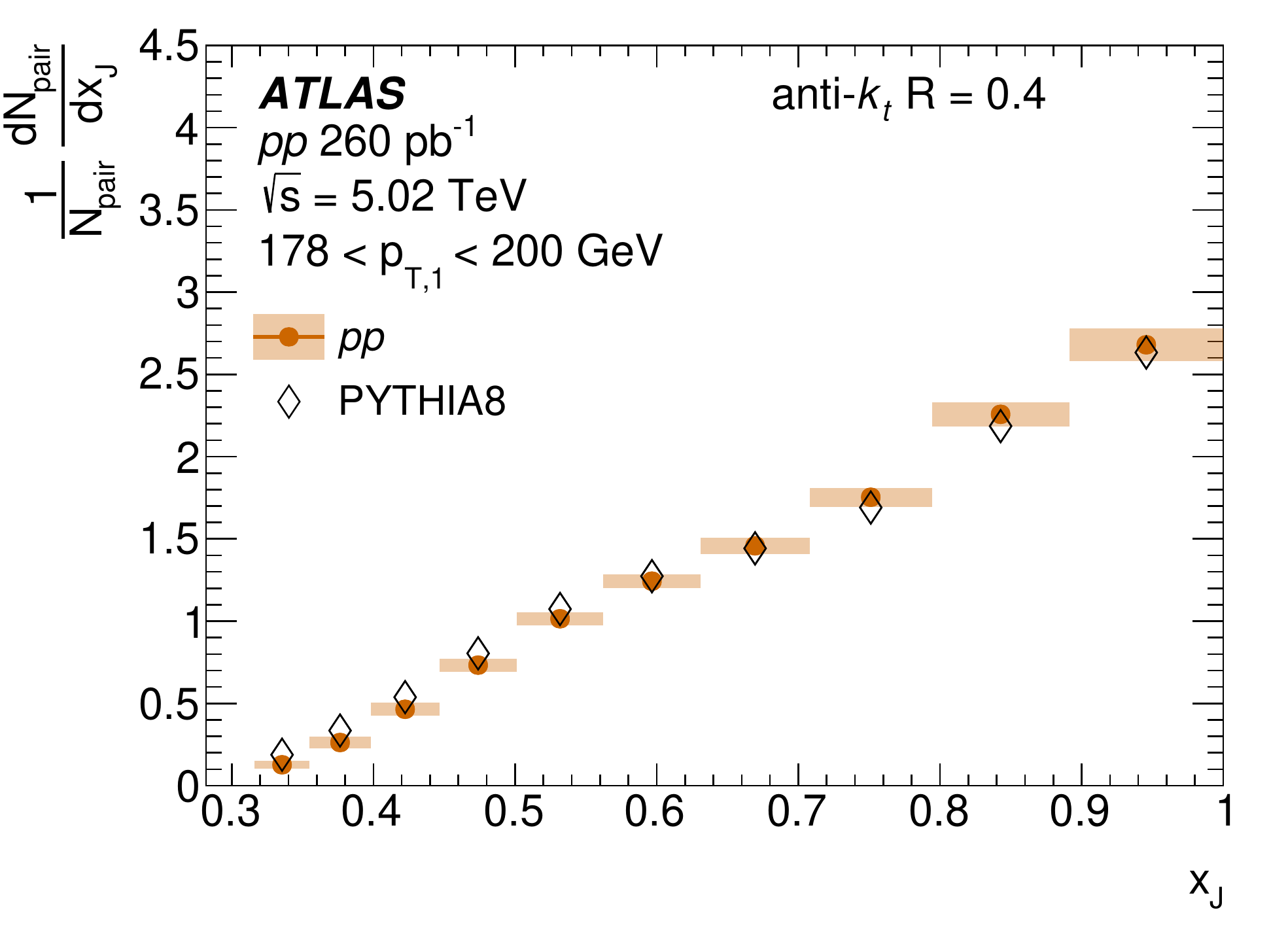}
}
\hfill
\subfloat{
\centering
\includegraphics[width=0.33\textwidth]{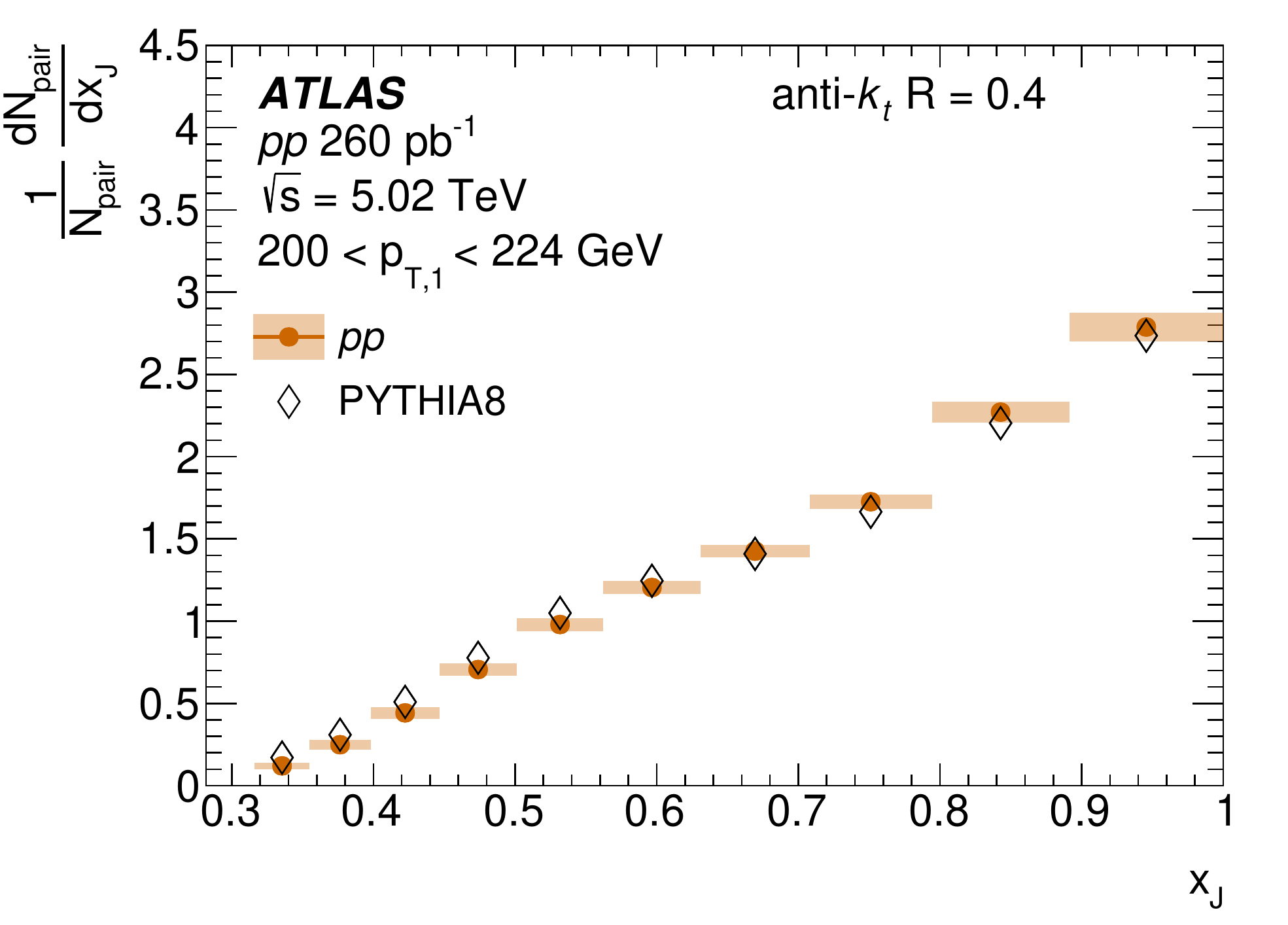}
}
\subfloat{
\centering
\includegraphics[width=0.33\textwidth]{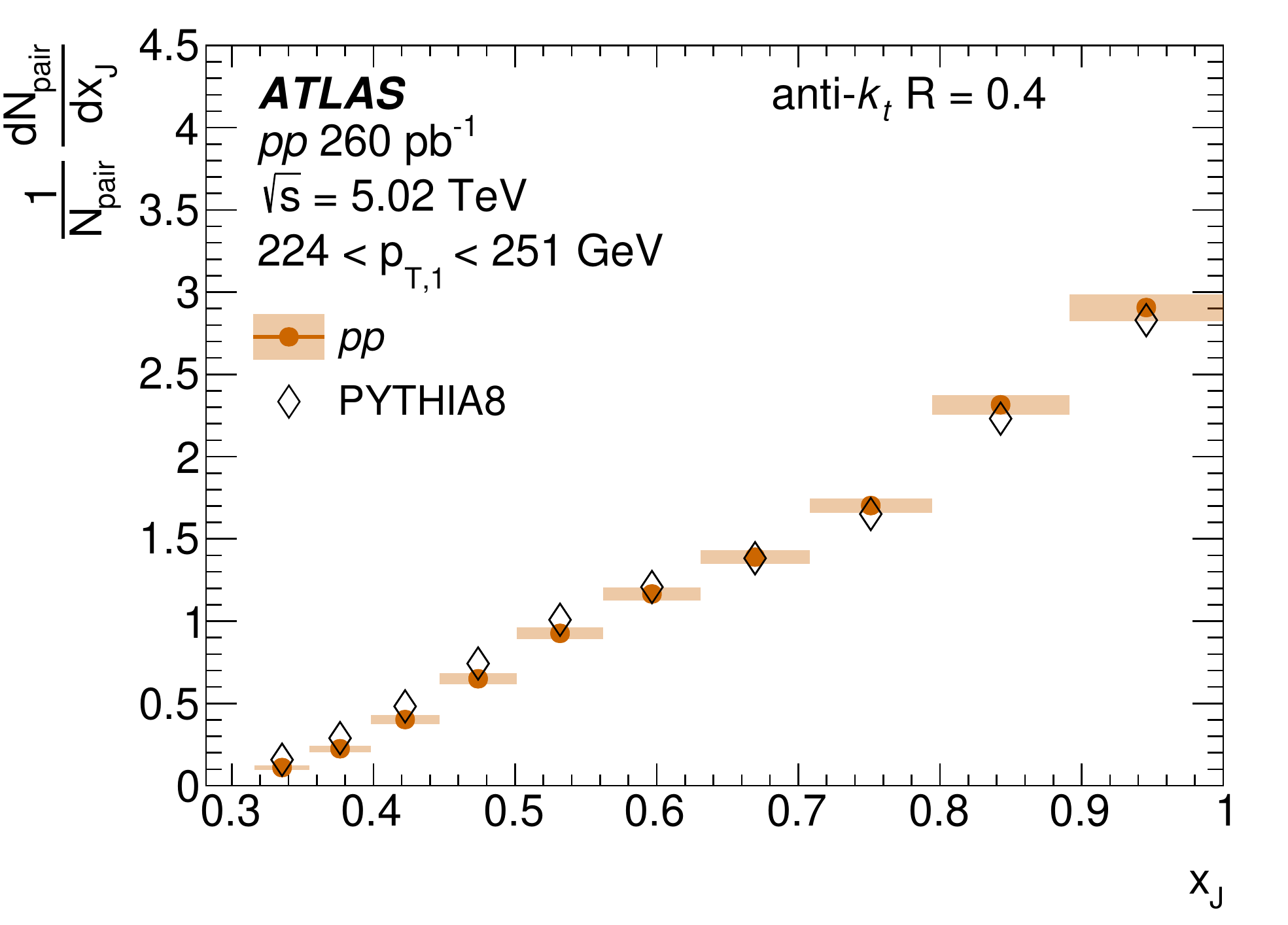}
}
\subfloat{
\centering
\includegraphics[width=0.33\textwidth]{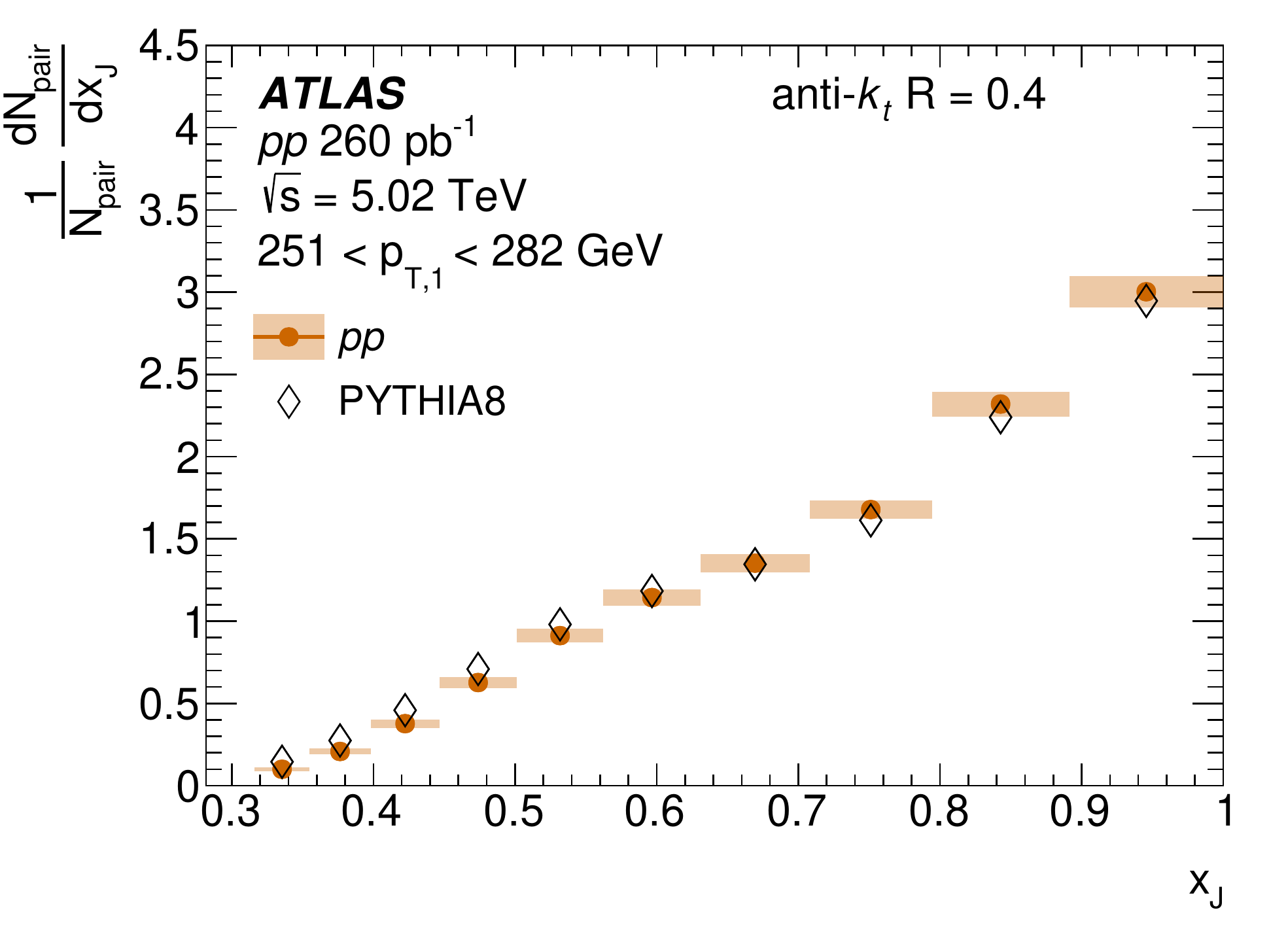}
}
\hfill
\subfloat{
\centering
\includegraphics[width=0.33\textwidth]{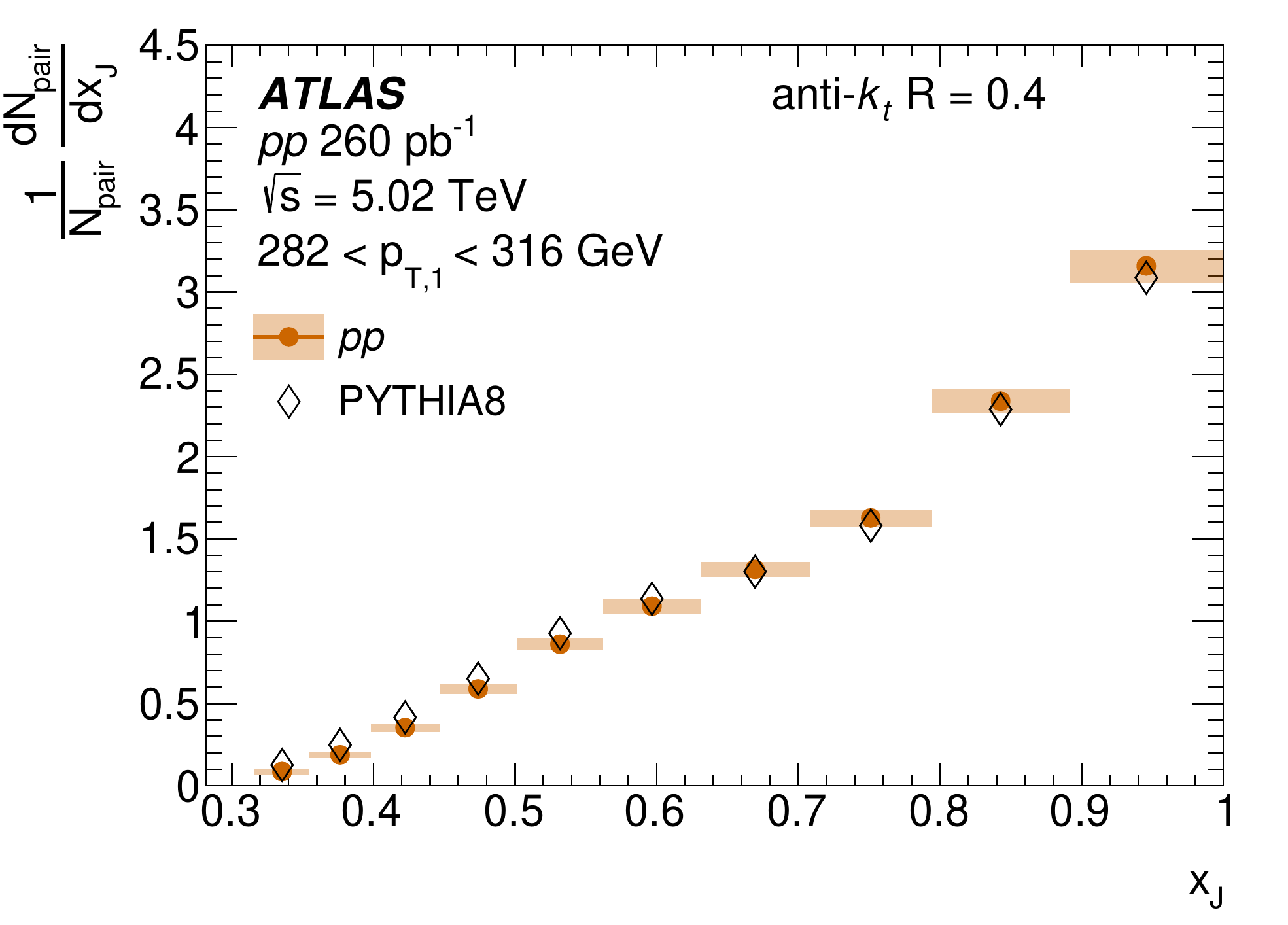}
}
\subfloat{
\centering
\includegraphics[width=0.33\textwidth]{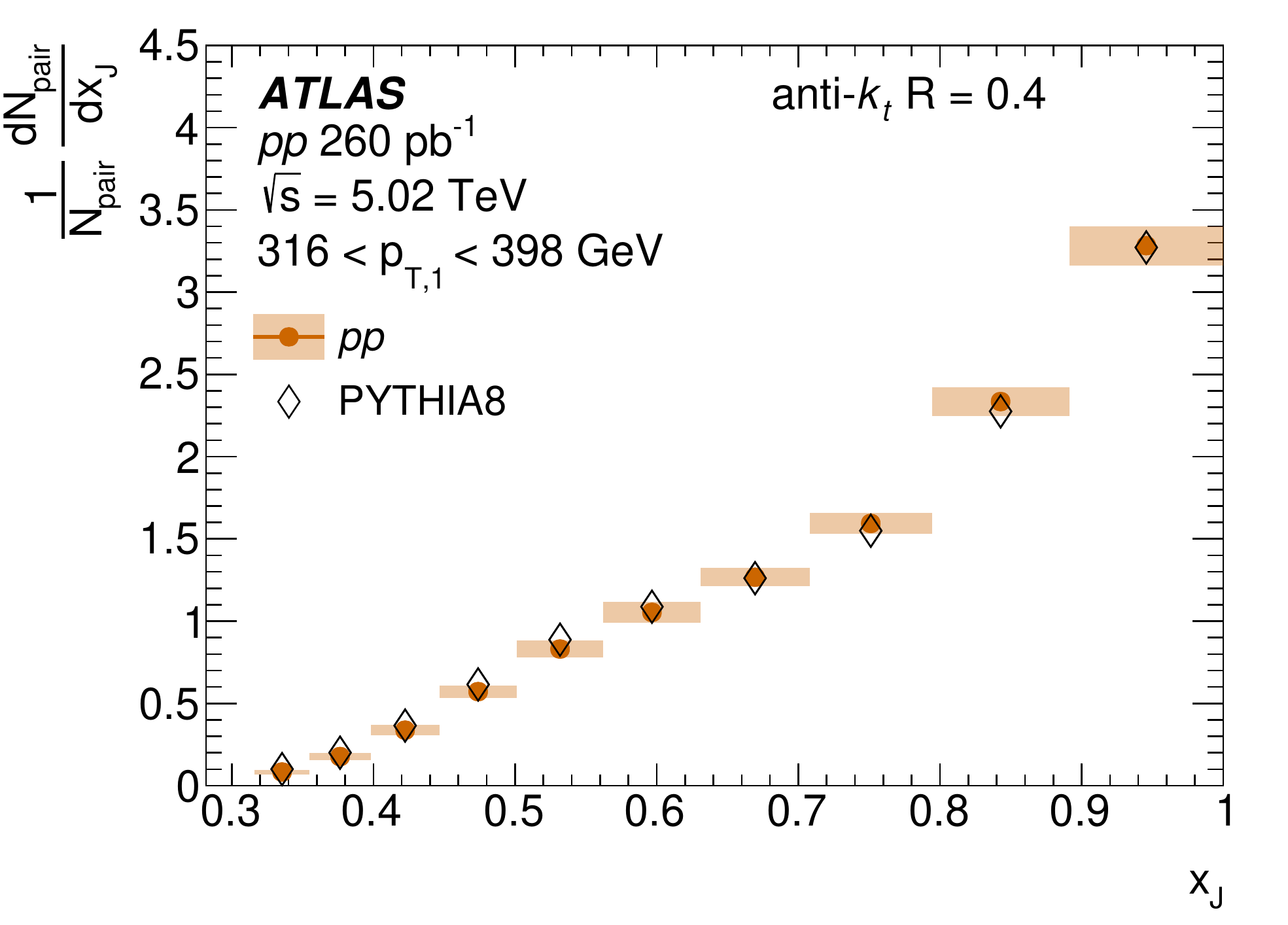}
}
\subfloat{
\centering
\includegraphics[width=0.33\textwidth]{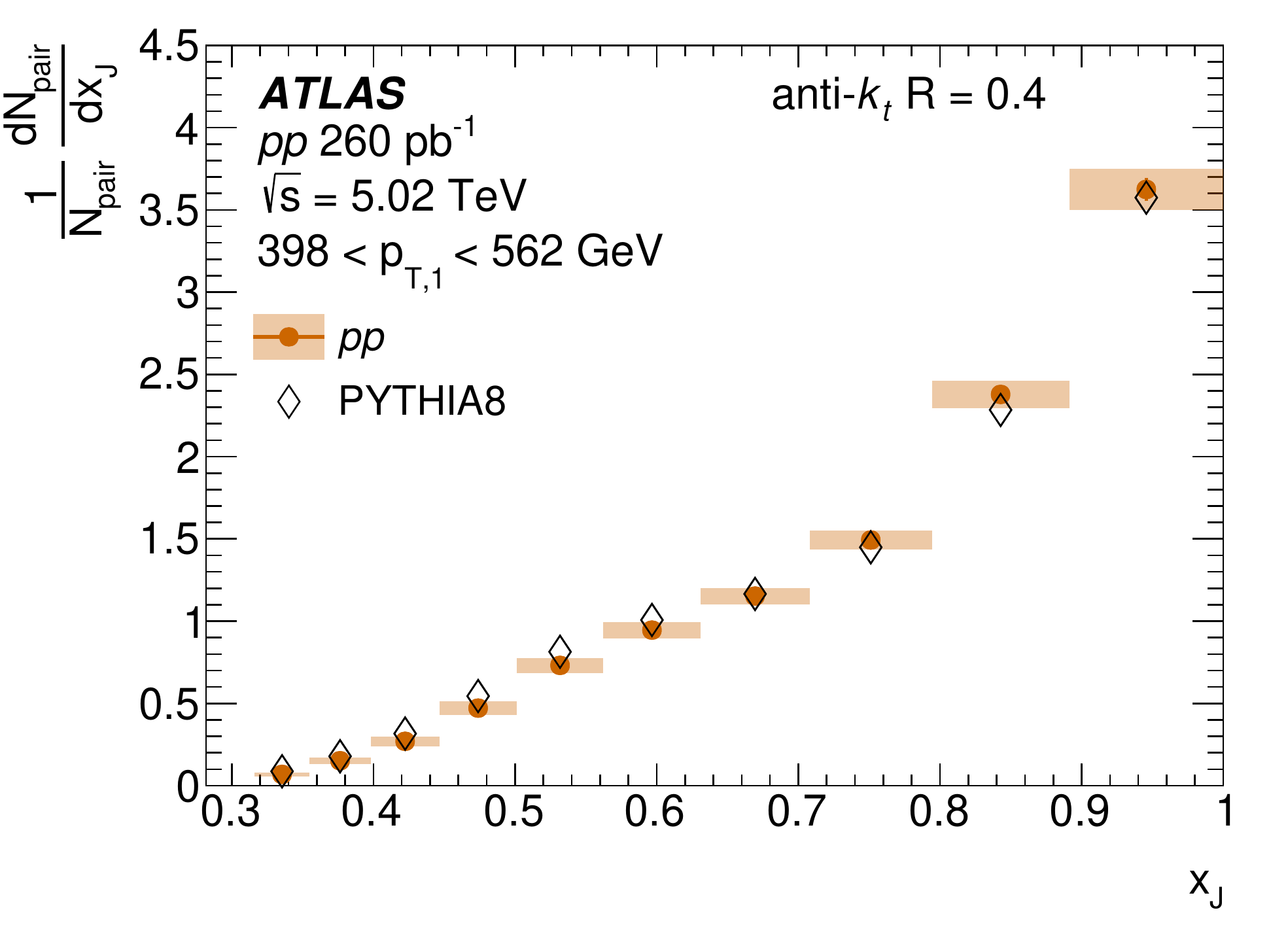}
}
\hfill
\caption{\label{fig:auxppresult}The dijet-yield-normalized \dNdxJ\ distributions in \pp\ collisions for twelve
\pta\ selections from 100~\GeV\ to 562~\GeV\
overlaid with the corresponding distributions from \pythiaeight. The statistical uncertainties
are shown as error bars (which are smaller than the markers in many cases), and the
systematic uncertainties are shown as boxes.}
\end{figure}
Figure~\ref{fig:auxppresult} shows the dijet-yield-normalized distributions for
$pp$ collisions, compared to \pythiaeight\ calculations.
\begin{figure}[t]
\centering
\subfloat{
\centering
\includegraphics[width=0.33\textwidth]{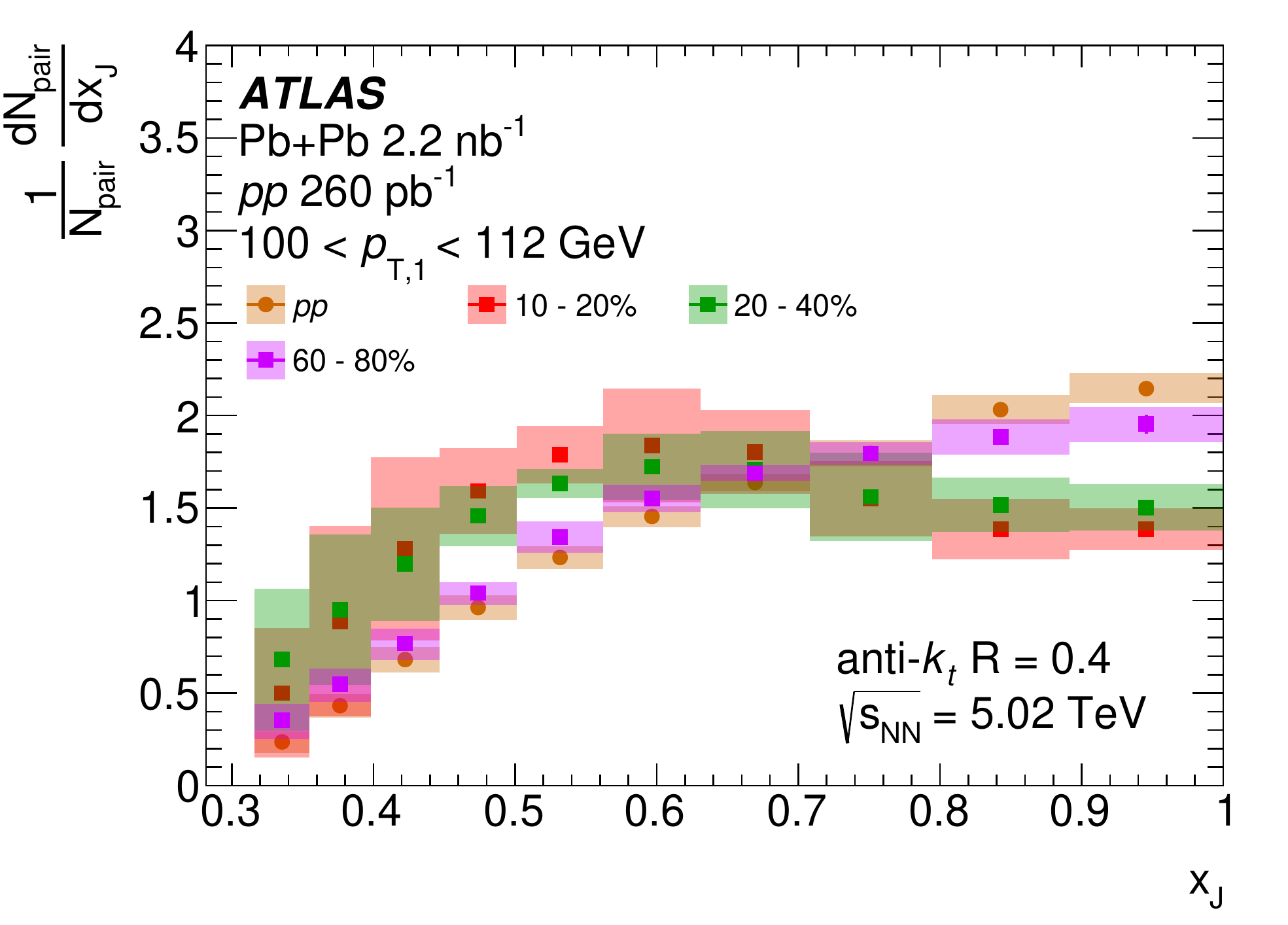}
}
\subfloat{
\centering
\includegraphics[width=0.33\textwidth]{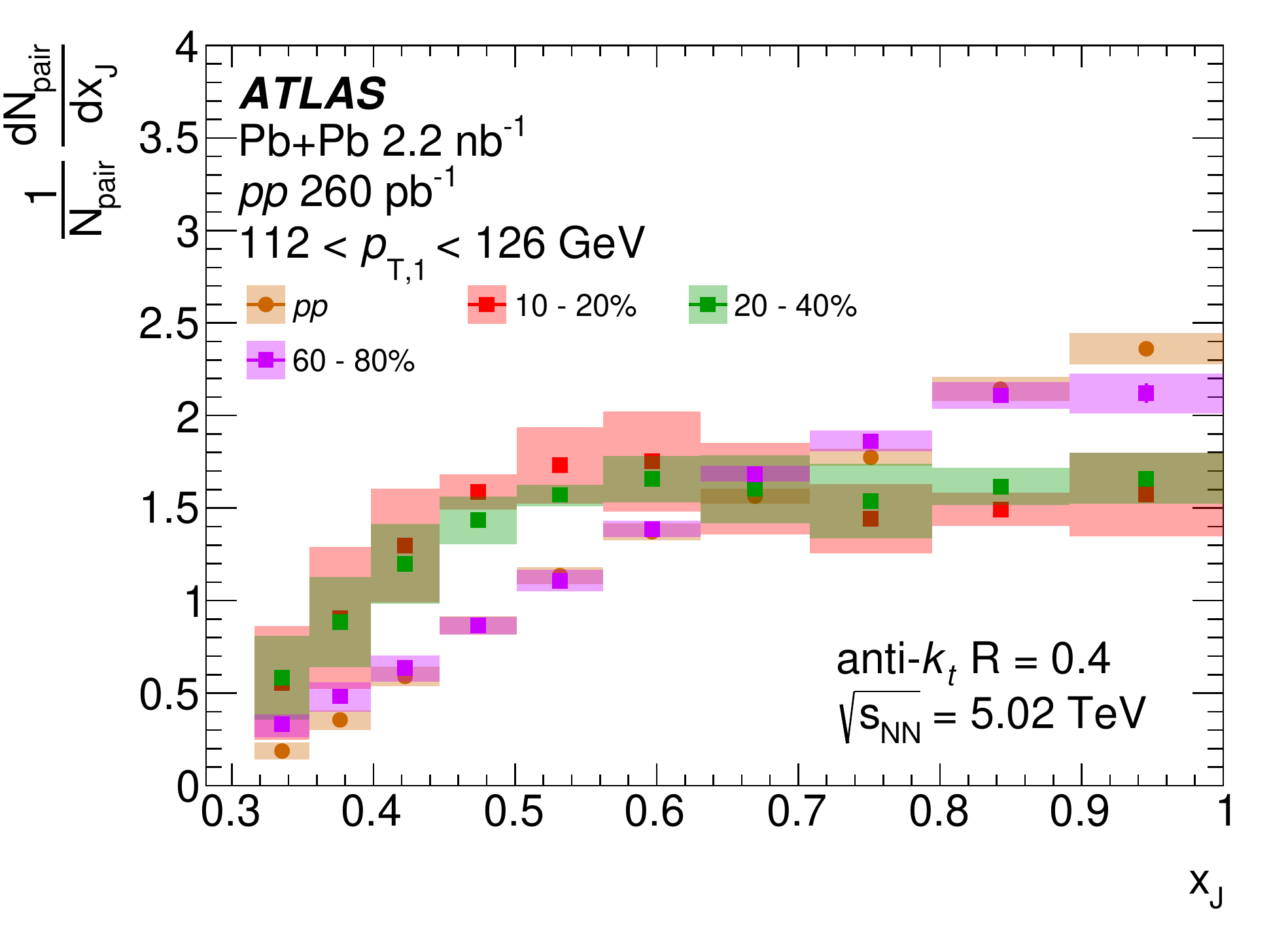}
}
\subfloat{
\centering
\includegraphics[width=0.33\textwidth]{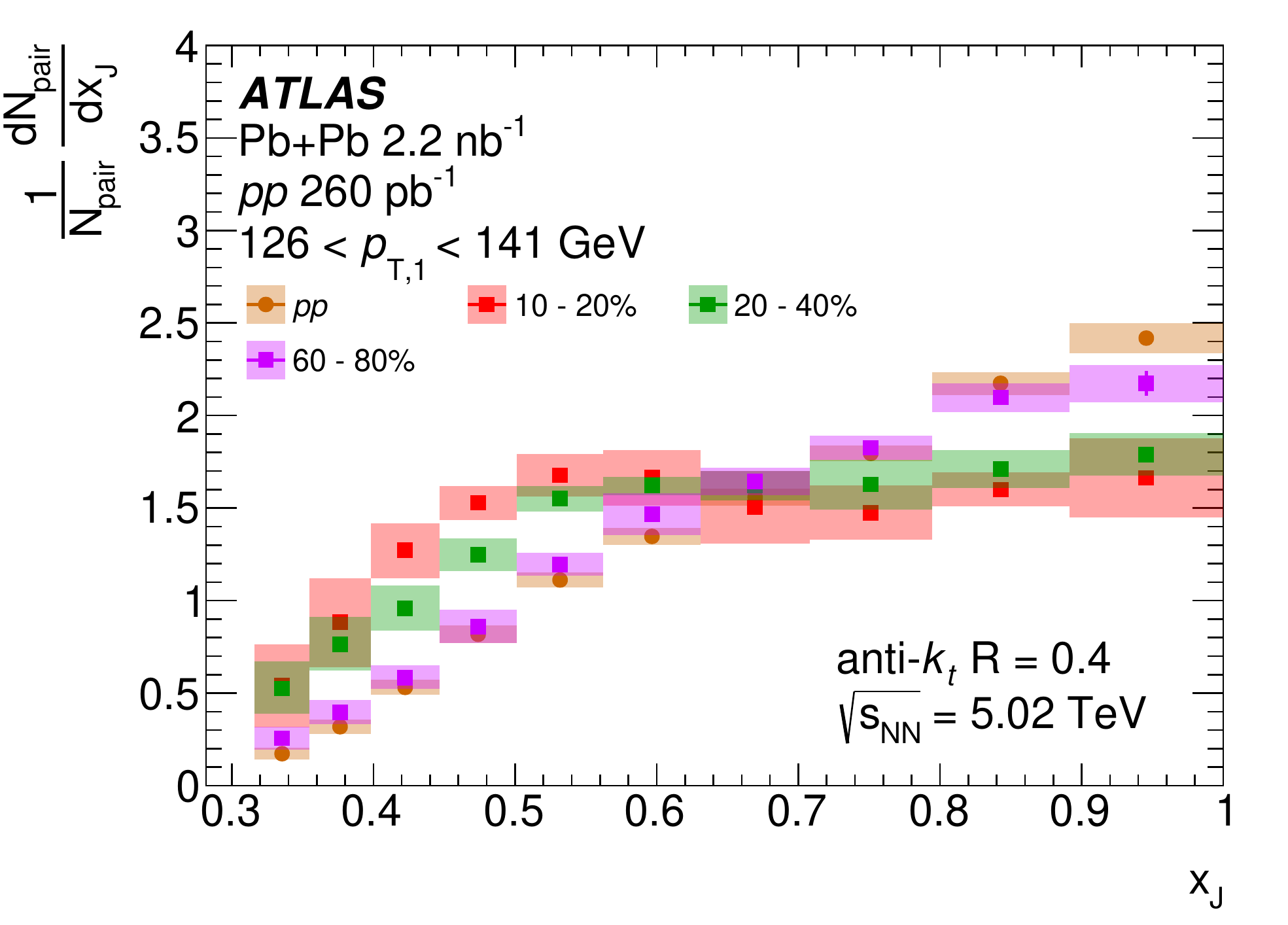}
}
\hfill

\subfloat{
\centering
\includegraphics[width=0.33\textwidth]{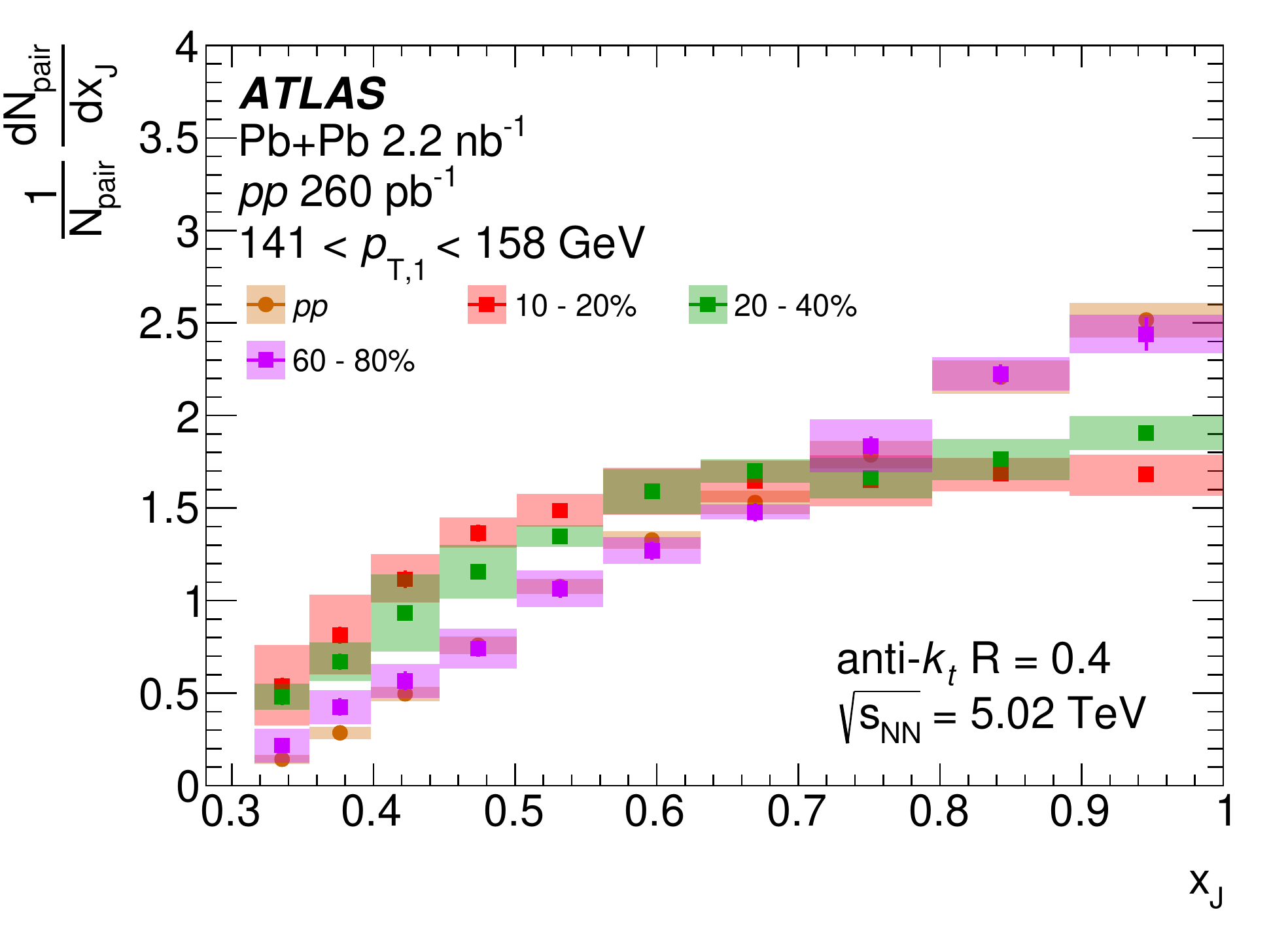}
}
\subfloat{
\centering
\includegraphics[width=0.33\textwidth]{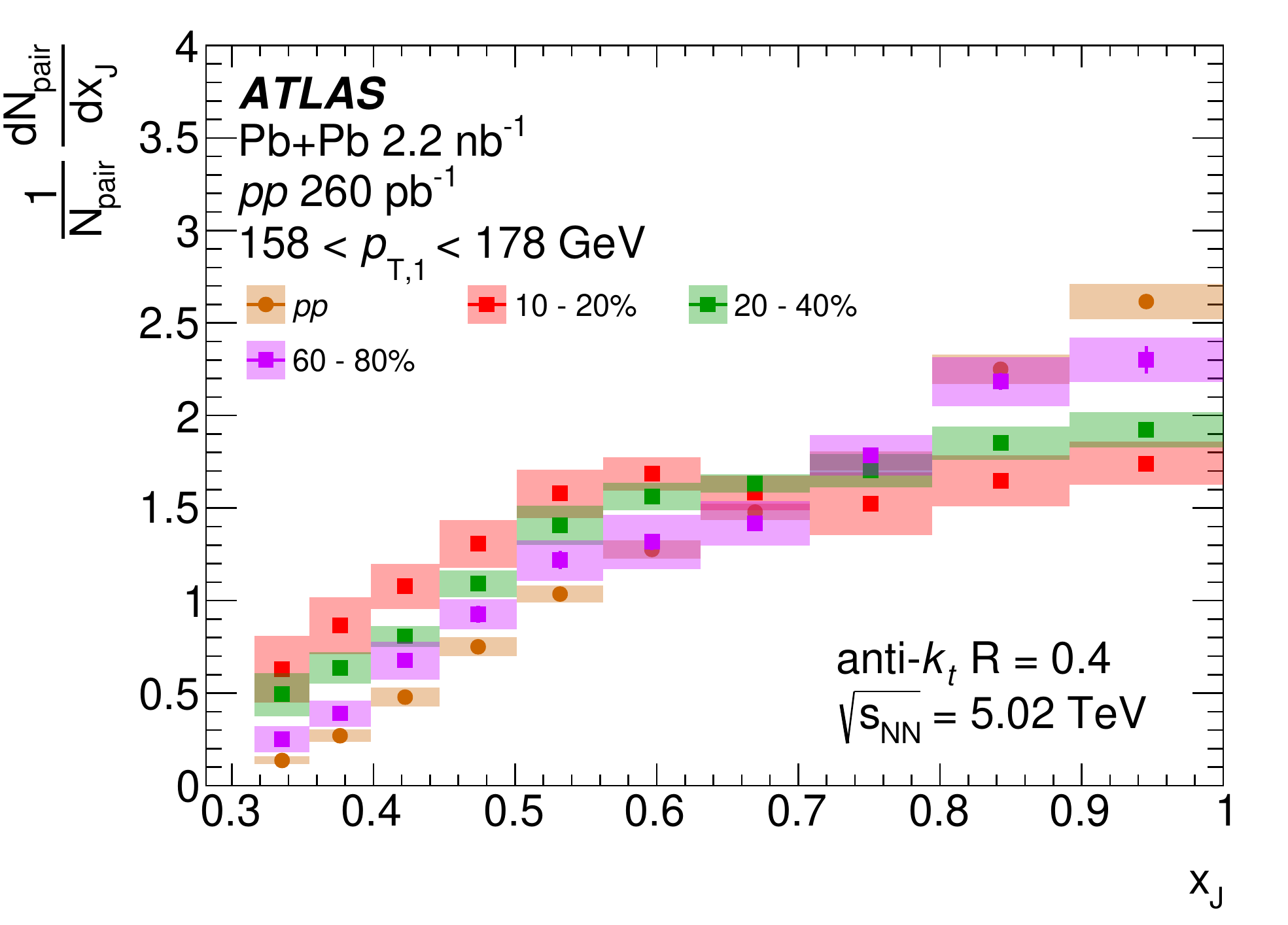}
}
\subfloat{
\centering
\includegraphics[width=0.33\textwidth]{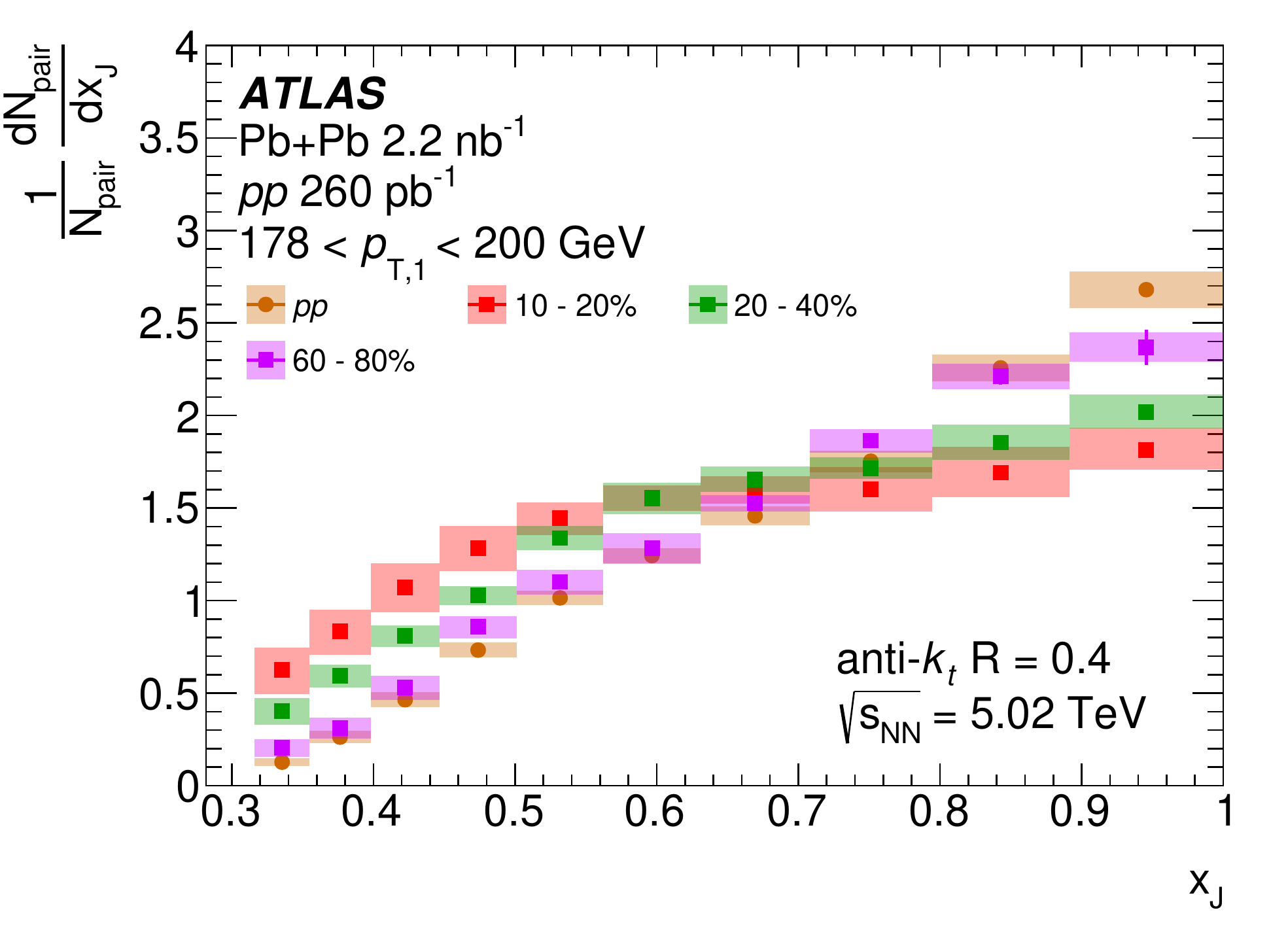}
}
\hfill
\subfloat{
\centering
\includegraphics[width=0.33\textwidth]{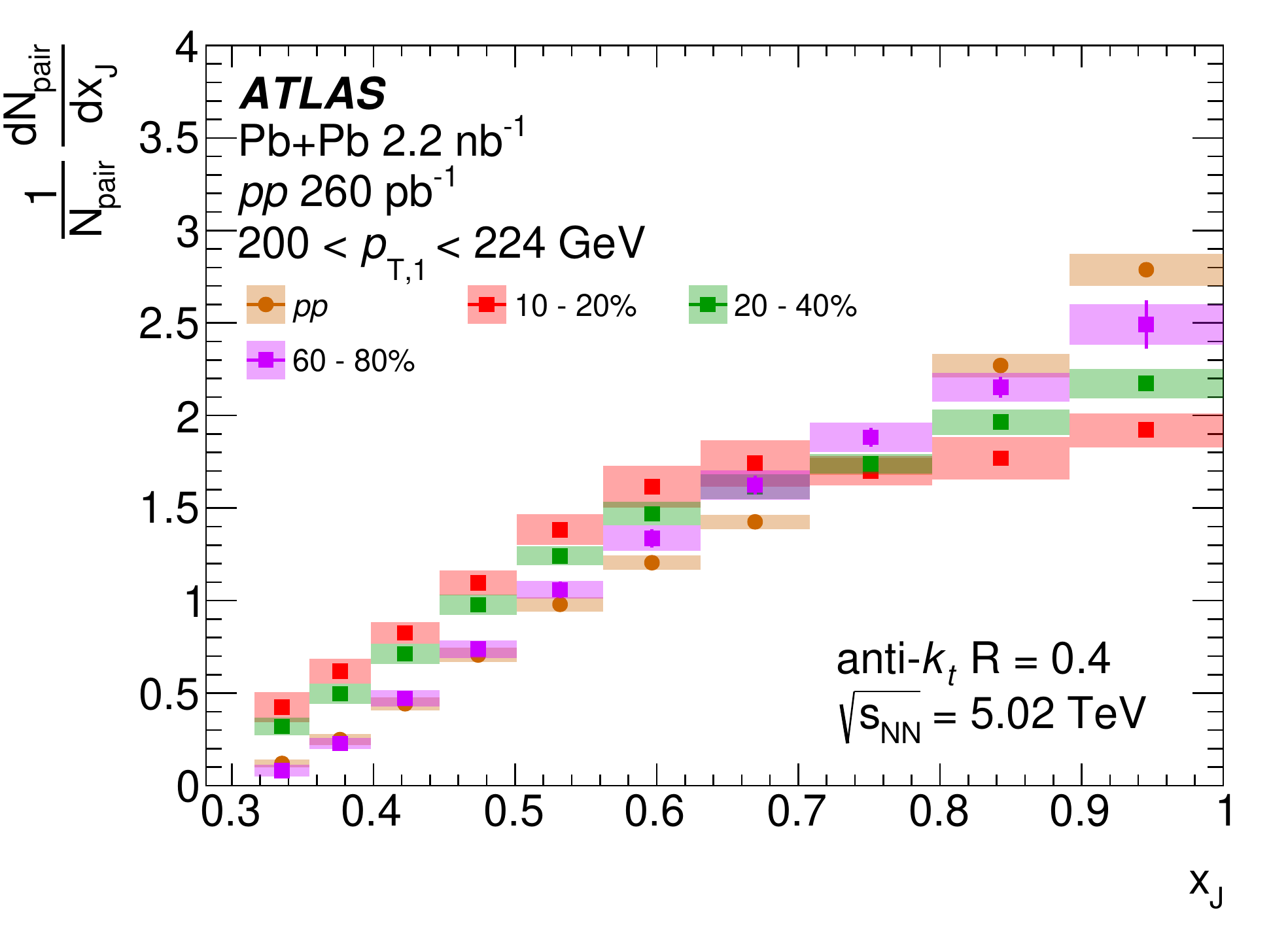}
}
\subfloat{
\centering
\includegraphics[width=0.33\textwidth]{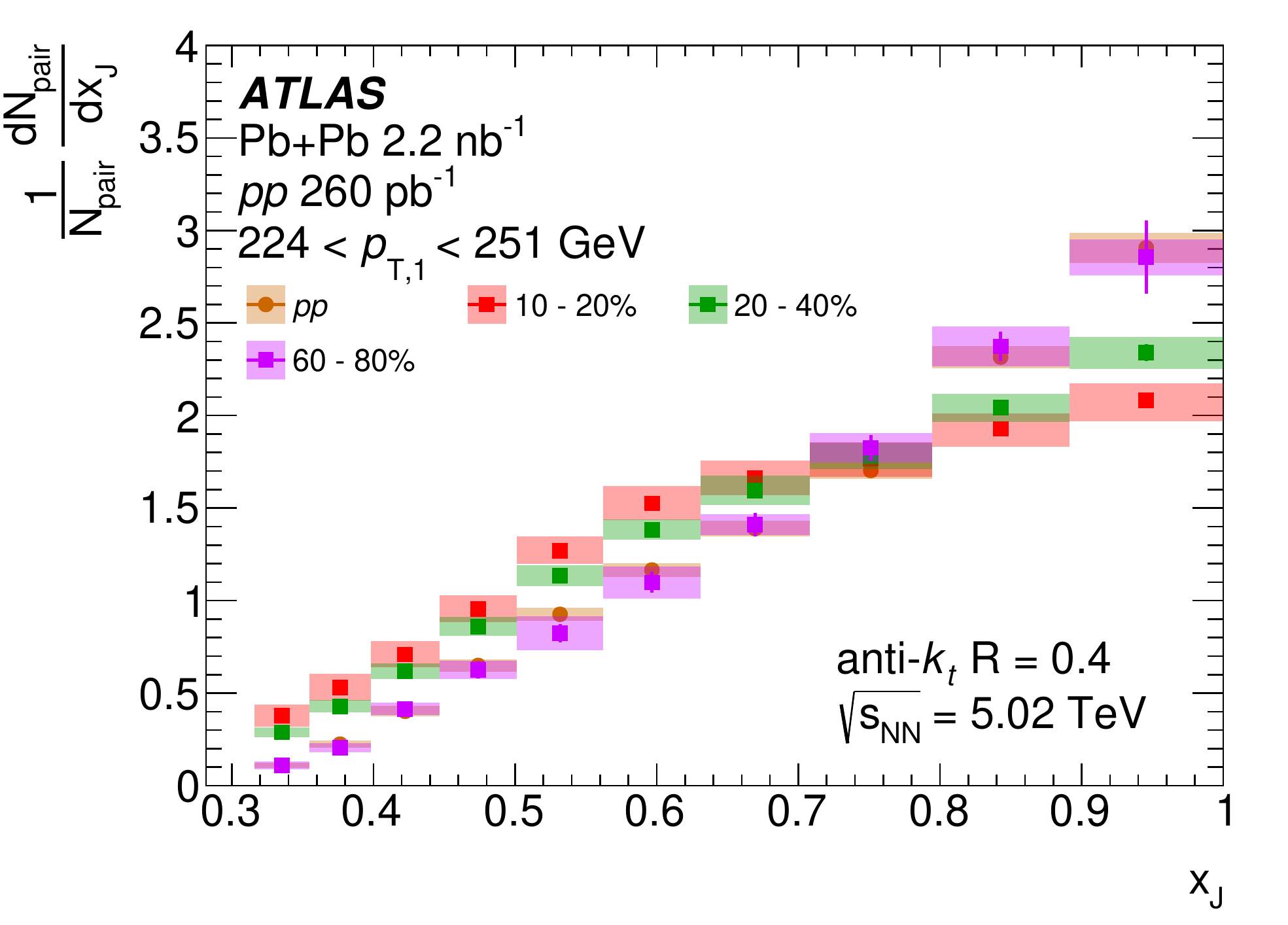}
}
\subfloat{
\centering
\includegraphics[width=0.33\textwidth]{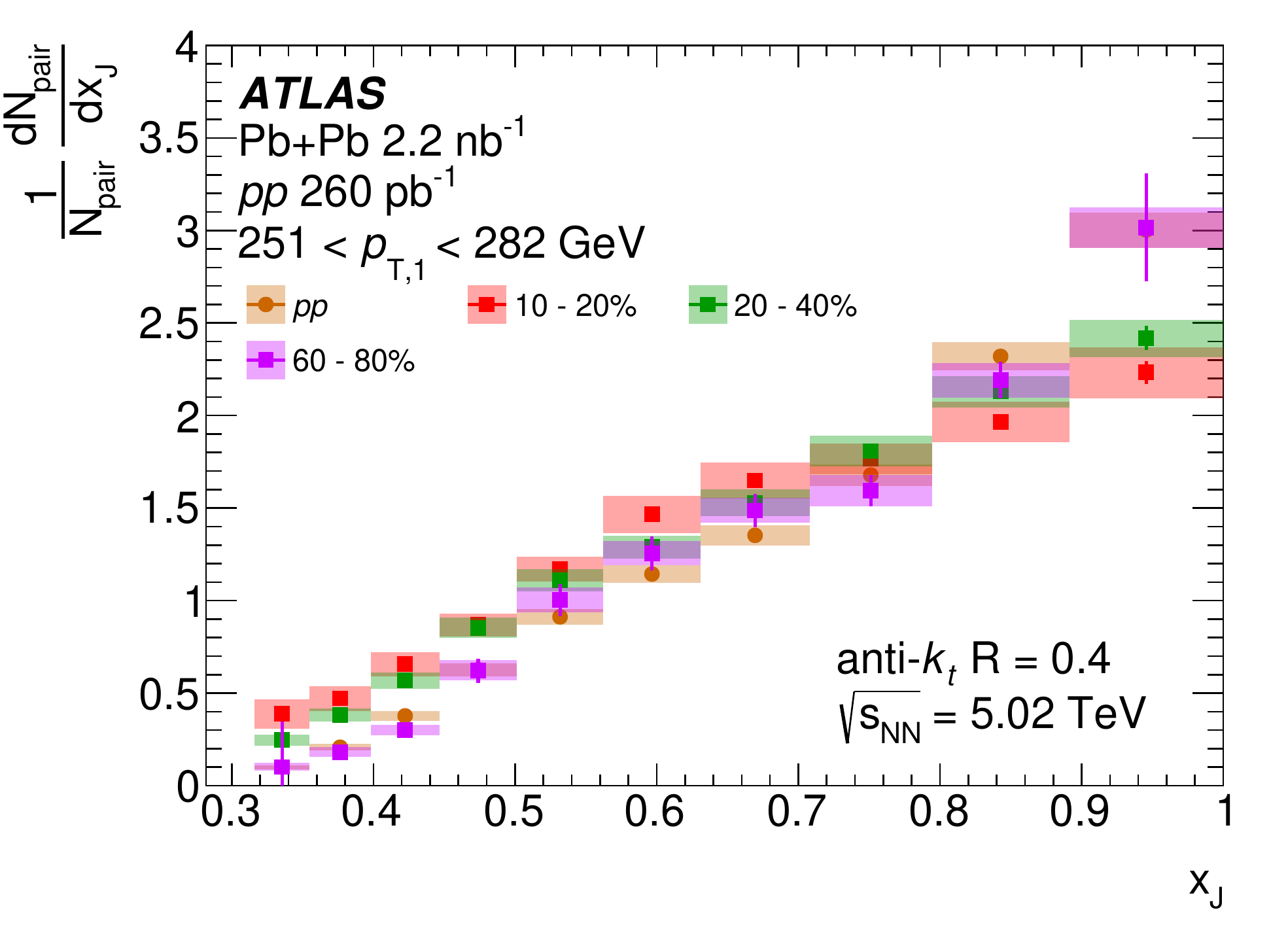}
}
\hfill
\subfloat{
\centering
\includegraphics[width=0.33\textwidth]{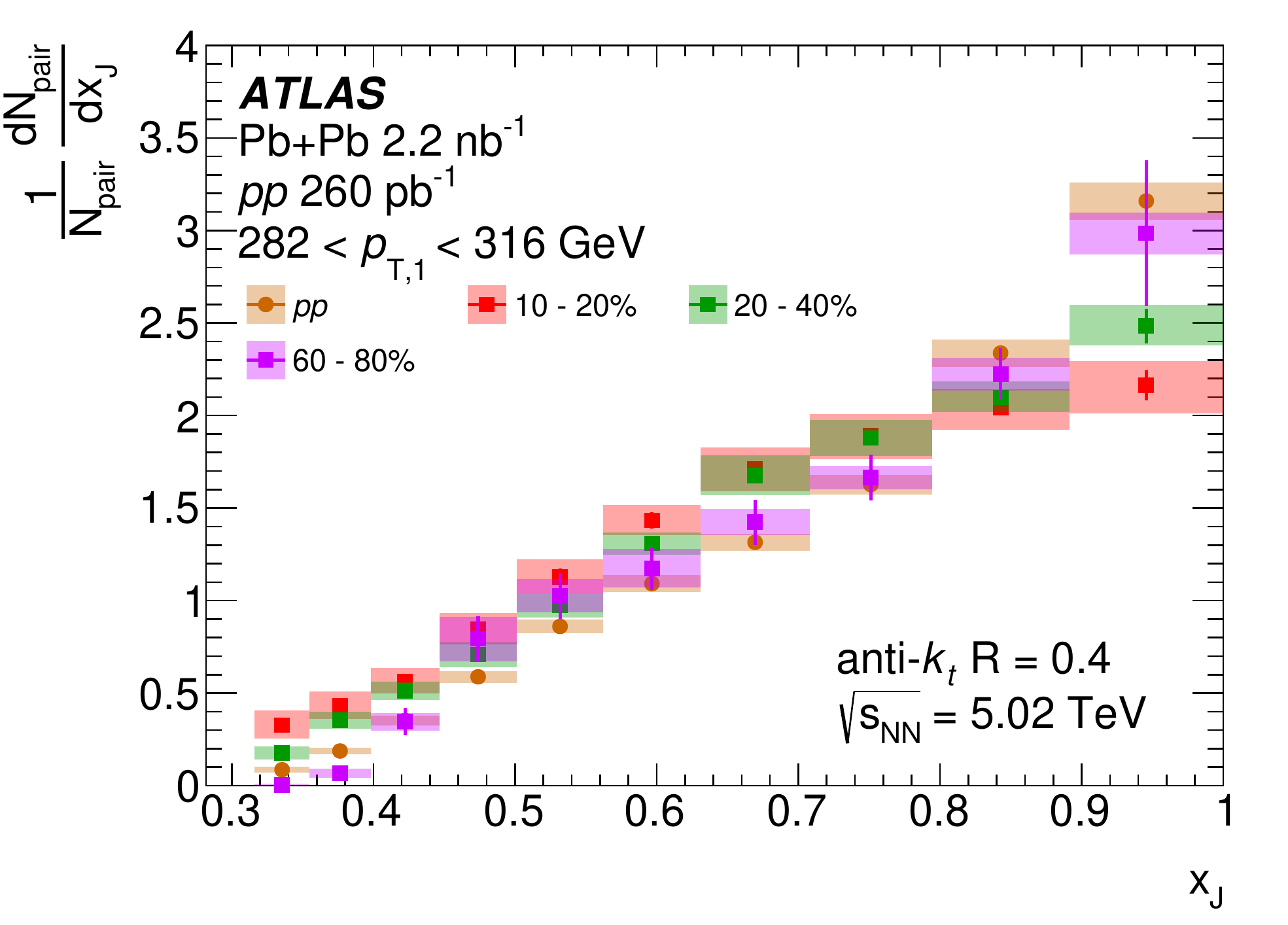}
}
\subfloat{
\centering
\includegraphics[width=0.33\textwidth]{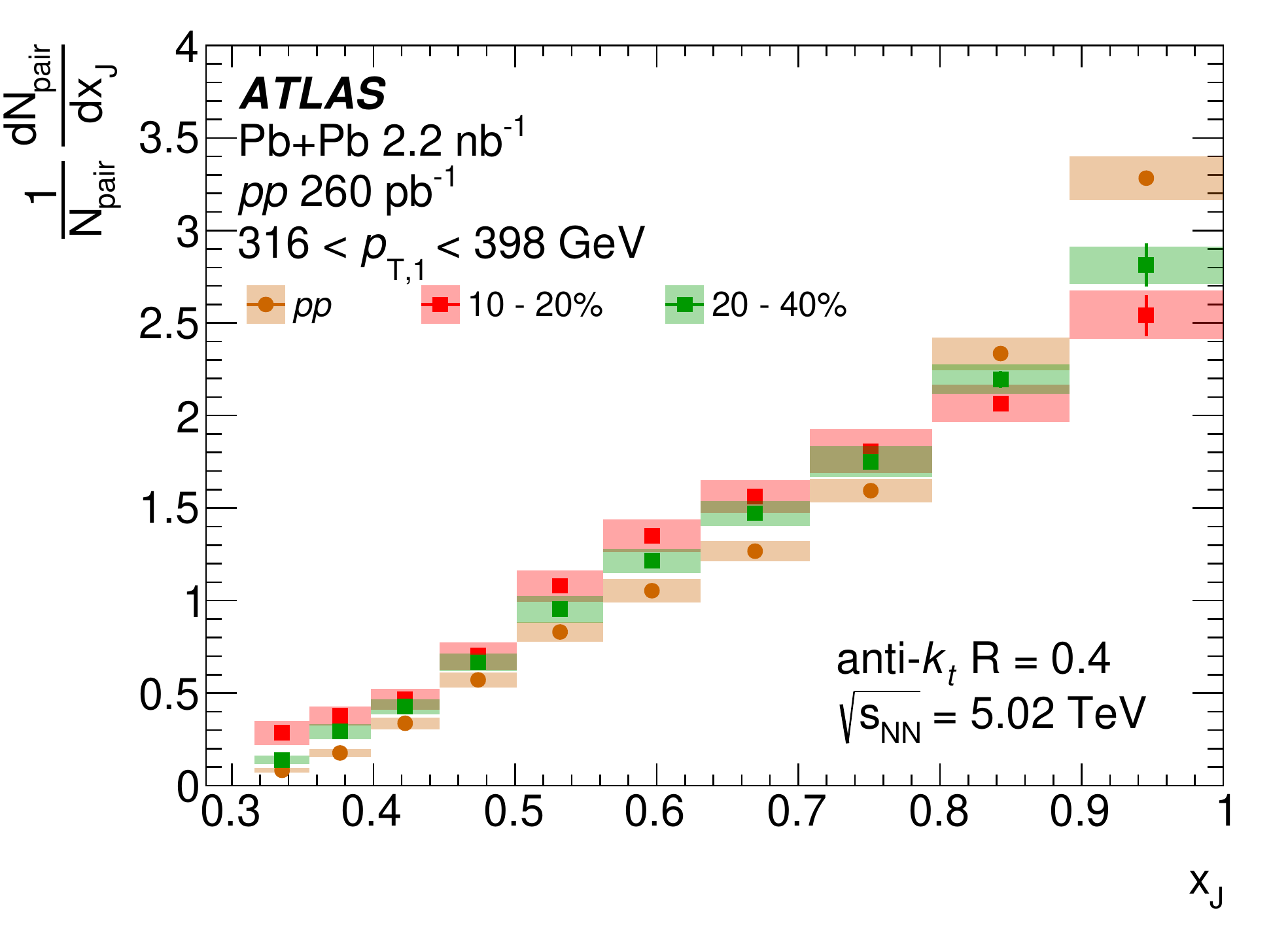}
}
\subfloat{
\centering
\includegraphics[width=0.33\textwidth]{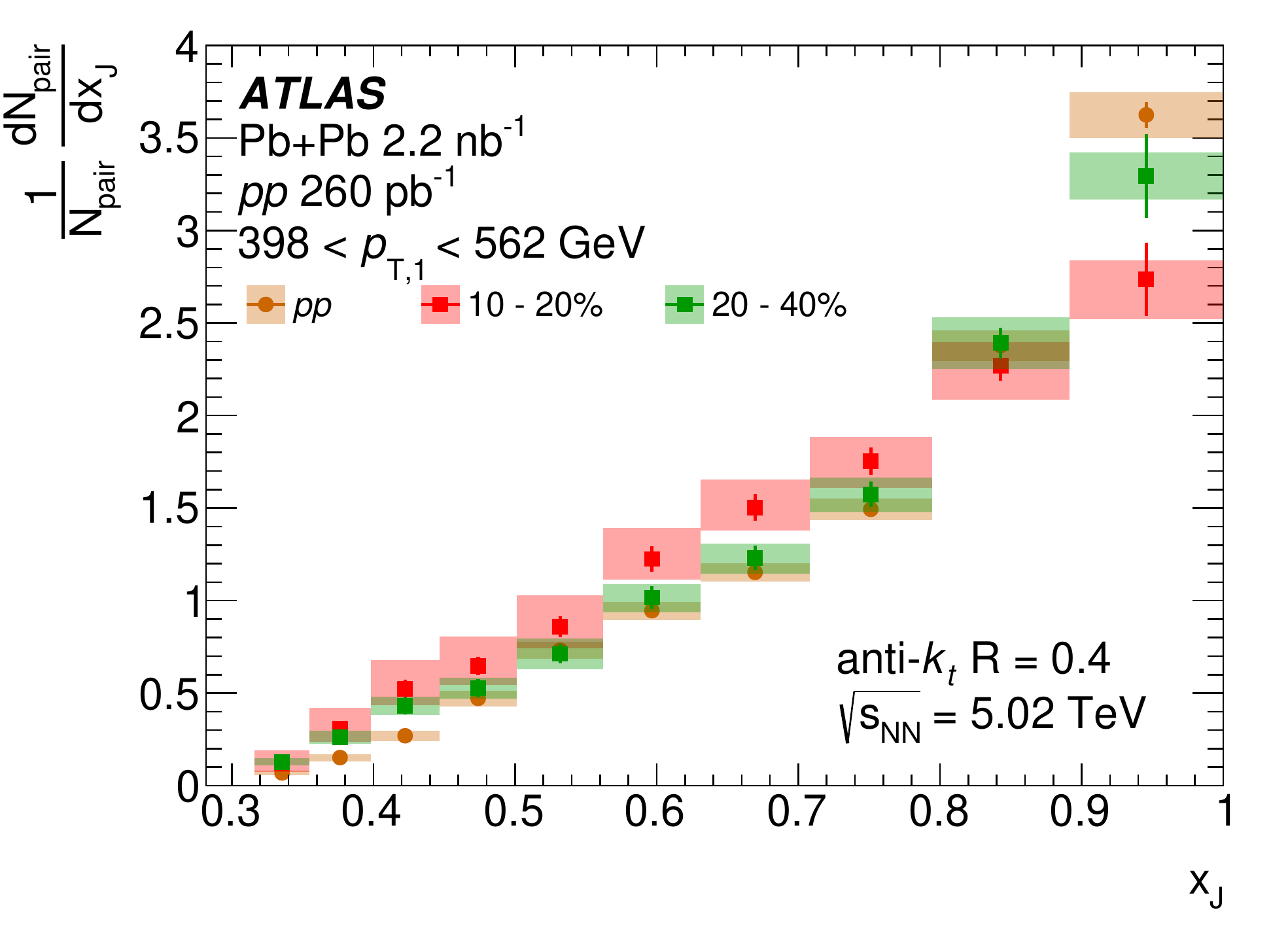}
}
\hfill
\caption{\label{fig:aux_selfnorm}The unfolded dijet-yield-normalized \xJ\ distributions in both \pbpb\ and \pp\ collisions for twelve
\ptlead\ selections from 100~\GeV\ to 562~\GeV.  The results in \pbpb\ collisions are shown in three centrality
selections, 10--20\%, 20--40\% and 60--80\%.  Statistical uncertainties are shown as
error bars and systematic uncertainties are shown as boxes.}
\end{figure}
Figure~\ref{fig:aux_selfnorm} shows unfolded \dNdxJ\ distributions for both
\pbpb\ and \pp\ collisions over twelve selections in \pta\ between 100 and 562~\GeV, for centrality intervals omitted
previously.
\begin{figure}[t]
\centering
\subfloat{
\centering
\includegraphics[width=0.33\textwidth]{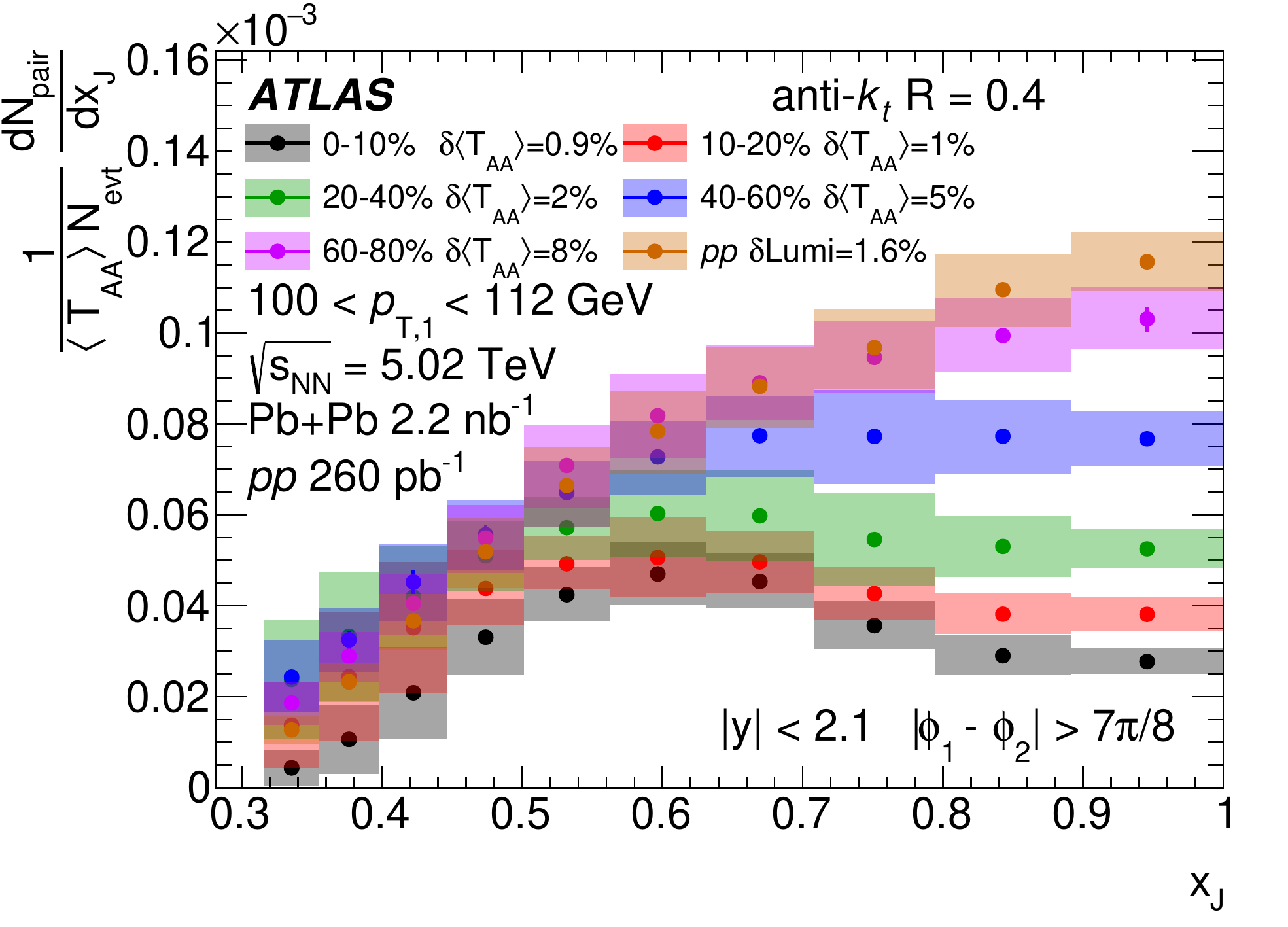}
}
\subfloat{
\centering
\includegraphics[width=0.33\textwidth]{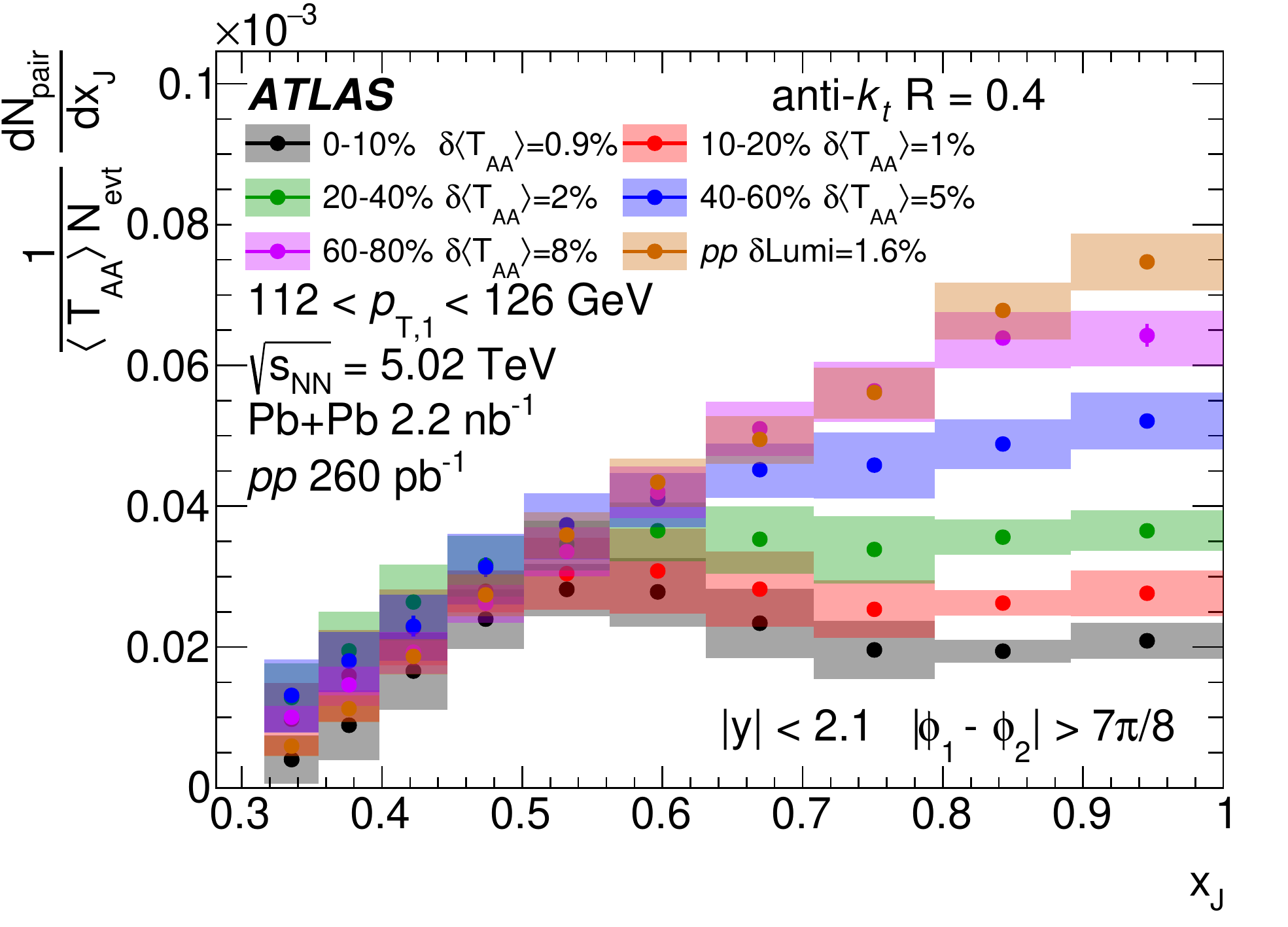}
}
\subfloat{
\centering
\includegraphics[width=0.33\textwidth]{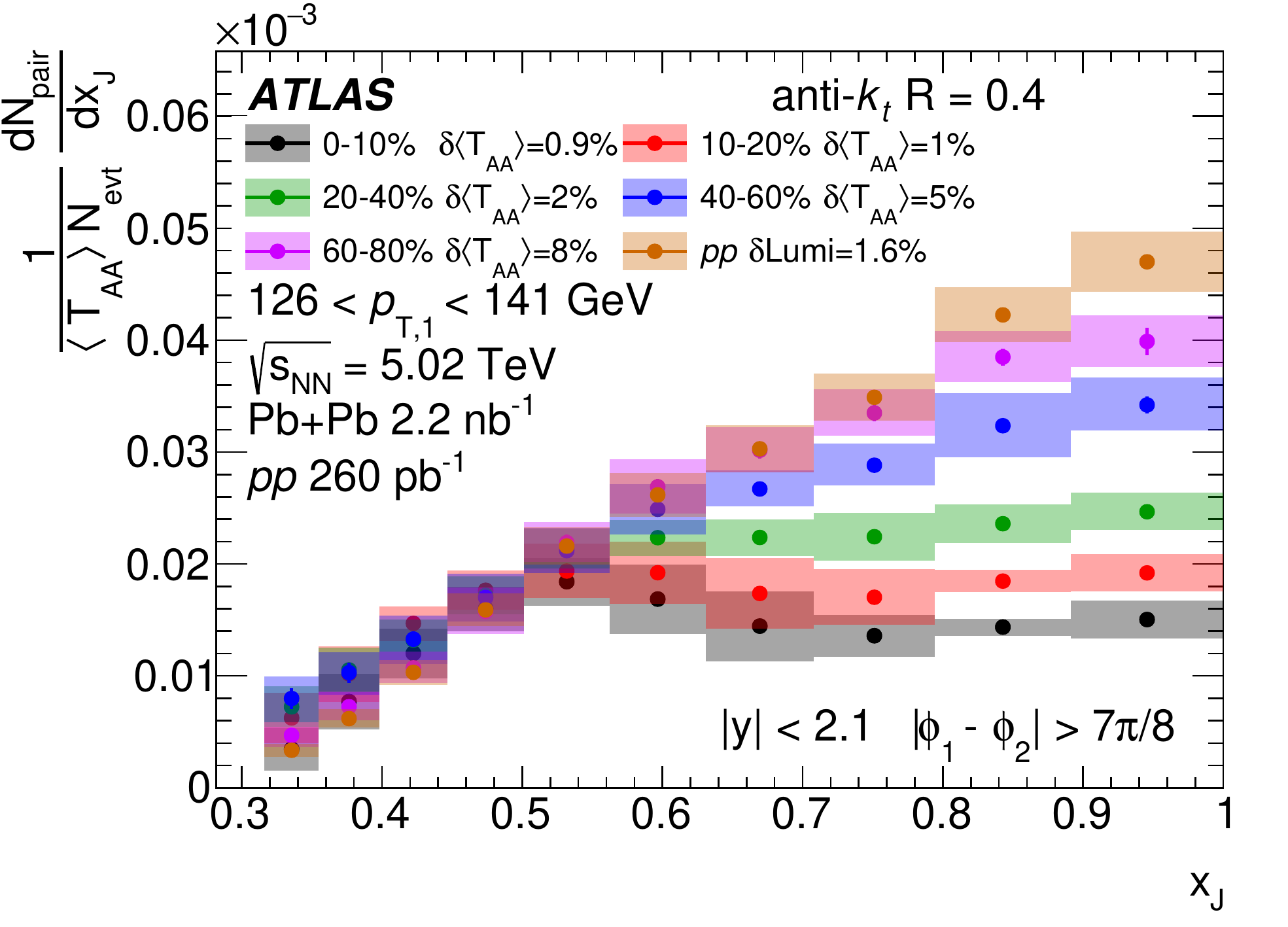}
}
\hfill

\subfloat{
\centering
\includegraphics[width=0.33\textwidth]{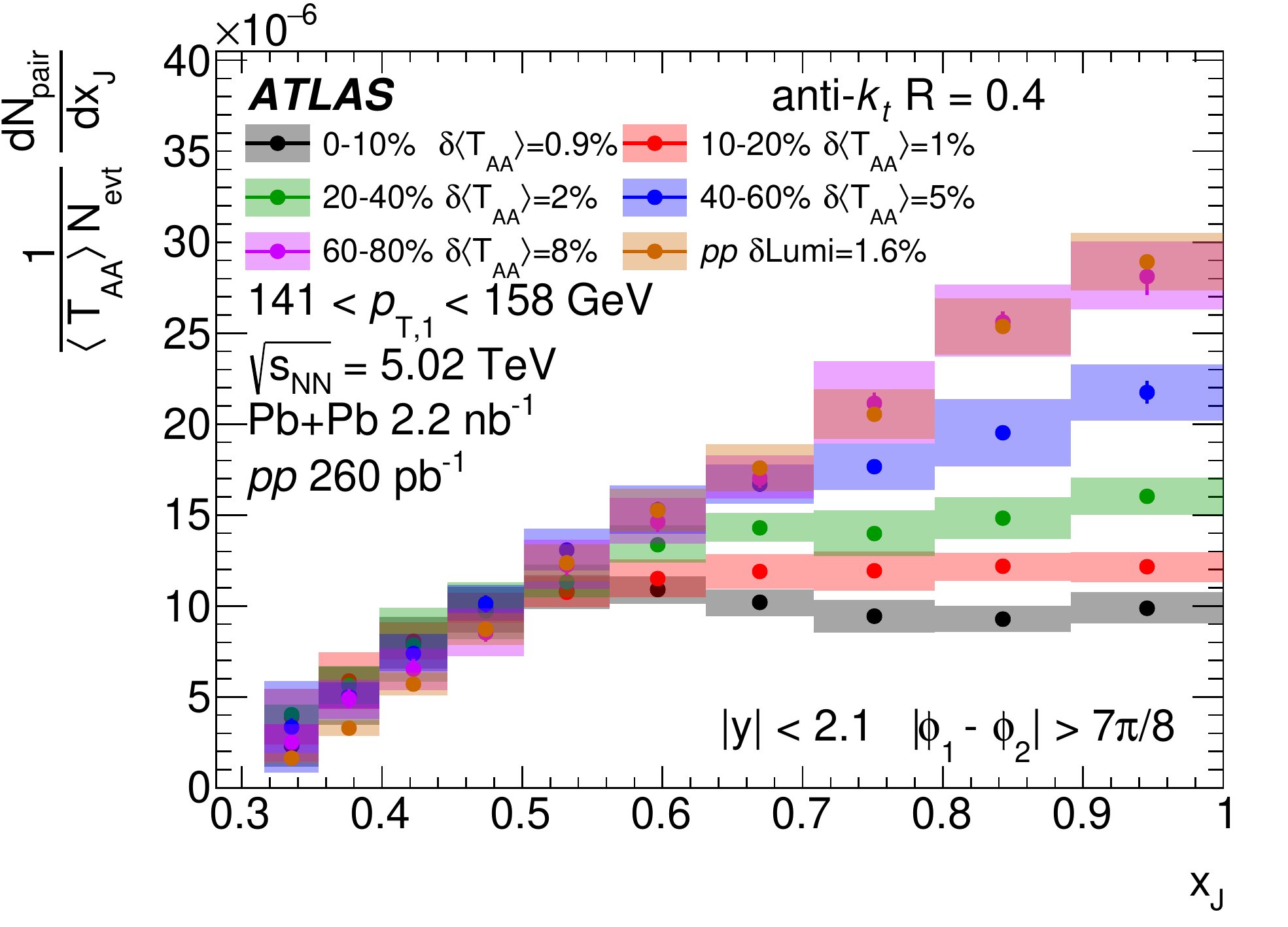}
}
\subfloat{
\centering
\includegraphics[width=0.33\textwidth]{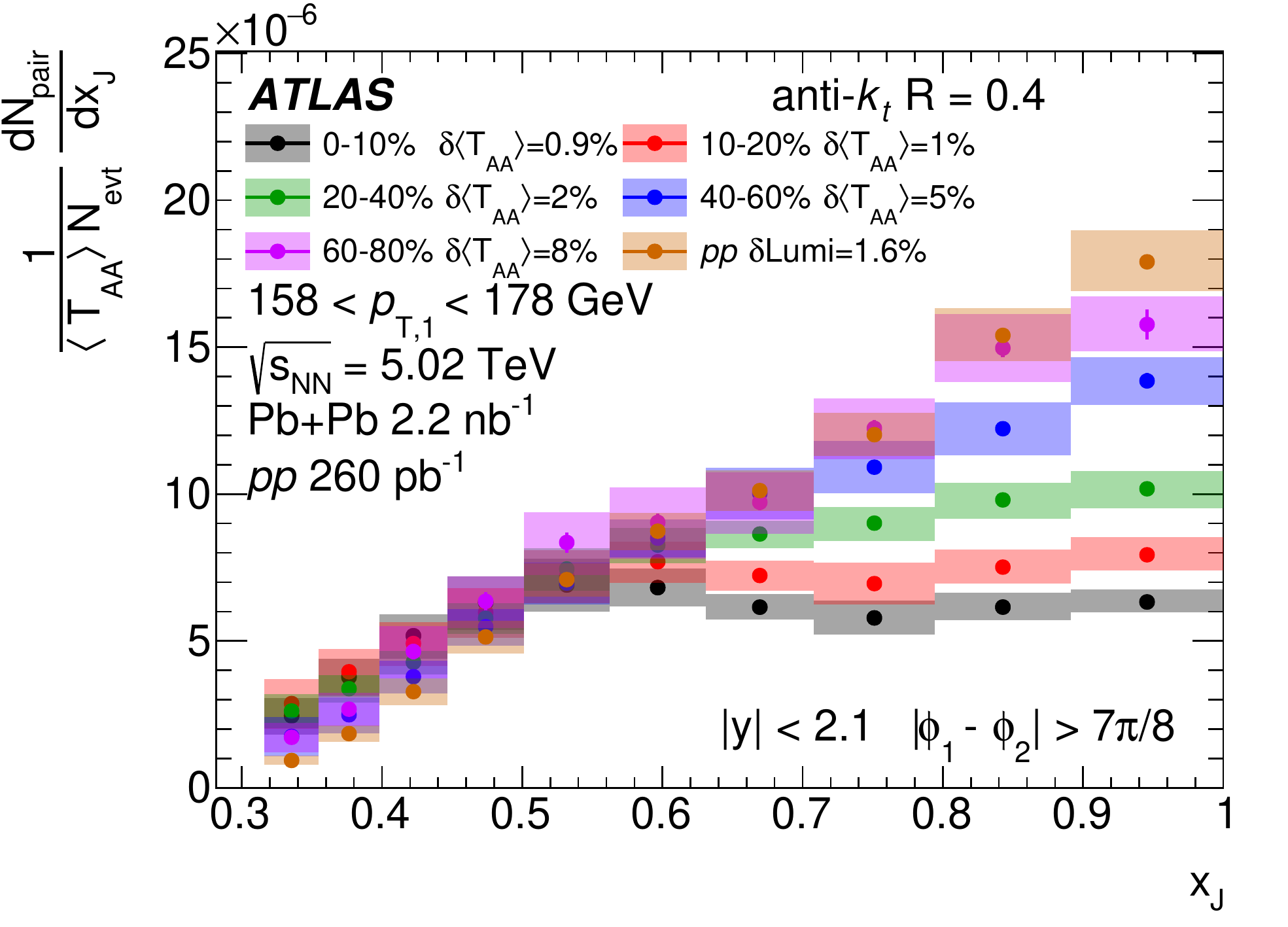}
}
\subfloat{
\centering
\includegraphics[width=0.33\textwidth]{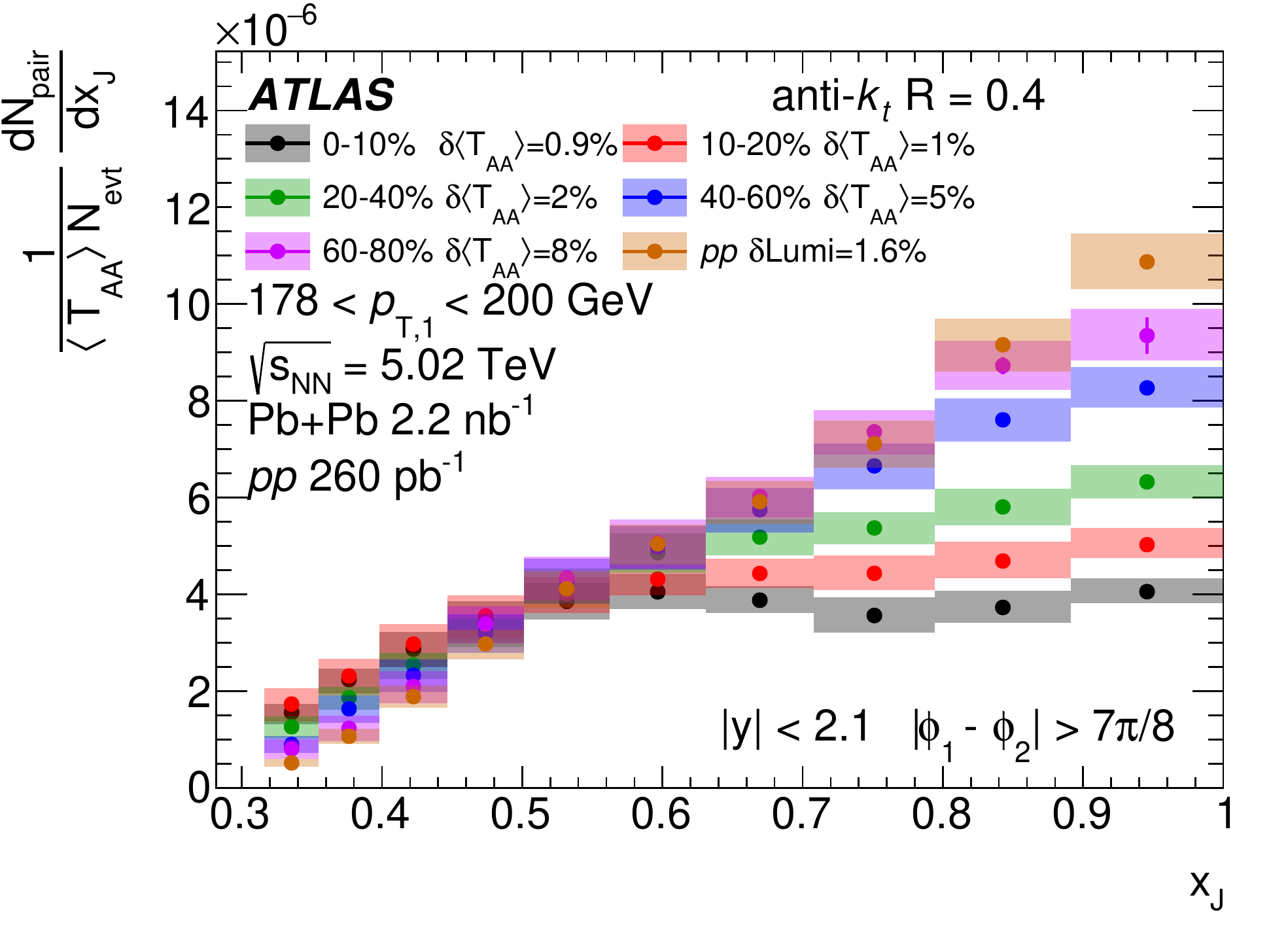}
}
\hfill
\subfloat{
\centering
\includegraphics[width=0.33\textwidth]{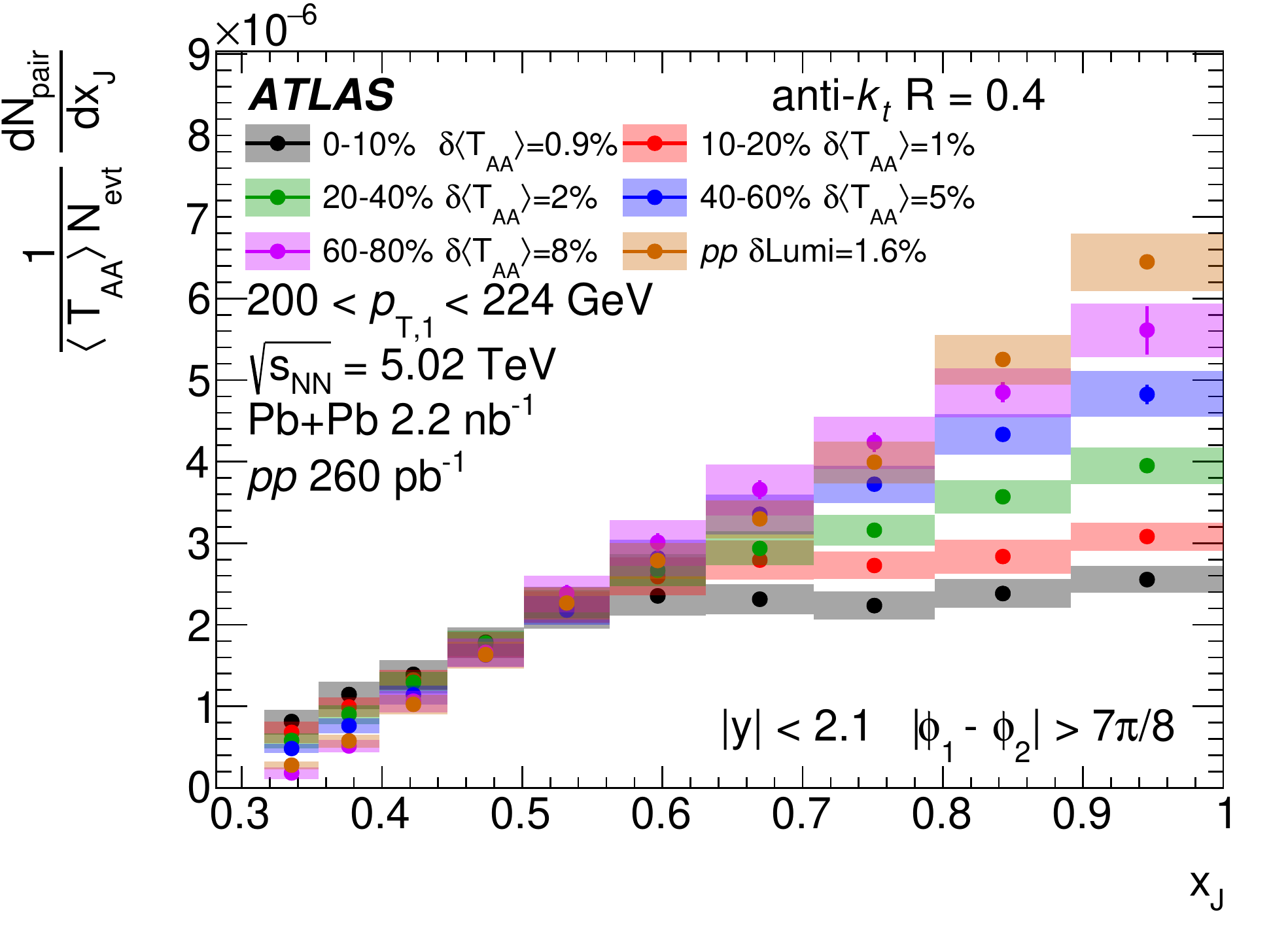}
}
\subfloat{
\centering
\includegraphics[width=0.33\textwidth]{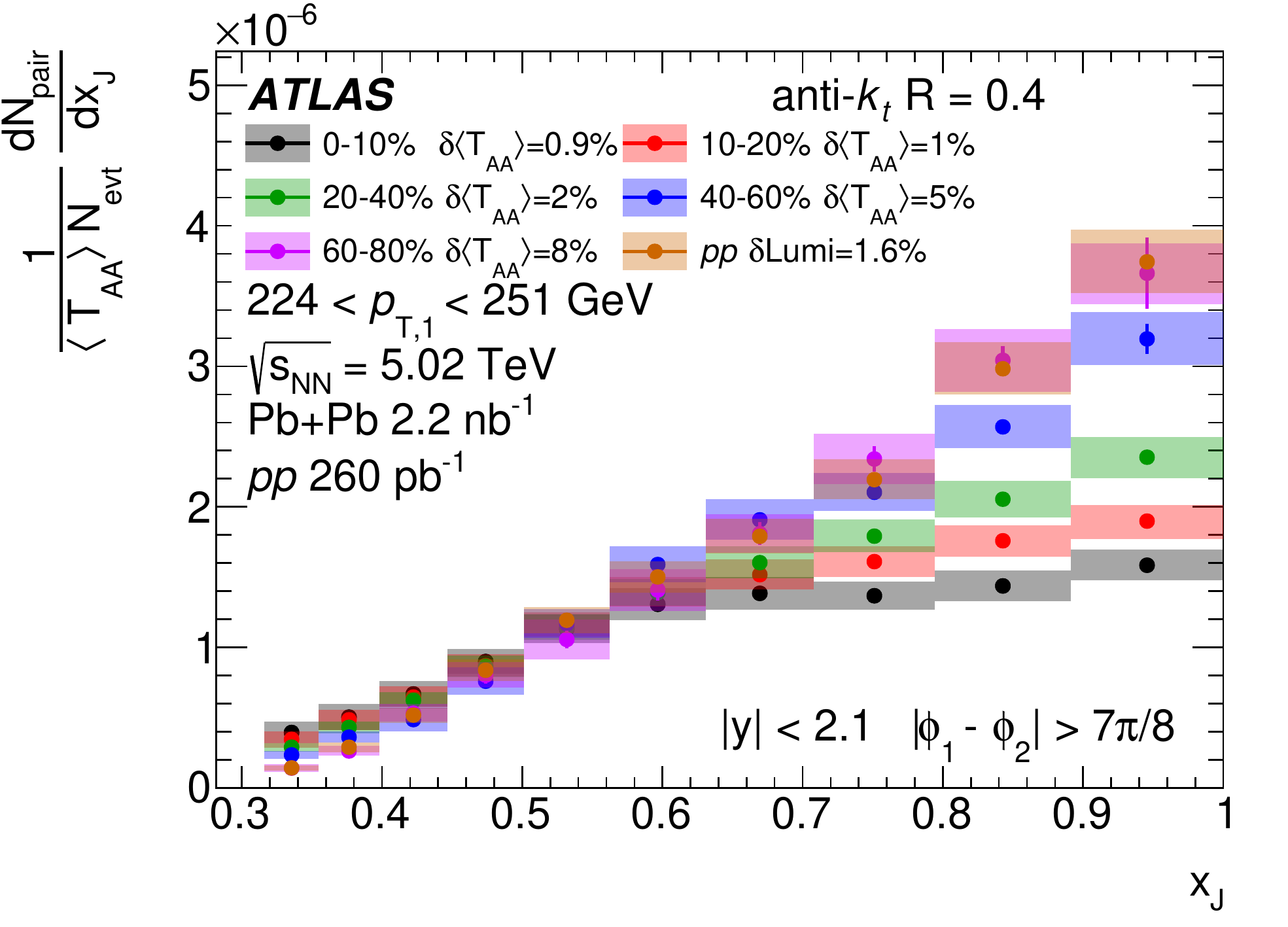}
}
\subfloat{
\centering
\includegraphics[width=0.33\textwidth]{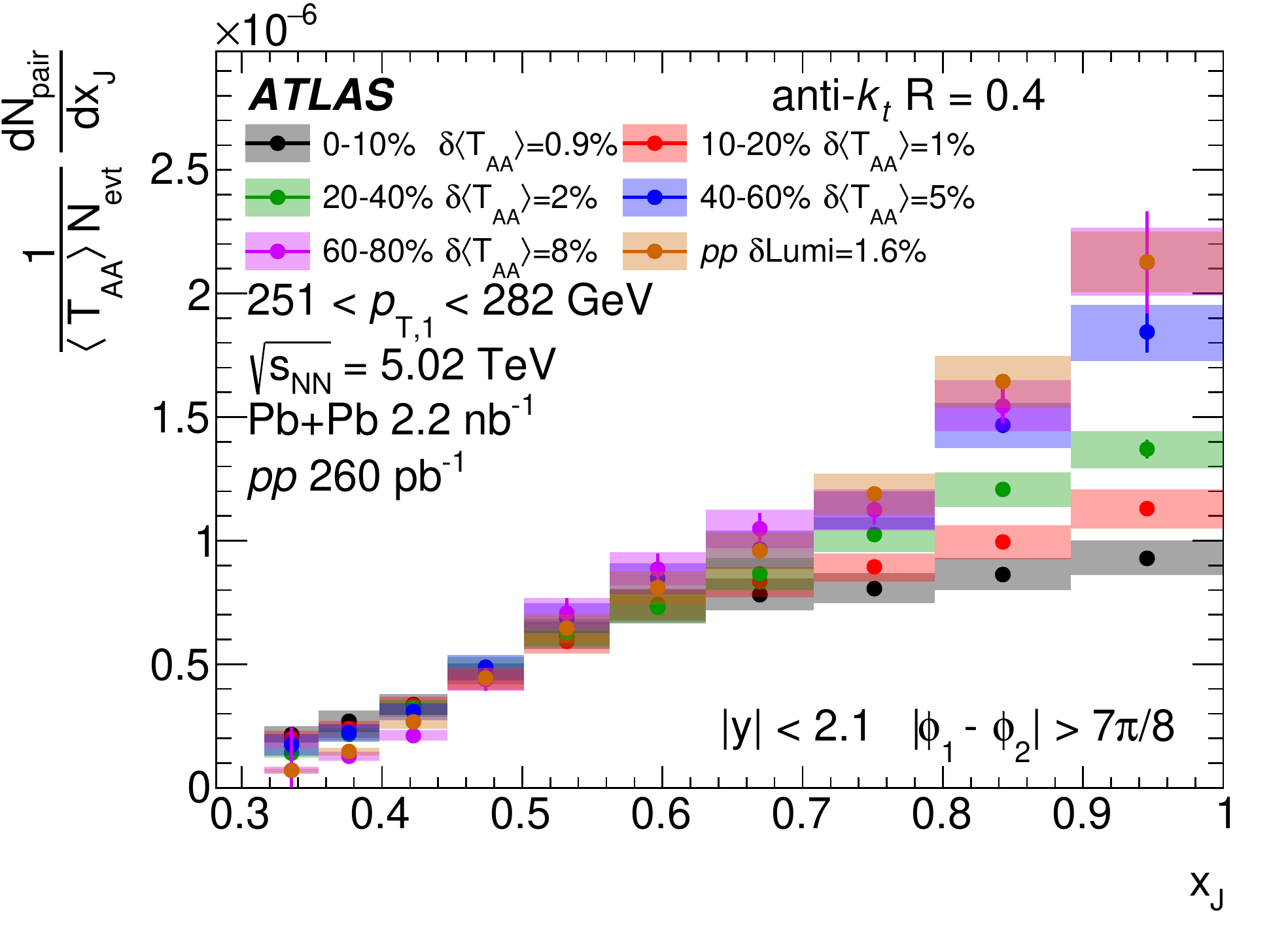}
}
\hfill
\subfloat{
\centering
\includegraphics[width=0.33\textwidth]{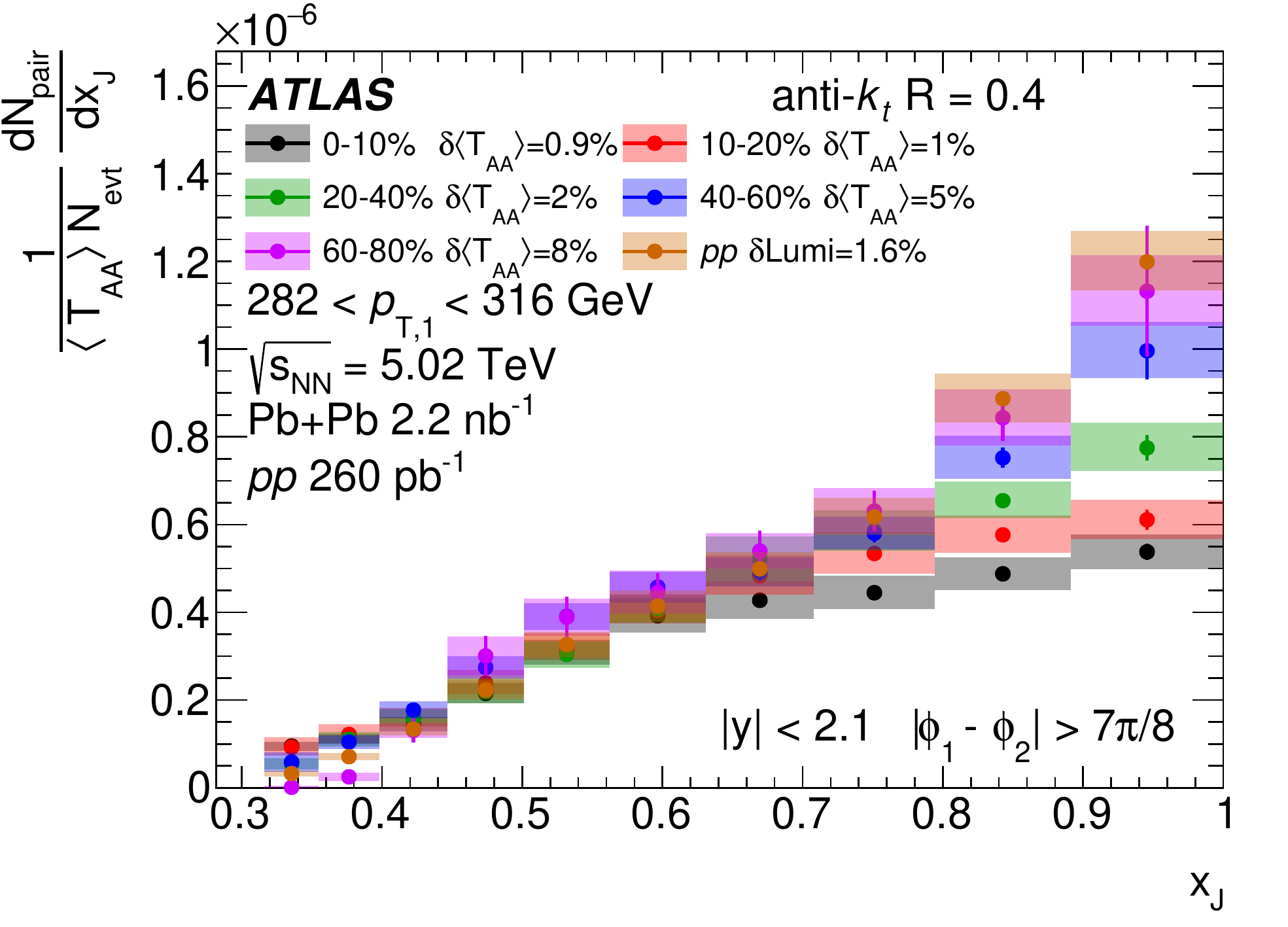}
}
\subfloat{
\centering
\includegraphics[width=0.33\textwidth]{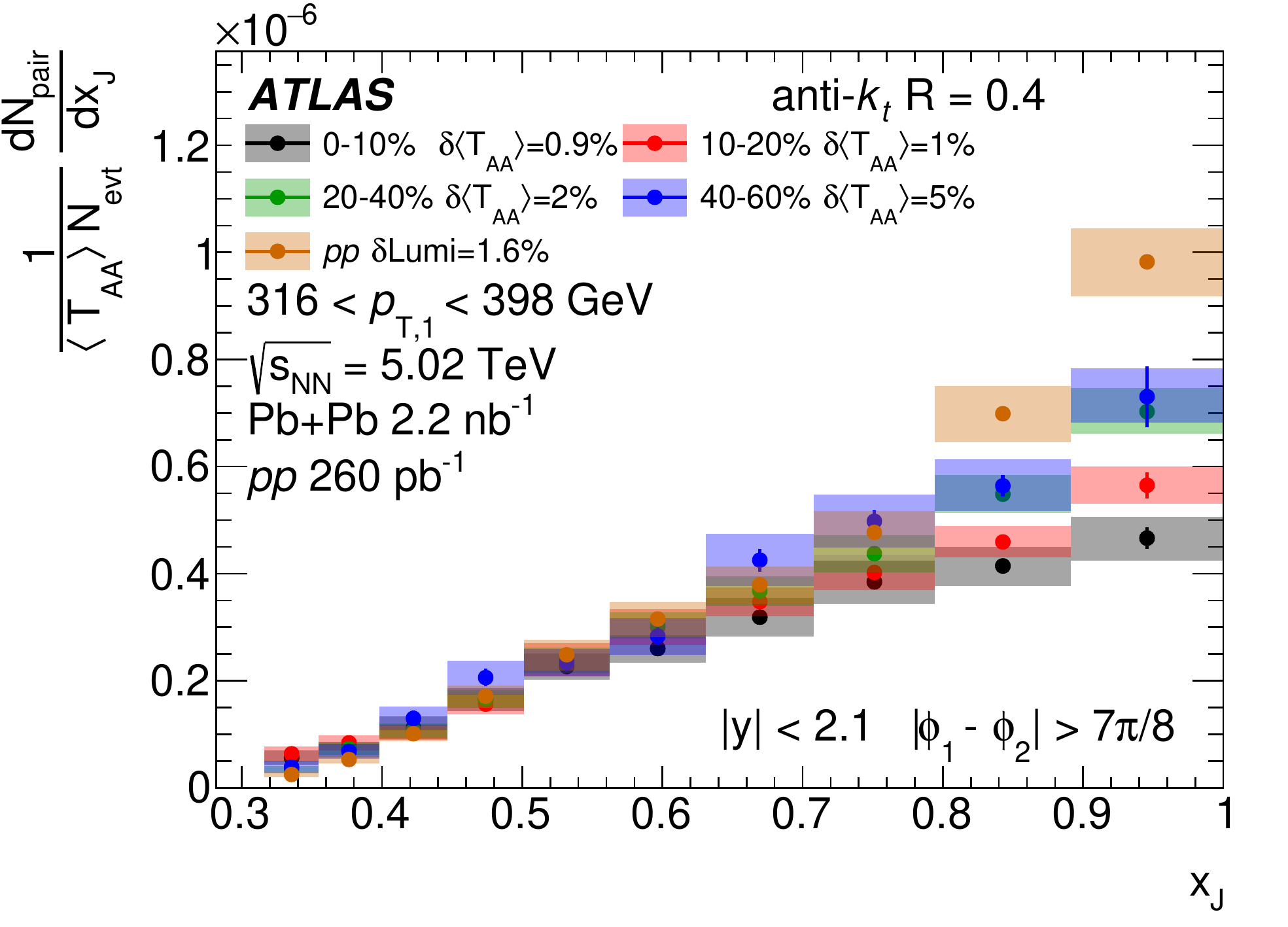}
}
\subfloat{
\centering
\includegraphics[width=0.33\textwidth]{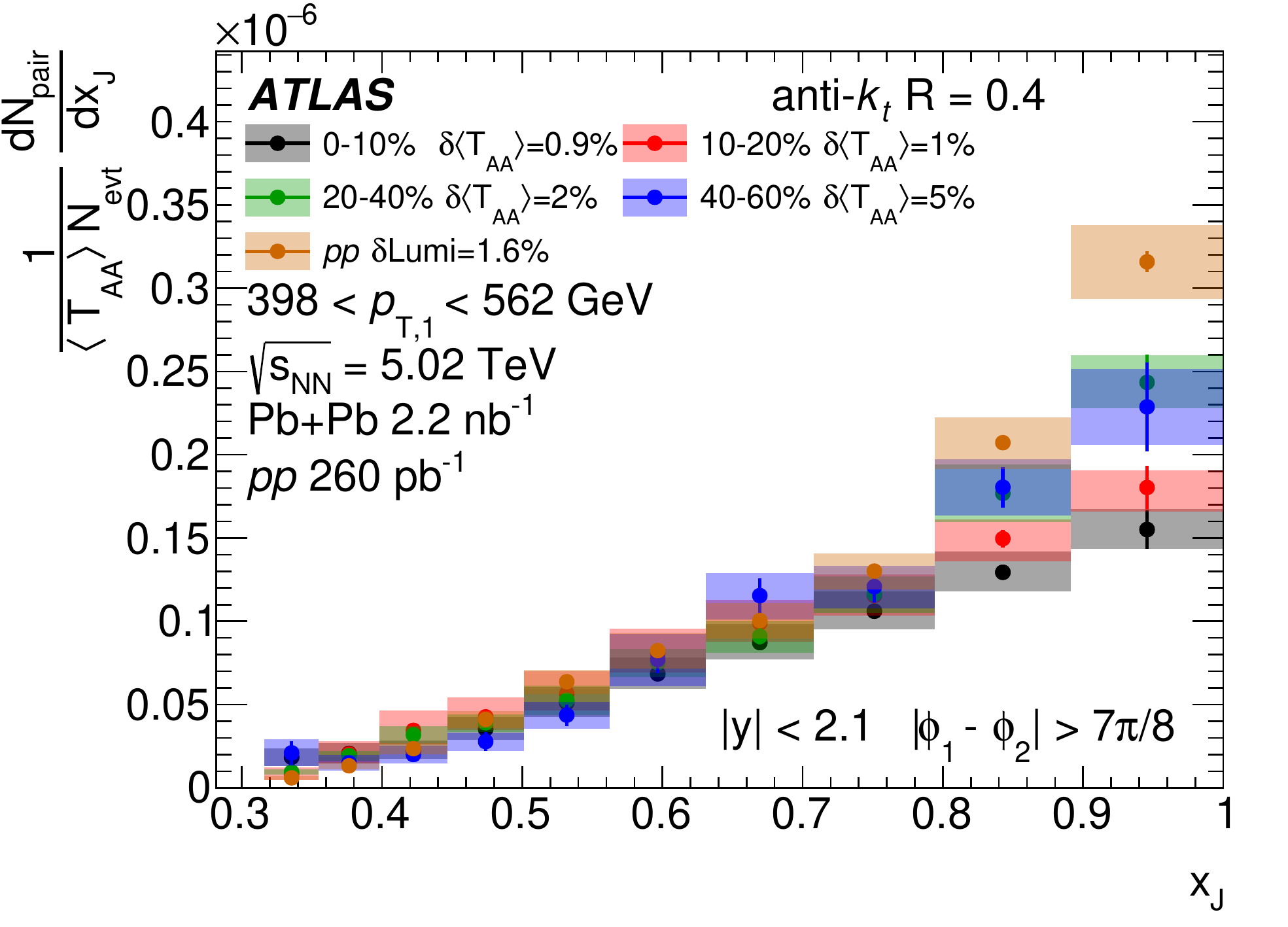}
}
\hfill
\caption{\label{fig:aux_per_event_xJ} Comparison of the absolutely normalized \xJ\ distributions in \pp\ collisions and five centrality selections in \pbpb\ collisions.}
\end{figure}
Figure~\ref{fig:aux_per_event_xJ} shows absolutely normalized \xJ\ distributions for five
centrality intervals for both \pbpb\ and \pp\ collisions for 12 selections of \pta\ between 100 and 562~\GeV.
\begin{figure}[ht]
\centering
\subfloat{
\centering
\includegraphics[width=0.6\textwidth]{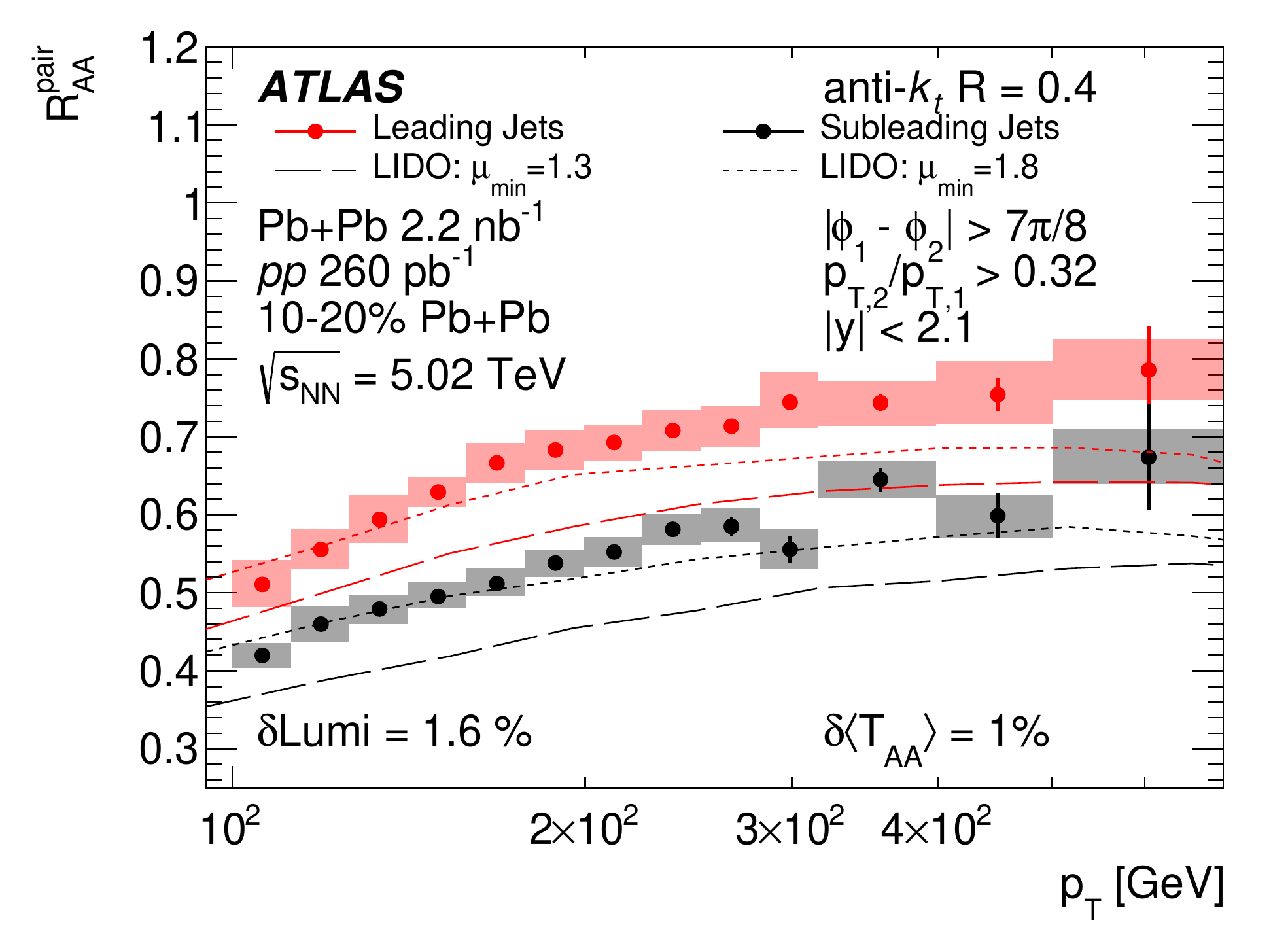}
}
\hfill
 
\subfloat{
\centering
\includegraphics[width=0.6\textwidth]{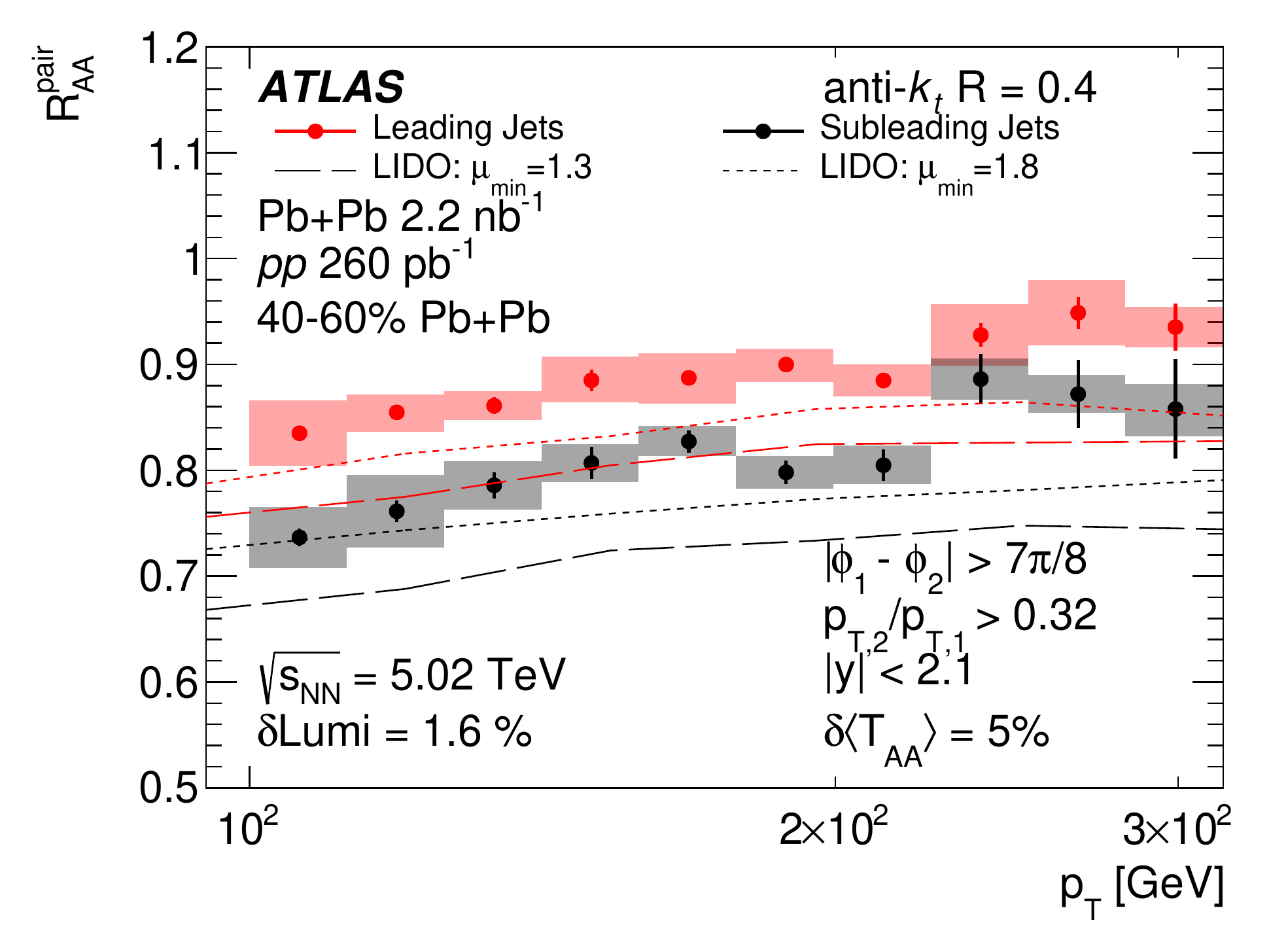}
}
 
\caption{\label{fig:aux_leadsublead_raa}The \RAAlead\ and \RAAsublead\
distributions measured for \pbpb\ centrality selections
compared with predictions from LIDO~\cite{Ke:2020clc}. }
\end{figure}
Finally, Figure~\ref{fig:aux_leadsublead_raa} provides \RAAlead\ and \RAAsublead\
distributions measured in two additional centrality intervals.


\clearpage
 
\printbibliography
 
\clearpage
 
\begin{flushleft}
\hypersetup{urlcolor=black}
{\Large The ATLAS Collaboration}

\bigskip

\AtlasOrcid[0000-0002-6665-4934]{G.~Aad}$^\textrm{\scriptsize 101}$,
\AtlasOrcid[0000-0002-5888-2734]{B.~Abbott}$^\textrm{\scriptsize 119}$,
\AtlasOrcid[0000-0002-7248-3203]{D.C.~Abbott}$^\textrm{\scriptsize 102}$,
\AtlasOrcid[0000-0002-1002-1652]{K.~Abeling}$^\textrm{\scriptsize 55}$,
\AtlasOrcid[0000-0002-8496-9294]{S.H.~Abidi}$^\textrm{\scriptsize 29}$,
\AtlasOrcid[0000-0002-9987-2292]{A.~Aboulhorma}$^\textrm{\scriptsize 35e}$,
\AtlasOrcid[0000-0001-5329-6640]{H.~Abramowicz}$^\textrm{\scriptsize 150}$,
\AtlasOrcid[0000-0002-1599-2896]{H.~Abreu}$^\textrm{\scriptsize 149}$,
\AtlasOrcid[0000-0003-0403-3697]{Y.~Abulaiti}$^\textrm{\scriptsize 116}$,
\AtlasOrcid[0000-0003-0762-7204]{A.C.~Abusleme~Hoffman}$^\textrm{\scriptsize 136a}$
% ...
\bigskip
\\

% ...
$^{29}$Physics Department, Brookhaven National Laboratory, Upton NY; United States of America.\\
% ...
$^{35}$$^{(a)}$Facult\'e des Sciences Ain Chock, R\'eseau Universitaire de Physique des Hautes Energies - Universit\'e Hassan II, Casablanca;$^{(b)}$Facult\'{e} des Sciences, Universit\'{e} Ibn-Tofail, K\'{e}nitra;$^{(c)}$Facult\'e des Sciences Semlalia, Universit\'e Cadi Ayyad, LPHEA-Marrakech;$^{(d)}$LPMR, Facult\'e des Sciences, Universit\'e Mohamed Premier, Oujda;$^{(e)}$Facult\'e des sciences, Universit\'e Mohammed V, Rabat;$^{(f)}$Institute of Applied Physics, Mohammed VI Polytechnic University, Ben Guerir; Morocco.\\
% ...
$^{55}$II. Physikalisches Institut, Georg-August-Universit\"{a}t G\"ottingen, G\"ottingen; Germany.\\
% ...
$^{101}$CPPM, Aix-Marseille Universit\'e, CNRS/IN2P3, Marseille; France.\\
$^{102}$Department of Physics, University of Massachusetts, Amherst MA; United States of America.\\
% ...
$^{116}$Department of Physics, New York University, New York NY; United States of America.\\
% ...
$^{119}$Homer L. Dodge Department of Physics and Astronomy, University of Oklahoma, Norman OK; United States of America.\\
% ...
$^{136}$$^{(a)}$Departamento de F\'isica, Pontificia Universidad Cat\'olica de Chile, Santiago;$^{(b)}$Millennium Institute for Subatomic physics at high energy frontier (SAPHIR), Santiago;$^{(c)}$Instituto de Investigaci\'on Multidisciplinario en Ciencia y Tecnolog\'ia, y Departamento de F\'isica, Universidad de La Serena;$^{(d)}$Universidad Andres Bello, Department of Physics, Santiago;$^{(e)}$Instituto de Alta Investigaci\'on, Universidad de Tarapac\'a, Arica;$^{(f)}$Departamento de F\'isica, Universidad T\'ecnica Federico Santa Mar\'ia, Valpara\'iso; Chile.\\
% ...
$^{149}$Department of Physics, Technion, Israel Institute of Technology, Haifa; Israel.\\
$^{150}$Raymond and Beverly Sackler School of Physics and Astronomy, Tel Aviv University, Tel Aviv; Israel.\\
% ...
$^{a}$ Also Affiliated with an institute covered by a cooperation agreement with CERN.\\
$^{b}$ Also at Borough of Manhattan Community College, City University of New York, New York NY; United States of America.\\
$^{c}$ Also at Bruno Kessler Foundation, Trento; Italy.\\
$^{d}$ Also at Center for High Energy Physics, Peking University; China.\\
$^{e}$ Also at Centro Studi e Ricerche Enrico Fermi; Italy.\\
$^{f}$ Also at CERN, Geneva; Switzerland.\\
% ...

\end{flushleft}


\end{document}